\documentclass[a4paper,12pt]{article}
\usepackage[left=1in,top=.8in,right=.8in,bottom=.7in]{geometry}
\usepackage{amsmath}
\usepackage[utf8]{inputenc}
\usepackage{amssymb}
\usepackage{verbatim}
\makeatletter
\newenvironment{subxarray}{%
	\vcenter\bgroup
	\Let@ \restore@math@cr \default@tag
	\baselineskip\fontdimen10 \scriptfont\tw@
	\advance\baselineskip\fontdimen12 \scriptfont\tw@
	\lineskip\thr@@\fontdimen8 \scriptfont\thr@@
	\lineskiplimit\lineskip
	\ialign\bgroup\hfil
	$\m@th\scriptstyle##$&$\m@th\scriptstyle{}##$\hfil\crcr
}{%
	\crcr\egroup\egroup
}
\makeatother
\usepackage{multicol} 
\usepackage{framed}
\usepackage{nomencl}
\usepackage{ifthen}

\usepackage{graphicx}
\graphicspath{ {Pictures/} }
\usepackage{wrapfig}
\usepackage{epstopdf}
\usepackage{subcaption}
\usepackage{helvet}
\usepackage{footmisc}
\numberwithin{equation}{section}
\usepackage{hyperref}
\hypersetup{colorlinks=true,
	linkcolor=blue,
	filecolor=magenta,      
	urlcolor=blue}
\allowdisplaybreaks
\usepackage{url}            
\usepackage{booktabs}       
\usepackage{amsfonts}       
\usepackage{nicefrac}       
\usepackage{microtype}      
\usepackage[square,numbers]{natbib}
\usepackage{doi}
\usepackage{authblk}

\title{Stokes’ Second Problem and Oscillatory Couette Flow for  a Two-Layer Fluid: Analytical Solutions}
\author[1]{Moslem Uddin}
\author[1]{Abdullah Murad\footnote{Corresponding author. Email: murad-math@cu.ac.bd}}
\affil[1]{Department of Mathematics, University of Chittagong, Chittagong-4331, Bangladesh}

\begin{document}
\maketitle

\newbox\keywbox
\setbox\keywbox=\hbox{\bfseries Keywords:}%

\newcommand\keywords{%
\noindent\rule{\wd\keywbox}{0.25pt}\\\textbf{Keywords:}\ }
\begin{abstract}
The unsteady motion of a two-layer fluid induced by oscillatory motion of a flat plate along its length is mathematically analyzed. Two cases are considered: (i) the two-layer fluid is bounded only by the oscillating plate (Stokes’ second problem), (ii) the two-layer fluid is confined between two parallel plates, one of which oscillates while the other is held stationary (oscillatory Couette flow). In each of the Stokes' and Couette cases, both cosine and sine oscillations of the plate are considered. It is assumed that the fluids are immiscible, and that the flat interface between the fluids remains flat for all times. Solutions to the initial-boundary value problems are obtained using the Laplace transform method. Steady periodic and transient velocity fields are explicitly presented. Transient and steady-state shear stresses at the boundaries of the flows are calculated. The results derived in this paper retrieve previously known results for corresponding single-layer flows. Further, illustrative example of each of the Stokes' problem and the Couette flow is presented and discussed. Again, the results obtained could also be applicable to a problem of heat conduction in a composite solid with sinusoidal temperature variation on the surface. 

\keywords Stokes' second problem, Oscillatory Couette flow Two-layer fluid, The Laplace transform method, Transient velocity, Steady-state velocity
\end{abstract}

\section{Introduction}
\label{s1}

In fluid mechanics, Stokes’ second problem refers to the motion of a semi-infinite viscous incompressible fluid induced by an oscillating flat plate\cite{currie,schilchting}. However, Zeng and Weinbaum\cite{zeng_weinbaum_1995} has called it as Stokes’ first problem. In the Stokes’ problem, the fluid is bounded only by the oscillating plate. Again, when the fluid is bounded by two parallel plates, one of which oscillates while the other is held stationary, the problem is termed Couette flow\cite{KHALED2004795}. The study of Stokes’ second problem finds its applications in fields such as chemical engineering, medical and biomedical sciences, biomechanics, micro- and nano-technology, geophysical flows, and heat conduction problems\cite{ai05,liu,nalim,TSO2018e01085}. It is worth mentioning here that Stokes’ second problem has its counterparts in problems: acoustic streaming past an oscillating body, and settled boundary layer with fluctuating incident fluid velocity\cite{panton_1968}. Fluid flow due to oscillatory motion of a plate has drawn attention of many researchers. Here we mention the works of several of them. Erdogan\cite{ERDOGAN20001} and Fetecau \textit{et al.}\cite{FETECAU2008451} have dealt with Stokes’ second problem in-depth. They have presented complete solution to the problem, obtained by the Laplace transform method. It is to be noted here that the complete solution contains transient and steady-state solutions. The original solution of the now-classical Stokes’ second problem contains only steady-state solution, not valid for small values of time $t$. Again, Khaled and Vafai\cite{KHALED2004795} have studied Stokes’ second problem and oscillatory Couette flow with slip boundary condition on the plate(or plates). All the works mentioned above are for Newtonian fluids. For non-Newtonian fluid, in the present context, the works of Rajagopal\cite{rajagopal}, Ai and Vafai\cite{ai05}, and Asghar \textit{et al.}\cite{ASGHAR200275} are worth mentioning, among others. The object of the present paper is to explore the flow of a two-layer fluid induced by oscillatory motion of a flat plate. We consider that both the fluids are Newtonian. The two fluids are of different viscosities, densities, and thickness.

Flow of a two-layer fluid occurs in chemical engineering, lubricated piping, lithographic printing, and oil industry\cite{wang17,ng_wang,lenz_kumar_2007,sellier_lenz_2010,cosgrove_forbes_2012,Brauner2003,tf}. Most of the works in the literature that concern two-layer fluid flows have considered flows due to pressure gradient ( Poiseuille flows). Two-layer Poiseuille flows between parallel plates due to constant or time-dependent pressure gradient have been studied by Bird \textit{et. al.}\cite{tp}, Kapur and Shukla\cite{kapur}, Bhattacharyya\cite{bt} and Wang\cite{WANG2011032007}. Recently, Wang\cite{wang17} and Ng and Wang\cite{ng_wang} have examined starting Poiseuille flows of a two-layer fluid in a channel and in a circular tube, respectively. On the other hand, works studying a gravity-driven or shear-driven (wall-driven) two-layer flow are, so far as we are aware, only a few in the literature.

Panton\cite{doi:10.1002/9781118713075.ch7} has discussed the flow of two films of immiscible fluids due to gravity along an inclined plane. Again, Papanastasiou \textit{et al.}\cite{vff} have discussed a two-layer flow between parallel plates where the motion is caused by uniform motion of one of the plates. Recently, Ng\cite{Ng2017} has investigated change of Navier slip length with respect to time in starting flows using a two-layer flow model between parallel plates, where the flow is caused by impulsive motion of one of the plates along its length. In fact, the flow model is an extension of classical Stokes’ first problem\cite{schilchting} for a single-layer fluid to the case of unsteady Couette flow of a two-layer fluid due to sudden motion of one of the plates. In order to obtain the velocity fields for the two layers of fluids, he has utilized the result given in\cite{carslaw} concerning heat conduction in a composite solid.

The works mentioned above have motivated us to discuss the present problems concerning the flow of a two-layer fluid caused by oscillatory motion of a flat plate. We believe the current study will help further our understanding of the flow of a two-layer fluid caused by oscillatory motion of a wall in an engineering application. Here we note what follows. The current study could be applicable to a case where the interface is flat or the deviation of the interface from flat shape is small. Moreover, the analytical results presented in this paper may be used for validation purpose of future numerical works dealing with problems similar to the current ones but consider wavy interface between the fluids. Note that in the current study we have considered flat interface between the fluids. Again, this work may provide a basis for future researches on Stokes’ second problem and oscillatory Couette flow for two-layer fluids where one or both the  fluids are non-Newtonian. Furthermore, the current study is also applicable to a problem of heat conduction in a composite solid with the conditions as follows. The composite solid is initially at a uniform zero temperature and then suddenly, the surface of the solid comes into contact with a heat source with sinusoidal temperature variation. Note that there is an analogy between viscous diffusion in liquids and unsteady heat conduction in solids. Note also that a composite solid is formed by attaching together slaps of two different materials. Relevantly, Carslaw and Jaeger\cite{carslaw} have studied heat conduction in semi-infinite and finite composite solids where in each of the cases the surface of the solid suddenly comes into contact with a heat source with constant temperature of certain amount. They have also examined heat-conduction in a single-layer solid where the surface of the solid suddenly comes into contact with a heat source with sinusoidal temperature variation. It is worth mentioning here that Parasnis\cite{parasnis} has investigated steady-state heat conduction in a semi-infinite composite solid where the surface temperature varies sinusoidally with time $t$. But, as far as we are aware, the literature lacks any study of heat conduction in a finite composite solid with sinusoidal temperature variation on the surface. Further, the literature lacks any exhaustive work dealing with heat conduction in a semi-infinite composite solid with sinusoidal temperature variation on the surface.

\begin{figure}[h]
	\centering
	\includegraphics[width=0.7\linewidth]{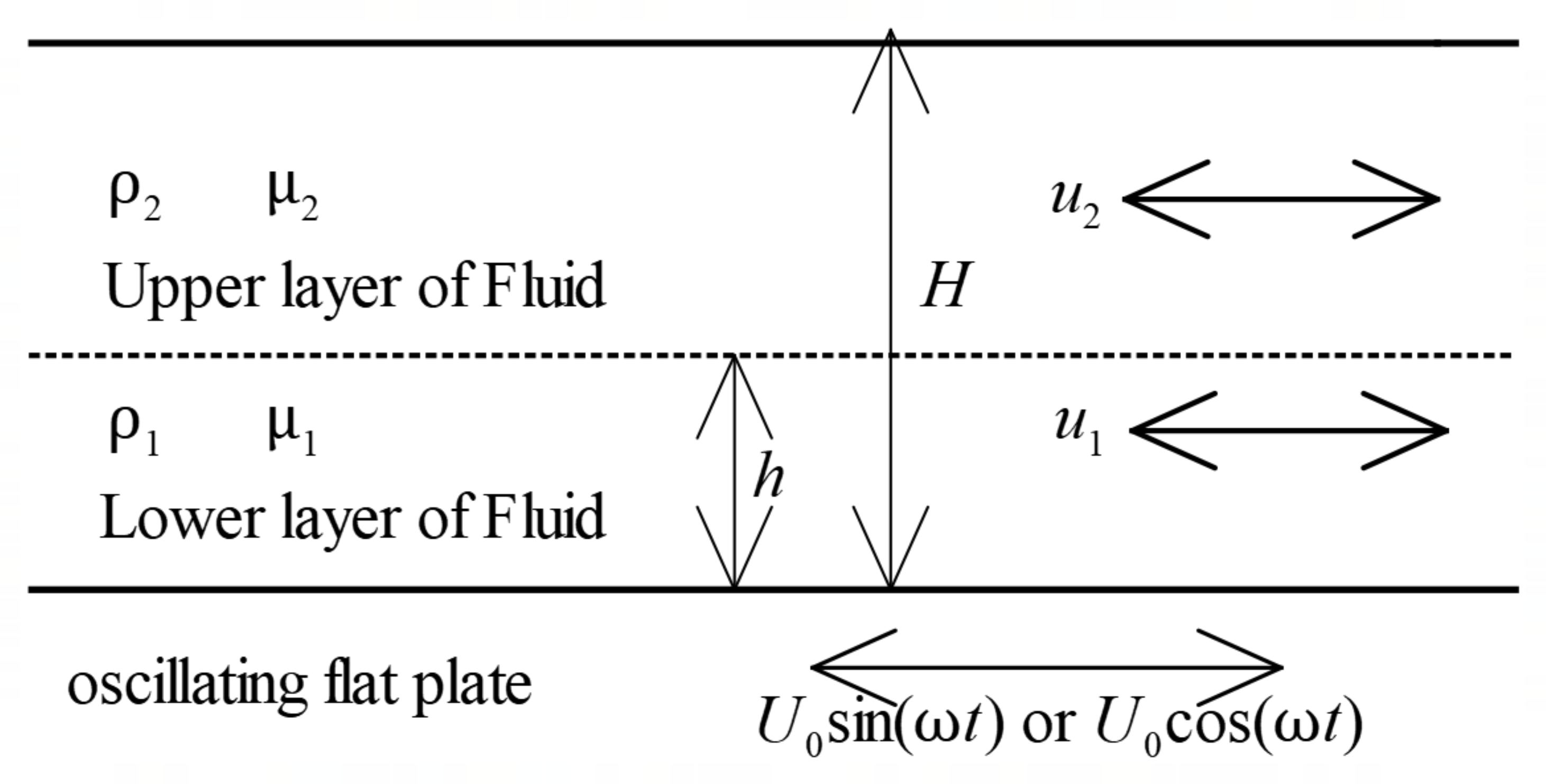}
	\caption{Schematic diagram for oscillatory Couette flow for a two-layer fluid. The solid lines represent the plates, and the broken line is the interface of the fluids.}
	
	\label{f9c}
\end{figure}

In this work, we derive exact solutions for two cases of unsteady motion of a two-layer fluid induced by sinusoidal oscillation of a flat plate. We consider two cases: (i) the two-layer fluid is bounded only by the oscillating plate (Stokes’ second problem) (see sketch in Fig. \ref{f9}), (ii) the two-layer fluid is confined between two parallel plates, one of which oscillates while the other is held stationary (oscillatory Couette flow)  (see sketch in Fig. \ref{f9c}). In each of the cases, we consider both cosine and sine oscillations of the plate. The fluids of the two layers have different viscosities, densities, and thickness. We assume that the fluids are immiscible, and that the flat interface of the fluids is stable. We utilize the Laplace transform method to solve the initial-boundary value problems related to the two cases mentioned here. For both the cases, we present analytical results for velocity fields for starting and steady periodic flows. The result for a starting flow is the sum of transient solution and steady-state solution and valid for small values of time $t$. The transient disappears gradually (or rapidly) as time progresses. Whereas the steady-state solution represents the time periodic motion of the fluid and is valid for large values of time $t$. We calculate transient and steady-state shear-stresses at the boundaries of the flows.  We recover related previously known results for single-layer flows from the results obtained in this study. We present and discuss illustrative example of each of the Stokes' problem and the Couette flow.

The remaining part of the paper is organized into four sections. Section \ref{s2} deals with the Stokes’ second problem case, while section \ref{s3} concerns the case of oscillatory Couette flow. Section \ref{s4} presents results and illustrative examples. And section \ref{s5} concludes the paper.
\section{Stokes’ second problem for a two-layer fluid}
\label{s2}
\subsection{Mathematical Formulation}
Consider two superposed  layers of two immiscible fluids of different viscosities and densities over a flat plate that coincides with the $x$-$z$ plane of the Cartesian co-ordinate system $(x,y,z)$. Suppose that the lower fluid occupies the region $0\le y\le h$, $h$ being a positive real number. And the upper fluid fills the region $h\le y<\infty$. The $y$-axis is the coordinate normal to the plate. We consider that the fluids and the plate are initially at rest and then the plate starts to oscillate parallel to itself, along $x$-axis, with velocity $U_0\cos(\omega t)$ or $U_0\sin(\omega t)$, where $U_0$, $\omega$, and $t$ being the plate velocity amplitude, frequency of oscillations, and the time, respectively. We assume that the flow is two-dimensional, and there is no body force. The motion of fluids  is only due to oscillatory motion of the plate. The velocity fields for the lower and upper fluids are governed by the reduced Navier-Stokes equations:
\begin{figure}[t]
	\centering
	\includegraphics[width=0.7\linewidth]{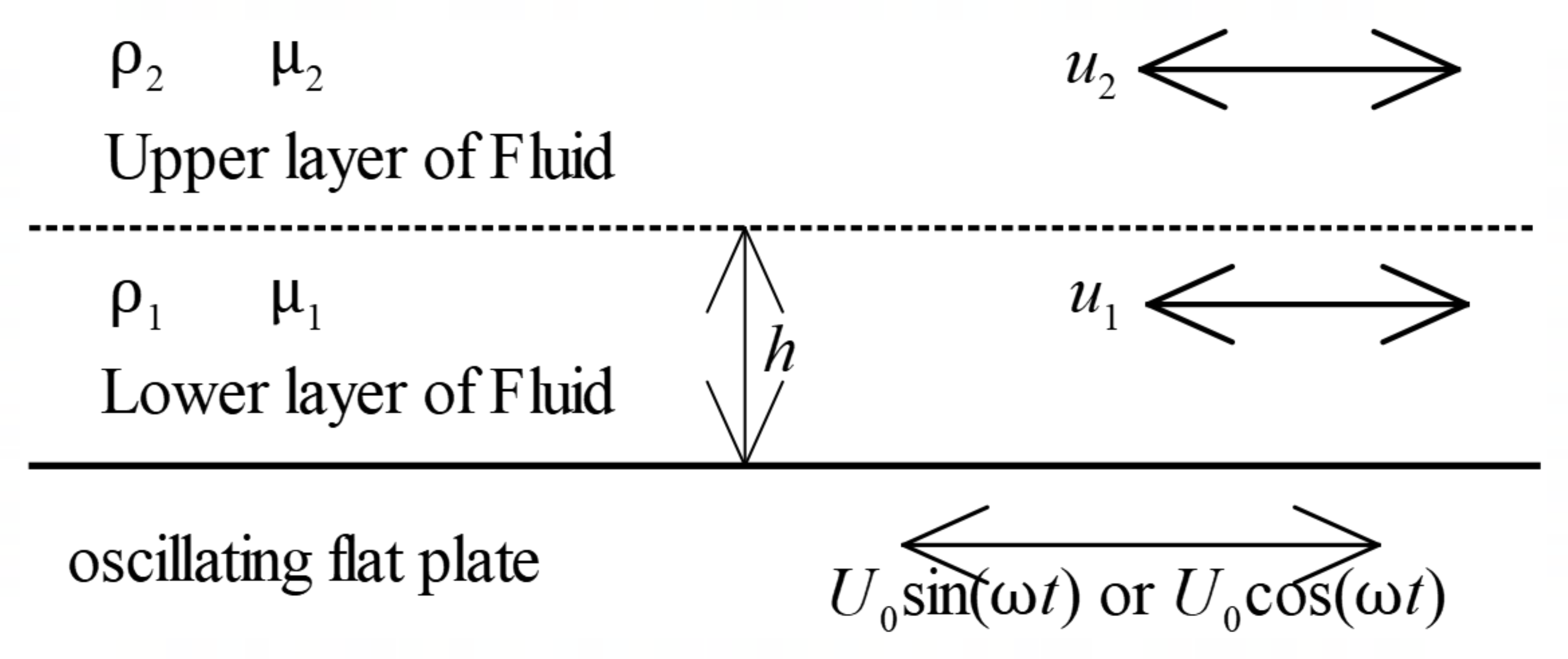}
	\caption{Schematic diagram for Stokes' second problem for a two-layer fluid. The solid line represents the plate, and the broken line is the interface of the fluids.}
	
	\label{f9}
\end{figure}
\begin{align}
\frac{\partial u_1}{\partial t}&=\nu_1\frac{\partial^2 u_1}{\partial y^2},\label{421} \\
\frac{\partial u_2}{\partial t}&=\nu_2\frac{\partial^2 u_2}{\partial y^2}\label{422},
\end{align}
respectively, where $\nu_1$ is the kinematic viscosity of the lower fluid, and $\nu_2$ is that for the upper fluid. The kinematic viscosity of a fluid  is defined by $\nu=\dfrac{\mu}{\rho}$, where,  $\rho$  is the fluid density, and $\mu$ is the  dynamic viscosity or simply the viscosity of the fluid. We write $\mu_1$ and $\rho_1$ for dynamic viscosity and density of the lower fluid, and $\mu_2$ and $\rho_2$ for the corresponding quantities of the upper fluid. In equations \eqref{421} and \eqref{422}, $u_1(y,t)$ and $u_2(y,t)$ are velocities in the $x$-direction. Here, we consider no-slip boundary condition at the plate. Further, we consider continuity of velocity and that of shear stress at the interface of the fluids. Accordingly, the associated initial and boundary conditions are:
\refstepcounter{equation}\label{eqn1}%
\refstepcounter{equation}\label{eqn2}%
\begin{align}
&u_1(0,t)=U_0\cos(\omega t) \quad\text{or}\quad u_1(0,t)=U_0\sin(\omega t)\qquad\text{for $t> 0$}\tag{\ref*{eqn1},\,\ref*{eqn2}},\label{423}\\
&u_1(y,0)=0,\label{424}\\
&u_2(y,0)=0,\label{425}\\
&u_1(h,t)=u_2(h,t),\label{426}\\
&\mu_1\frac{\partial u_1}{\partial y}\Big\vert_{y=h}=\mu_2\frac{\partial u_2}{\partial y}\Big\vert_{y=h},\label{427}\\
&u_2(y\rightarrow\infty,t)=0.\label{428}
\end{align}
\subsection{Solution}
\subsubsection{Solution for the cosine oscillations of the plate}\label{221}
\subsubsection*{Calculation of velocity fields}
In order to obtain the velocity fields for the lower and upper fluids, we need to find solution to the initial-boundary value problem consists of governing equations \eqref{421} and \eqref{422} and initial and boundary conditions \eqref{eqn1}, and \eqref{424}-\eqref{428}. 

We employ the Laplace transform method to solve the mathematical problem. The method provides complete, analytical solution to a initial-boundary value problem that is valid for small and large values of time $t$. The Laplace transform of a given function $u(y,t)$ is defined by
\begin{equation}
\mathcal{L}(u(y,t))=\tilde{u}(y,s)=\int_{0}^{\infty}u\exp(-s t) dt,
\end{equation}
where $s$ is the transform variable and $\exp(-st)$ is the kernel of the transform. For time $t>0$, the transform may be inverted using the following inversion formula: 
\begin{equation}
	u(y,t)=\mathcal{L}^{-1}(\tilde{u}(y,s))=\dfrac{1}{2\pi i}\int_{\gamma-i\infty}^{\gamma+i\infty}\tilde{u}(y,s)\exp(s t) ds,
\end{equation}
where $\gamma$ is an arbitrary constant, and it must be greater than the real part of each of the singularities of $\tilde{u}(y,s)$.

We take the Laplace transforms of equations \eqref{421} and \eqref{422}, yielding
\begin{align}
&\tilde{u_1}^{\prime\prime}-\dfrac{s}{\nu_1}\tilde{u_1}=0,\label{1a}\\
&\tilde{u_2}^{\prime\prime}-\dfrac{s}{\nu_2}\tilde{u_2}=0,\label{1b}
\end{align}
respectively. Note that the initial conditions \eqref{424} and \eqref{425} have been utilized to obtain the transforms. Here, primes stand for differentiation with respect to $y$. The transformations of the boundary conditions \eqref{eqn1}, and \eqref{426}-\eqref{428} result in
\begin{align}
&\tilde{u_1}(0,s)=U_0\dfrac{s}{s^2+\omega^2},\label{423b}\\
&\tilde{u_1}(h,s)=\tilde{u_2}(h,s),\label{426b}\\
&\mu_1\tilde{u_1}^{\prime}(h,s)=\mu_2\tilde{u_2}^{\prime}(h,s),\label{427b}\\
&\tilde{u_2}(y\rightarrow\infty,s)=0\label{428b},
\end{align}
respectively.
The solutions of equations \eqref{1a} and \eqref{1b} subject to boundary conditions \eqref{423b}-\eqref{428b} are
\begin{align}
&\tilde{u_1}(y,s)=\frac{sU_0}{(s^{2}+\omega^{2})}\bigg[-\sum_{m=1}^{\infty}M^m\exp\big(-a_1\sqrt{s}\big)+\sum_{m=0}^{\infty}M^m\exp\big(-a_2\sqrt{s}\big)\bigg],\label{431}\\
&\tilde{u_2}(y,s)=\frac{sU_0}{(s^2+\omega^2)}\bigg[\sum_{m=0}^{\infty}(1-M)M^m\exp\big(-a_3\sqrt{s}\big)\bigg],\label{432}
\end{align}
respectively, where
\begin{align}
&M=\dfrac{\alpha-1}{\alpha+1}\label{2a},\\ 
&a_1=\dfrac{(2mh-y)}{\sqrt{\nu_1}}\label{2a1},\\  &a_2=\dfrac{(2mh+y)}{\sqrt{\nu_1}}\label{2a2},\\
&a_3=\big(\dfrac{(y-h)\sqrt{\nu_1}+(2m+1)h\sqrt{\nu_2}}{\sqrt{\nu_1\nu_2}}\big)\label{2a3},
\end{align}
with 
\begin{align}
 &\alpha=\dfrac{\mu_2}{\mu_1}\sqrt{\dfrac{\nu_1}{\nu_2}}\label{2b}.
\end{align}

The Laplace transforms \eqref{431} and \eqref{432} can be inverted to obtain the velocity fields for the lower and upper fluids,  $u_1(y,t)$ and $u_2(y,t)$, respectively. The velocity fields for the lower and upper fluids  are 
\begin{align}
u_1(y,t)=&U_0\bigg[-\sum_{m=1}^{\infty}M^m\exp(-a_1\sqrt{\frac{\omega}{2}})\cos (\omega t-a_1\sqrt{\frac{\omega}{2}})\notag\\&+\sum_{m=0}^{\infty}M^m\exp(-a_2\sqrt{\frac{\omega}{2}})\cos (\omega t-a_2\sqrt{\frac{\omega}{2}})\bigg]\notag\\&+\biggl\{-\frac{U_0}{\pi}\int_{0}^{\infty}\frac{-\sum\limits_{m=1}^{\infty}M^m\sigma\exp (-\sigma t)\sin (a_1\sqrt{\sigma})+\sum\limits_{m=0}^{\infty}M^m\sigma\exp (-\sigma t)\sin (a_2\sqrt{\sigma})}{\sigma^{2}+\omega^{2}} d\sigma\biggr\},\label{43}\\
u_2(y,t)=&U_0\bigg[\sum_{m=0}^{\infty}(1-M)M^m\exp(-a_3\sqrt{\frac{\omega}{2}})\cos (\omega t-a_3\sqrt{\frac{\omega}{2}})\bigg]\notag\\&+\biggl\{-\frac{U_0}{\pi}\int_{0}^{\infty}\frac{\sum\limits_{m=0}^{\infty}(1-M)M^m\sigma\exp (-\sigma t)\sin (a_3\sqrt{\sigma})}{\sigma^{2}+\omega^{2}} d\sigma\biggr\}.\label{43a}
\end{align}
Here, $M$, $a_1$, $a_2$, and $a_3$ are as defined in \eqref{2a} and \eqref{2a1}-\eqref{2a3}, respectively. We note that to invert the Laplace transforms \eqref{431} and \eqref{432} term by term, we have utilized the following result: 
\begin{equation}
	\mathcal{L}^{-1}\bigg(\frac{s\exp(-a\sqrt{s})}{s^{2}+\omega^{2}}\bigg)=\exp(-a\sqrt{\frac{\omega}{2}})\cos (\omega t-a\sqrt{\frac{\omega}{2}})-\frac{1}{\pi}\int_{0}^{\infty}\frac{\sigma\exp (-\sigma t)\sin (a\sqrt{\sigma})}{\sigma^{2}+\omega^{2}} d\sigma,\label{438}
\end{equation}
where $a>0$. The result is given in\cite{sos} as an exercise problem. In order to make this study as self-contained as possible, we have derived the result in detail in \ref{a}.

If we let $t\rightarrow\infty$ into expression \eqref{43}, the part within the curly brackets tends to zero. Therefore, the part within the curly brackets of the expression represents the transient velocity field for the lower fluid. And the remaining part of the expression corresponds to steady periodic velocity field for the fluid. Again, the part within the curly brackets of expression \eqref{43a} represents transient velocity field for the upper fluid as it approaches zero as we let $t\rightarrow\infty$ into the expression. The remaining part of the expression corresponds to steady periodic velocity field for the fluid.
\subsubsection*{Calculation of wall shear stress}
The velocity fields for both the lower and upper fluids have been determined. We now are interested in calculating wall shear stress. We know that the shear stress can be obtained using Newton's law of fluid friction:
\begin{align}
\tau(y,t)=\mu\dfrac{\partial u(y,t)}{\partial y},\label{43e}
\end{align}
where $\tau$ is the shear stress, $\mu$ is the viscosity (dynamic viscosity) of the fluid, and $u(y,t)$ is the velocity field.

The shear stress at the plate can be determined by putting $y=0$ in the expression obtained by plugging expression \eqref{43}  in the formula given by equation \eqref{43e}, which leads to
\begin{align}
\tau_{1w}(0,t)=&\mu_1 U_0\bigg[\sqrt{\frac{\omega}{\nu_1}}\bigg(\sum_{m=1}^{\infty}M^m\exp(-a_4\sqrt{\frac{\omega}{2}})\cos (\omega t-a_4\sqrt{\frac{\omega}{2}}-\frac{3\pi}{4})\notag\\&+\sum_{m=0}^{\infty}M^m\exp(-a_4\sqrt{\frac{\omega}{2}})\cos (\omega t-a_4\sqrt{\frac{\omega}{2}}-\frac{3\pi}{4})\bigg)\bigg]\notag\\&+\biggl\{-\frac{\mu_1U_0}{\pi}\int_{0}^{\infty}\frac{\sqrt{\dfrac{\sigma}{\nu_1}}\bigg(\sum\limits_{m=1}^{\infty}M^m\sigma\exp (-\sigma t)\cos (a_4\sqrt{\sigma})+\sum\limits_{m=0}^{\infty}M^m\sigma\exp (-\sigma t)\cos (a_4\sqrt{\sigma})\bigg)}{\sigma^{2}+\omega^{2}} d\sigma\biggr\},\label{43bc}
\end{align}
where $M$ is as defined in\eqref{2a}, and 
\begin{align}
a_4=\dfrac{2mh}{\sqrt{\nu_1}}.\label{2a4}
\end{align}
If we let $t\rightarrow\infty$ into the wall shear stress given by equation  \eqref{43bc}, the part within the curly brackets approaches zero. So the part inside the curly brackets represents the transient wall shear stress. And the remaining part of expression  \eqref{43bc} corresponds to steady-state wall shear stress.
\subsubsection*{Special Case: single-layer limit}
When $h\rightarrow\infty$, $h$ being the thickness of the lower fluid, the two-layer problem reduces to classical Stokes’ second problem for a single-layer fluid with the cosine oscillations of the plate. If we let $h\rightarrow\infty$, $\mu_1=\mu_2=\mu$ (say the viscosity of the single-layer fluid), and $\nu_1=\nu_2=\nu$ (say the kinematic viscosity of the single-layer fluid) in the velocity field for the lower fluid, \eqref{43}, we find that each of the terms of the series in the expression becomes identically zero, except those that we get for $m=0$. Thus, for the case the velocity field for the lower fluid, \eqref{43}, becomes
\begin{align}
u_{cla}(y,t)=&U_0\bigg[\exp(-y\sqrt{\frac{\omega}{2\nu}})\cos (\omega t-y\sqrt{\frac{\omega}{2\nu}})-\frac{1}{\pi}\int_{0}^{\infty}\frac{\sigma\exp (-\sigma t)\sin (y\sqrt{\dfrac{\sigma}{\nu}})}{\sigma^{2}+\omega^{2}} d\sigma\bigg].\label{43av}
\end{align}
Expression \eqref{43av} is the velocity field for classical Stokes’ second problem when the plate oscillates as $U_0\cos(\omega t)$. The velocity field \eqref{43av} agrees with the result for the flow that can be easily obtained from the related result reported in \cite{KHALED2004795}.

Note that the preceding velocity field for a single-layer fluid can also be deduced from the velocity field for the upper fluid, \eqref{43a}, as a special case. If we let $h=0$ (meaning that the lower fluid ceases to exist),  $\mu_1=\mu_2=\mu$ (say the viscosity of the single-layer fluid), and $\nu_1=\nu_2=\nu$ (say the kinematic viscosity of the single-layer fluid) in expression \eqref{43a}, we obtain the result.
\subsubsection{Solution for the sine oscillations of the plate}
\subsubsection*{Calculation of velocity fields}
The velocity fields for the lower and upper fluids can be determined by solving the initial-boundary value problem consists of governing equations \eqref{421} and \eqref{422} and initial and boundary conditions \eqref{eqn2}-\eqref{428}. We note that the initial-boundary value problem is the same as the one we have dealt earlier in this section, except  that condition \eqref{eqn2} replaces condition \eqref{eqn1}. Therefore, to solve the mathematical problem in hand, we follow the same procedure adopted earlier. We obtain the velocity fields for the lower and upper fluids as follows:
\begin{align}
u_1(y,t)=&U_0\bigg[-\sum_{m=1}^{\infty}M^m\exp(-a_1\sqrt{\frac{\omega}{2}})\sin (\omega t-a_1\sqrt{\frac{\omega}{2}})\notag\\&+\sum_{m=0}^{\infty}M^m\exp(-a_2\sqrt{\frac{\omega}{2}})\sin (\omega t-a_2\sqrt{\frac{\omega}{2}})\bigg]\notag\\&+\biggl\{\frac{U_0\omega}{\pi}\int_{0}^{\infty}\frac{-\sum\limits_{m=1}^{\infty}M^m\exp (-\sigma t)\sin (a_1\sqrt{\sigma})+\sum\limits_{m=0}^{\infty}M^m\exp (-\sigma t)\sin (a_2\sqrt{\sigma})}{\sigma^{2}+\omega^{2}} d\sigma\biggr\},\label{4310}\\
u_2(y,t)=&U_0\bigg[\sum_{m=0}^{\infty}(1-M)M^m\exp(-a_3\sqrt{\frac{\omega}{2}})\sin (\omega t-a_3\sqrt{\frac{\omega}{2}})\bigg]\notag\\&+\biggl\{\frac{U_0\omega}{\pi}\int_{0}^{\infty}\frac{\sum\limits_{m=0}^{\infty}(1-M)M^m\exp (-\sigma t)\sin (a_3\sqrt{\sigma})}{\sigma^{2}+\omega^{2}} d\sigma\biggr\},\label{4310ac}
\end{align}
respectively. Here, $M$, $a_1$, $a_2$, and $a_3$ are as defined in \eqref{2a} and \eqref{2a1}-\eqref{2a3}.

It is to be noted here that in expression \eqref{4310}, the part within the curly brackets represents the transient velocity field for the lower fluid as it approaches zero as we let $t\rightarrow\infty$ into the expression. The remaining part of the expression represents the steady periodic velocity field for the fluid. Similarly, in expression \eqref{4310ac}, the part inside the curly brackets represents the transient velocity field for the upper fluid, and the remaining part of the expression represents the steady periodic velocity field for the fluid.

It is worth mentioning that Duffy\cite{dgd} has solved the initial-boundary value problem that we have tackled here, obtaining mathematical solution similar to the one reported here. He has indicated that the solutions can be used to investigate the physical problem concerning heat conduction in a two-layer solid body.
\subsubsection*{Calculation of wall shear stress}
We have calculated the velocity fields for the lower and upper fluids. We now intend to evaluate the shear stress at the plate. 
The shear stress at the plate can be found by substituting $y=0$ into the expression obtained by plugging expression \eqref{4310} in the formula given by equation \eqref{43e}, which leads to
\begin{align}
\tau_{1w}(0,t)=&\mu_1 U_0\bigg[\sqrt{\frac{\omega}{\nu_1}}\big(\sum_{m=1}^{\infty}M^m\exp(-a_4\sqrt{\frac{\omega}{2}})\sin (\omega t-a_4\sqrt{\frac{\omega}{2}}-\frac{3\pi}{4})\notag\\&+\sum_{m=0}^{\infty}M^m\exp(-a_4\sqrt{\frac{\omega}{2}})\sin (\omega t-a_4\sqrt{\frac{\omega}{2}}-\frac{3\pi}{4})\big)\bigg]\notag\\&+\biggl\{\mu_1 U_0\frac{\omega}{\pi}\int_{0}^{\infty}\frac{\sqrt{\dfrac{\sigma}{\nu_1}}\bigg(\sum\limits_{m=1}^{\infty}M^m\exp (-\sigma t)\cos (a_4\sqrt{\sigma})+\sum\limits_{m=0}^{\infty}M^m\exp (-\sigma t)\cos (a_4\sqrt{\sigma})\bigg)}{\sigma^{2}+\omega^{2}} d\sigma\biggr\},\label{43dc}
\end{align}
where $M$ and $a_4$ are as defined in \eqref{2a}  and \eqref{2a4}, respectively. Here, the part within the curly brackets represents transient shear stress at the plate as it tends to zero as we let $t\rightarrow\infty$ into the result for wall shear stress. The remaining part of the result represents steady-state shear stress at the plate. It is valid for large values of time $t$.

\subsubsection*{Special Case: single-layer limit}
Earlier in this section, we have deduced the velocity field for Stokes’ second problem for a single-layer fluid, \eqref{43av}, as a special case. The velocity field corresponds to the case where the plate oscillates as $U_0  \cos(\omega t)$ (the cosine oscillations). We follow the same procedure to deduce the velocity field for Stokes’ second problem for a single-layer fluid related to the sine oscillations of the plate from the velocity field for the lower fluid, \eqref{4310}. The result is
\begin{align}
u_{cla}(y,t)=&U_0\bigg[\exp(-y\sqrt{\frac{\omega}{2\nu}})\sin (\omega t-y\sqrt{\frac{\omega}{2\nu}})+\frac{\omega}{\pi}\int_{0}^{\infty}\frac{\exp (-\sigma t)\sin (y\sqrt{\dfrac{\sigma}{\nu}})}{\sigma^{2}+\omega^{2}} d\sigma\bigg].\label{43avc}
\end{align}
The velocity field \eqref{43avc} agrees with the result for the flow that can be easily obtained from the related result reported in \cite{KHALED2004795}.

Note that the preceding velocity field for a single-layer fluid can also be deduced from the velocity field for the upper fluid, \eqref{4310ac}, by adopting the procedure outlined earlier in this section.
\section{Oscillatory Couette flow for a two-layer fluid}
\label{s3}
\subsection{Mathematical Formulation}
Consider two layers of two immiscible fluids confined between two parallel plates which are at a distance $H$ apart. The fluids are of different viscosities, densities, and thicknesses. The lower plate coincides with the $x$-$z$ plane of Cartesian coordinate system $(x,y,z)$. We suppose that the lower fluid fills the region $0\le y\le h$, $h$ being a positive real number. And the upper fluid occupies the region $h\le y \le H$. The $y$-axis is the co-ordinate normal to the plates. The fluids and the plates are initially at rest and then the lower plate starts to oscillate parallel to itself, along $x$-axis,  with velocity $U_0 \cos(\omega t)$ or $U_0 \sin(\omega t)$, where $U_0$, $\omega$, and  $t$ being the plate velocity amplitude, frequency of oscillations, and time, respectively. It is considered that the flow is two-dimensional, and there is no body force. The motion of fluids are caused only by the oscillatory motion of the lower plate. The velocity fields for the lower and upper fluids are governed  by the reduced Navier-Stokes equations \eqref{421} and \eqref{422}, respectively. We consider the no-slip boundary condition at the plates. Besides, we consider continuity of velocity and that of shear stress at the interface of the fluids. Accordingly, the boundary and initial conditions are 
\refstepcounter{equation}\label{eqn3}
\refstepcounter{equation}\label{eqn4}
\begin{align}
&u_1(0,t)=U_0 \cos(\omega t) \quad \text{or}\quad u_1(0,t)=U_0 \sin(\omega t)\quad \text{for} \quad t>0\tag{\ref*{eqn3},\,\ref*{eqn4}},\label{323a}\\
&u_1(y,0)=0\label{324a}\\
&u_2(y,0)=0\label{325a}\\
&u_1(h,t)=u_2(h,t)\label{326a}\\
&\mu_1\frac{\partial u_1}{\partial y}\Big\vert_{y=h}=\mu_2\frac{\partial u_2}{\partial y}\Big\vert_{y=h}\label{327a}\\
&u_2(H,t)=0\label{328a}
\end{align}
\subsection{Solution}
\subsubsection{Solution for the cosine oscillations of the plate}\label{321}
\subsubsection*{Calculation of velocity fields}
In order to determine the velocity fields for the lower and upper fluids, we need to obtain solution to the initial-boundary value problem consists of governing equations \eqref{421} and \eqref{422}, and initial and boundary conditions \eqref{eqn3}, and \eqref{324a}-\eqref{328a}. The Laplace transforms of equations \eqref{421} and \eqref{422} are equations \eqref{1a} and \eqref{1b}, respectively. We note that initial conditions \eqref{324a} and \eqref{325a} have been utilized to obtain the transforms. Again, the transforms of the boundary conditions \eqref{eqn3}, \eqref{326a}, and \eqref{327a} are equations \eqref{423b}-\eqref{427b}, respectively. And the transform of the boundary condition \eqref{328a} is
\begin{align}
\tilde{u_2}(H,s)=0\label{423bc1}.
\end{align} 

The solutions of equations \eqref{1a} and \eqref{1b} subject to boundary conditions \eqref{423b}-\eqref{427b} and \eqref{423bc1} are
\begin{align}
\tilde{u_1}(y,s)=&-\dfrac{U_0 s}{s^{2}+\omega^{2}}\bigg[\dfrac{\big\{\sinh(p)\sinh(r-q)+\alpha\cosh(p)\cosh(r-q)\big\}\sinh(\sqrt{\dfrac{s}{\nu_1}}y)}{F_1(s)}\notag\\&-\cosh(\sqrt{\dfrac{s}{\nu_1}}y)\bigg],\label{12a}\\
\tilde{u_2}(y,s)=&\dfrac{U_0 s}{s^{2}+\omega^{2}}\bigg[\dfrac{\sinh(r-\sqrt{\dfrac{s}{\nu_2}}y)}{F_1(s)}\bigg],\label{12b}
\end{align} 
respectively, where
\begin{align}
&p=\sqrt{\dfrac{s}{\nu_1}}h,\label{3a}\\ 
&q=\sqrt{\dfrac{s}{\nu_2}}h,\label{3b} \\
&r=\sqrt{\dfrac{s}{\nu_2}}H,\label{3c}\\
&F_1(s)=\cosh(p)\sinh(r-q)+\alpha\sinh(p)\cosh(r-q)\label{3d},
\end{align}
and $\alpha$ is as defined in \eqref{2b}.

Each of equations \eqref{12a} and \eqref{12b} has simple poles at $s=i\omega$ and $s=-i\omega$. Also, each of these equations has infinite number of poles which lie on the negative real axis at $s=-k_m^2$, where $k_m$ is a real number and $m$ is the index number (an integer) of the pole. Here, $k_m$ can be obtained from the following equation:
\begin{equation}
\alpha\tan(k_m\dfrac{h}{\sqrt{\nu_1}})=-\tan(k_m\dfrac{(H-h)}{\sqrt{\nu_2}}).\label{3e}
\end{equation}
The steady periodic velocity fields for the lower and upper fluids are related to the simple poles at $s=i\omega$ and $s=-i\omega$, whereas the poles located at $s=-k_m^2$ are responsible for the transient  velocity fields. 

The Laplace inverse for $\tilde{u_1}(y,s)$, \eqref{12a}, and $\tilde{u_2}(y,s)$, \eqref{12b}, can be computed, respectively, from the following relations: 
\begin{align}
&u_1(y,t)=\sum_{m=1}^{\infty} Res[\tilde{u_1}(y,s)]_{-k_m^{2}}+Res[\tilde{u_1}(y,s)]_{i\omega}+Res[\tilde{u_1}(y,s)]_{-i\omega},\label{u1}\\
&	u_2(y,t)=\sum_{m=1}^{\infty} Res[\tilde{u_2}(y,s)]_{-k_m^{2}}+Res[\tilde{u_2}(y,s)]_{i\omega}+Res[\tilde{u_2}(y,s)]_{-i\omega},\label{u2}
\end{align}
where $Res$ stands for the residue.

 The steady periodic velocity field for the lower fluid can be found by evaluating residues at $s=i\omega$ and $s=-i\omega$, which yields
\begin{align}
u_{1s}(y,t)=&\dfrac{U_0}{A^{2}+B^{2}}\bigg[[-(g_1(y)A+g_2(y)B)+\cosh(\sqrt{\dfrac{\omega}{2\nu_1}}y)\cos(\sqrt{\dfrac{\omega}{2\nu_1}}y)(A^{2}+B^{2})]\cos(\omega t)-\notag\\&[g_1(y)B-g_2(y)A+\sinh(\sqrt{\dfrac{\omega}{2\nu_1}}y)\sin(\sqrt{\dfrac{\omega}{2\nu_1}}y)(A^{2}+B^{2})]\sin(\omega t)\bigg].\label{a1}
\end{align}
In a similar way the steady periodic velocity field for the upper fluid can be determined, which is
\begin{align}
	u_{2s}(y,t)=&\dfrac{U_0}{A^{2}+B^{2}}\big[(g_3(y)A+g_4(y)B)\cos(\omega t)-(g_4(y)A-g_3(y)B)\sin(\omega t)\big]\label{a2}.
\end{align}
In expressions \eqref{a1} and \eqref{a2}, the constants $A$ and $B$ are defined as follows:
\begin{align}
A=&\alpha \cos (a) \sinh (a) \cos (b-c) \cosh (b-c)-\alpha \sin (a) \cosh (a) \sin (b-c) \sinh (b-c)\notag\\&-\cos (a) \cosh (a) \cos (b-c) \sinh (b-c)+\sin (a) \sinh (a) \sin (b-c) \cosh (b-c),\label{6c}\\
B=&\alpha \cos (a) \sinh (a) \sin (b-c) \sinh (b-c)+\alpha \sin (a) \cosh (a) \cos (b-c) \cosh (b-c)\notag\\&-\sin (a) \sinh (a) \cos (b-c) \sinh (b-c)-\cos (a) \cosh (a) \sin (b-c) \cosh (b-c),\label{6d}
\end{align}
with
\begin{align}
	&a=\sqrt{\dfrac{\omega}{2\nu_1}}h,\label{7a}\\ &b=\sqrt{\dfrac{\omega}{2\nu_2}}h,\label{7b}\\ &c=\sqrt{\dfrac{\omega}{2\nu_2}}H,\label{7c}
\end{align}
and $\alpha$ is as defined in equation \eqref{2b}.

The functions $g_1(y)$, $g_2(y)$, $g_3(y)$, and $g_4(y)$ are defined as follows:
\begin{align}
g_1(y)=&-\sin (e) \cosh (e)\bigg[\alpha \sin (a) \sinh (a) \cos (b-c) \cosh (b-c)\notag\\&+\alpha \cos (a) \cosh (a) \sin (b-c) \sinh (b-c)-\cos (a) \sinh (a) \sin (b-c) \cosh (b-c)\notag\\&-\sin (a) \cosh (a) \cos (b-c)\sinh (b-c)\bigg]+\notag\\&\cos (e) \sinh (e)\bigg[\alpha \cos (a) \cosh (a) \cos (b-c) \cosh (b-c)-\alpha \sin (a) \sinh (a)  \sin (b-c) \sinh (b-c)\notag\\&-\cos (a) \sinh (a) \cos (b-c) \sinh (b-c)+\sin (a) \cosh (a) \sin (b-c) \cosh (b-c)\bigg],\label{6e}\\
 g_2(y)=&\sin (e) \cosh (e)\bigg[\alpha \cos (a) \cosh (a) \cos (b-c) \cosh (b-c)\notag\\&-\alpha \sin (a) \sinh (a) \sin (b-c) \sinh (b-c)-\cos (a) \sinh (a)\cos (b-c) \sinh (b-c)\notag\\&+\sin (a) \cosh (a) \sin (b-c) \cosh (b-c)\bigg]+\cos (e) \sinh (e)\bigg[\alpha \sin (a) \sinh (a) \cos (b-c) \cosh (b-c)\notag\\&+\alpha \cos (a) \cosh (a) \sin (b-c) \sinh (b-c)\notag\\&-\cos (a) \sinh (a) \sin (b-c) \cosh (b-c)-\sin (a) \cosh (a) \cos (b-c) \sinh (b-c)\bigg],\label{6f}\\
 g_3(y)=&\cos (c-d) \sinh (c-d),\label{6a}\\
 g_4(y)=&\cosh (c-d) \sin (c-d),\label{6b}
\end{align}
with 
\begin{align}
 &d=\sqrt{\dfrac{\omega}{2\nu_2}}y,\label{7d}\\
&e=\sqrt{\dfrac{\omega}{2\nu_1}}y.\label{7e}
\end{align}

Again, the transient velocity field for the lower fluid can be obtained by calculating  the residues at all $s=-k_m^2$, which results in
\begin{align}
u_{1t}(y,t)=&-\sum_{m=1}^{\infty}\dfrac{2U_0  k_m^3}{k_m^4+\omega^2}\bigg[\dfrac{F_2(k_m)\sin(k_m\dfrac{y}{\sqrt{\nu_1}})-F_3(k_m)\cos(k_m\dfrac{y}{\sqrt{\nu_1}})}{F_4(k_m)}\bigg]\exp(-k_m^2 t).\label{5}
\end{align} 
We follow the similar procedure to determine the transient  velocity field for the upper fluid, yielding 
\begin{align}
	u_{2t}(y,t)=&\sum_{m=1}^{\infty}\dfrac{2U_0  k_m^3}{k_m^4+\omega^2}\bigg[\dfrac{\sin(k_m\dfrac{(H-y)}{\sqrt{\nu_2}})}{F_4(k_m)}\bigg]\exp(-k_m^2 t).\label{6}
\end{align} 
In expressions \eqref{5} and \eqref{6},	
\begin{align}
F_2(k_m)=&-\sin(k_m\dfrac{h}{\sqrt{\nu_1}})\sin(k_m\dfrac{(H-h)}{\sqrt{\nu_2}})+\alpha\cos(k_m\dfrac{h}{\sqrt{\nu_1}})\cos(k_m\dfrac{(H-h)}{\sqrt{\nu_2}})\label{e1},\\
F_3(k_m)=&\cos(k_m\dfrac{h}{\sqrt{\nu_1}})\sin(k_m\dfrac{(H-h)}{\sqrt{\nu_2}})+\alpha\sin(k_m\dfrac{h}{\sqrt{\nu_1}})\cos(k_m\dfrac{(H-h)}{\sqrt{\nu_2}})\label{e2},\\
 F_4(k_m)=&\bigg[\dfrac{(H-h)}{\sqrt{\nu_2}}\big\{\cos(k_m\dfrac{h}{\sqrt{\nu_1}})\cos(k_m\dfrac{(H-h)}{\sqrt{\nu_2}})-\alpha\sin(k_m\dfrac{h}{\sqrt{\nu_1}})\sin(k_m\dfrac{(H-h)}{\sqrt{\nu_2}})\big\}\notag\\&+\dfrac{h}{\sqrt{\nu_1}}\big\{-\sin(k_m\dfrac{h}{\sqrt{\nu_1}})\sin(k_m\dfrac{(H-h)}{\sqrt{\nu_2}})+\alpha\cos(k_m\dfrac{h}{\sqrt{\nu_1}})\cos(k_m\dfrac{(H-h)}{\sqrt{\nu_2}})\big\}\bigg].\label{e3}
\end{align}

Now, in accordance with \eqref{u1} and \eqref{u2}, the complete velocity field for each of the lower and upper fluids is the sum of the respective steady periodic and transient velocity fields. Therefore, the complete velocity fields for the lower and the upper fluids are
\begin{align}
&u_1(y,t)=u_{1s}(y,t)+u_{1t}(y,t),\\
&u_2(y,t)=u_{2s}(y,t)+u_{2t}(y,t),
\end{align}
respectively. Here, $u_{1s}(y,t)$, $u_{1t}(y,t)$, $u_{2s}(y,t)$, and $u_{2t}(y,t)$ are given by \eqref{a1}, \eqref{5}, \eqref{a2}, and \eqref{6}, respectively.
\subsubsection*{Calculation of wall shear stresses}
The velocity fields for both the lower and upper fluids have been explicitly obtained. We now are interested in evaluating the shear stresses at the plates. The steady-state and transient shear stresses in the lower fluid can be calculated by plugging the expressions \eqref{a1} and \eqref{5} in the formula given by the equation \eqref{43e}, respectively, obtaining
\begin{align}
\tau_{1s}(y,t)=&\dfrac{\mu_1U_0}{A^{2}+B^{2}}\bigg[\cos(\omega t)[-(g_1^\prime(y) A+g_2^\prime(y) B)+\sqrt{\dfrac{\omega}{2\nu_1}}(\sinh(e)\cos(e)-\sin(e)\cosh(e))(A^{2}+B^{2})]\notag\\&-\sin(\omega t)[g_1^\prime(y) B-g_2^\prime(y) A+\sqrt{\dfrac{\omega}{2\nu_1}}(\cosh(e)\sin(e)+\sinh(e)\cos(e))(A^{2}+B^{2})]\bigg],\label{c1}\\
\tau_{1t}(y,t)=&-\sum_{m=1}^{\infty}(\dfrac{2\mu_1U_0 k_m^4}{\sqrt{\nu_1}(k_m^4+\omega^2)})\bigg[\dfrac{F_2(k_m)\cos(k_m\dfrac{y}{\sqrt{\nu_1}})+F_3(k_m)\sin(k_m\dfrac{y}{\sqrt{\nu_1}})}{F_4(k_m)}\bigg]\exp(-k_m^2 t)\label{c2},
\end{align}
respectively. Here, the constants $A$ and $B$ are as defined in \eqref{6c} and \eqref{6d}. The functions $g_1(y)$ and $g_2(y)$ are as defined in \eqref{6e} and \eqref{6f}. And $e$ is as defined in \eqref{7e}. Also, $F_2(k_m)$, $F_3(k_m)$, and $F_4(k_m)$ are as defined in \eqref{e1}-\eqref{e3}, respectively. Note that here primes denote differentiation with respect to $y$.

Again, we can evaluate the steady-state and transient shear stresses in the upper fluid by plugging, respectively, the expressions \eqref{a2} and \eqref{6} in the formula given by equation \eqref{43e}, yielding
\begin{align}
	\tau_{2s}(y,t)=&\dfrac{\mu_2U_0}{A^{2}+B^{2}}\bigg[(g_3^\prime(y) A+g_4^\prime(y) B)\cos(\omega t)-(g_4^\prime(y) A-g_3^\prime(y) B)\sin(\omega t)\bigg],\label{a2c}\\
	\tau_{2t}(y,t)=&-\sum_{m=1}^{\infty}\dfrac{2\mu_2U_0  k_m^4}{\sqrt{\nu_2}(k_m^4+\omega^2)}\bigg[\dfrac{\cos(k_m\dfrac{(H-y)}{\sqrt{\nu_2}})}{F_4(k_m)}\bigg]\exp(-k_m^2 t),\label{6c1}
\end{align}
respectively. Here, the functions $g_3(y)$ and $g_4(y)$ are as defined in \eqref{6a} and \eqref{6b}.

The steady periodic and transient shear stresses at the oscillating plate can be evaluated by substituting  $y=0$ into expressions \eqref{c1} and \eqref{c2}, respectively, obtaining

\begin{align}
\tau_{1ws}(0,t)=&\dfrac{\mu_1U_0}{A^{2}+B^{2}}\sqrt{\dfrac{\omega}{2\nu_1}}\bigg[\cos(\omega t)[-((-K_1+K_2) A+(K_1+K_2) B)]\notag\\&-\sin(\omega t)[(-K_1+K_2) B-(K_1+K_2) A]\bigg]\label{sp1},\\
\tau_{1wt}(0,t)=&-\sum_{m=1}^{\infty}(\dfrac{2\mu_1U_0k_m^4}{\sqrt{\nu_1}(k_m^4+\omega^2)})\bigg[\dfrac{F_2(k_m)}{F_4(k_m)}\bigg]\exp(-k_m^2 t)\label{sp2},
\end{align}
respectively. Here, the constants $K_1$ and $K_2$ are defined as follows:
\begin{align}
	K_1=&\alpha \sin (a) \sinh (a) \cos (b-c) \cosh (b-c)+\alpha \cos (a) \cosh (a) \sin (b-c) \sinh (b-c)\notag\\&-\cos (a) \sinh (a) \sin (b-c) \cosh (b-c)-\sin (a) \cosh (a) \cos (b-c)\sinh (b-c)\label{349},\\
	K_2=&\alpha \cos (a) \cosh (a) \cos (b-c) \cosh (b-c)-\alpha \sin (a) \sinh (a)  \sin (b-c) \sinh (b-c)\notag\\&-\cos (a) \sinh (a) \cos (b-c) \sinh (b-c)+\sin (a) \cosh (a) \sin (b-c) \cosh (b-c),\label{350}
\end{align}
where $\alpha$, $a$, $b$, and $c$ are as defined in equations \eqref{2b} and \eqref{7a}-\eqref{7c}, respectively.

Again, the steady periodic and transient shear stresses at the stationary plate can be calculated by putting $y=H$ in expressions \eqref{a2c} and \eqref{6c1}, respectively, yielding
\begin{align}
	\tau_{2ws}(H,t)=&-\dfrac{\mu_2U_0}{A^{2}+B^{2}}\sqrt{\dfrac{\omega}{2\nu_2}}\bigg[(A+ B)\cos(\omega t)-( A-B)\sin(\omega t)\bigg],\label{a2cc}\\
	\tau_{2wt}(H,t)=&-\sum_{m=1}^{\infty}\dfrac{2\mu_2U_0  k_m^4}{\sqrt{\nu_2}(k_m^4+\omega^2)}\bigg[\dfrac{\exp(-k_m^2 t)}{F_4(k_m)}\bigg],\label{6c1c}
\end{align}
respectively.
\subsubsection*{Special case: single-layer limit}
The two-layer fluid flow problem reduces to a single-layer one when $h$, the thickness of the lower fluid, becomes equal to $H$ or zero. We note that $H$ is the distance between the plates. If we let $h=H$, $\mu_1=\mu_2=\mu$ (say the viscosity of the single-layer fluid), and $\nu_1=\nu_2=\nu$ (say the kinematic viscosity of the single-layer fluid) in the steady periodic velocity field for lower fluid, \eqref{a1}, we obtain steady-state velocity field for the single-layer fluid flow. Similarly, the transient velocity field for the single-layer fluid flow can be obtained from expression \eqref{5}. The results for the steady-periodic and transient velocity fields are
\begin{align}
u_{cssc}(y,t)=&\dfrac{U_0}{A_1^{2}+B_1^{2}}\big[(f_1(y)A_1+f_2(y)B_1)\cos(\omega t)-(f_2(y)A_1-f_1(y)B_1)\sin(\omega t)\big],\label{a2sc1}\\
 u_{ctsc}(y,t)=&\sum_{m=1}^{\infty}\dfrac{2U_0 k_m^3 \sqrt{\nu}}{k_m^4+\omega^2}\bigg[\dfrac{\sin(k_m\dfrac{(H-y)}{\sqrt{\nu}})}{H\cos(k_m\dfrac{(H)}{\sqrt{\nu}})}\bigg]\exp(-k_m^2 t),\label{6tc1}
\end{align}
respectively. Here, the functions $f_1(y)$ and $f_2(y)$ are defined as follows:  
\begin{align}
&f_1(y)=\cos (\sqrt{\dfrac{\omega}{2\nu}}(H-y)) \sinh (\sqrt{\dfrac{\omega}{2\nu}}(H-y)),\label{6a1}\\
&f_2(y)=\cosh (\sqrt{\dfrac{\omega}{2\nu}}(H-y)) \sin (\sqrt{\dfrac{\omega}{2\nu}}(H-y)).\label{6b1}
\end{align} 
The constants $A_1$ and $B_1$ are defined as the following:
\begin{align}
	&A_1=\cos(\sqrt{\dfrac{\omega}{2\nu}}H)\sinh(\sqrt{\dfrac{\omega}{2\nu}}H),\label{6a2}\\
	&B_1=\sin(\sqrt{\dfrac{\omega}{2\nu}}H)\cosh(\sqrt{\dfrac{\omega}{2\nu}}H).\label{6b2}
\end{align}
We note that $k_m$ can be found by the following relation: 
\begin{equation}
k_m=m\dfrac{\sqrt{\nu}}{H}\pi, \quad m=1,2,3,..., \label{km}
\end{equation}
which is deduced from equation \eqref{3e}.

The complete velocity field for the single-layer fluid is the sum of the steady-state velocity field, \eqref{a2sc1} and the transient velocity field, \eqref{6tc1}. The transient dies out as the time $t$ progresses.

We note that some hints on the above deductions for the single-layer fluid are given in \ref{ab}, which a reader might find helpful.

Note that the steady-state and transient velocity fields for a single-layer flow, \eqref{a2sc1} and \eqref{6tc1}, can also be deduced from steady-state and transient velocity fields for the upper fluid, \eqref{a2} and \eqref{6}, respectively. In order to obtain the results, we need to let $h=0$(meaning that the lower fluid ceases to exist), $\mu_1=\mu_2=\mu$ (say the viscosity of the single-layer fluid), and $\nu_1=\nu_2=\nu$ (say the kinematic viscosity of the single-layer fluid) in expressions \eqref{a2} and \eqref{6}. Some helpful hints about the deductions are given in \ref{ab}.

It is important to be noted here that the complete velocity field for the single-layer fluid found here as a special case has not, so far as we are aware, been previously reported in the literature.

We can now compute the shear stresses at the oscillating and fixed plates for the single-layer fluid. We can calculate steady-state and transient shear stresses at the oscillating plate by substituting $y=0$ into the expressions obtained by plugging expressions \eqref{a2sc1} and \eqref{6tc1} in formula \eqref{43e}, respectively. Again, if we put $y=H$ in the expressions, it will result in steady periodic and transient shear stresses at the fixed plate. The steady periodic and transient shear stresses at the oscillating plate are 
\begin{align}
\tau_{cssc}(0,t)=&\dfrac{\mu U_0}{A_1^{2}+B_1^{2}}\big[(A_2A_1+B_2B_1)\cos(\omega t)-(B_2A_1-A_2B_1)\sin(\omega t)\big],\label{k}\\
\tau_{ctsc}(0,t)=&-\sum_{m=1}^{\infty}\dfrac{2\mu U_0 k_m^4 }{H(k_m^4+\omega^2)}\exp(-k_m^2 t),
\end{align}
respectively. The constants $A_2$ and $B_2$ are  defined as follows:
\begin{align}
	&A_2=\sqrt{\dfrac{\omega}{2\nu}}\bigg[\sin(\sqrt{\dfrac{\omega}{2\nu}}H)\sinh(\sqrt{\dfrac{\omega}{2\nu}}H)-\cos(\sqrt{\dfrac{\omega}{2\nu}}H)\cosh(\sqrt{\dfrac{\omega}{2\nu}}H)\bigg],\label{scws1}\\
	&B_2=-\sqrt{\dfrac{\omega}{2\nu}}\bigg[\sin(\sqrt{\dfrac{\omega}{2\nu}}H)\sinh(\sqrt{\dfrac{\omega}{2\nu}}H)+\cos(\sqrt{\dfrac{\omega}{2\nu}}H)\cosh(\sqrt{\dfrac{\omega}{2\nu}}H)\bigg].\label{scws2}
\end{align}

Again, the steady-state and transient shear stresses at the stationary plate are
\begin{align}
	\tau_{cssc}(H,t)=&-\dfrac{\mu U_0}{A_1^{2}+B_1^{2}}\sqrt{\dfrac{\omega}{2\nu}}\big[(A_1+B_1)\cos(\omega t)-(A_1-B_1)\sin(\omega t)\big],\\
	\tau_{ctsc}(H,t)=&-\sum_{m=1}^{\infty}\dfrac{2\mu U_0 k_m^4 }{k_m^4+\omega^2}\bigg[\dfrac{1}{H\cos(k_m\dfrac{(H)}{\sqrt{\nu}})}\bigg]\exp(-k_m^2 t),
\end{align}
respectively.
\subsubsection{Solution for the sine oscillations of the plate}\label{322}
\subsubsection*{Calculation of velocity fields}
In order to obtain the velocity fields for the lower and upper fluids, we need to obtain solution to the initial-boundary value problem consists of governing equations \eqref{421} and \eqref{422} and initial and boundary conditions \eqref{eqn4}-\eqref{328a}. We note that the mathematical problem is the same as the one we have tackled earlier in this section, except that condition \eqref{eqn4} replaces condition \eqref{eqn3}. Therefore, to deal with the initial-boundary value problem in hand, we adopt the same procedure that we have followed earlier. 

The steady periodic and transient velocity fields for the lower fluid are
\begin{align}
u_{1s}(y,t)=&\dfrac{U_0}{A^{2}+B^{2}}\bigg[[g_1(y)B-g_2(y)A+\sinh(e)\sin(e)(A^{2}+B^{2})]\cos(\omega t)+\notag\\&[-(g_1(y)A+g_2(y)B)+\cosh(e)\cos(e)(A^{2}+B^{2})]\sin(\omega t)\bigg],\label{a1s}\\
	u_{1t}(y,t)=&\sum_{m=1}^{\infty}\dfrac{2U_0 \omega k_m}{k_m^4+\omega^2}\bigg[\dfrac{F_2(k_m)\sin(k_m\dfrac{y}{\sqrt{\nu_1}})-F_3(k_m)\cos(k_m\dfrac{y}{\sqrt{\nu_1}})}{F_4(k_m)}\bigg]\exp(-k_m^2 t),\label{5t}
\end{align}
respectively. And the steady periodic and transient velocity fields for the upper fluid are
\begin{align}
	u_{2s}(y,t)=&\dfrac{U_0}{A^{2}+B^{2}}\big[(g_4(y)A-g_3(y)B)\cos(\omega t)+(g_3(y)A+g_4(y)B)\sin(\omega t)\big],\label{a2s}\\
	u_{2t}(y,t)=&-\sum_{m=1}^{\infty}\dfrac{2U_0 \omega k_m}{k_m^4+\omega^2}\bigg[\dfrac{\sin(k_m\dfrac{(H-y)}{\sqrt{\nu_2}})}{F_4(k_m)}\bigg]\exp(-k_m^2 t),\label{6t}
\end{align}
respectively. 

In the above expressions the constants $A$ and $B$ are as defined in equations \eqref{6c} and \eqref{6d}. The functions $g_1(y)$, $g_2(y)$, $g_3(y)$, $g_4(y)$, and $e$ are as defined in equations \eqref{6e}-\eqref{6b}, and \eqref{7e}. Also $F_2(k_m)$, $F_3(k_m)$, and $F_4(k_m)$ are as defined in equations \eqref{e1}-\eqref{e3}.

Note that the complete velocity field for each of the lower and upper fluids is the sum of the corresponding steady periodic and transient velocity fields. Accordingly, the complete velocity fields for the lower and upper fluids are  
\begin{align}
	&u_{1}(y,t)=u_{1s}(y,t)+u_{1t}(y,t),\\
	&u_{2}(y,t)=u_{2s}(y,t)+u_{2t}(y,t),
\end{align} 
respectively. Here, $u_{1s} (y,t)$, $u_{1t} (y,t)$, $u_{2s} (y,t)$, and $u_{2t}(y,t)$ are given by equations \eqref{a1s}- \eqref{6t}, respectively.
\subsubsection*{Calculation of wall shear stresses}
As the velocity fields for both the lower and upper fluids have been obtained, we can now compute the shear stresses at the oscillating and fixed plates. The steady periodic and transient shear stresses in the lower fluid can be found by substituting expressions \eqref{a1s} and \eqref{5t} into formula \eqref{43e}, respectively. The results are
\begin{align}
\tau_{1s}(y,t)=&\dfrac{\mu_1U_0}{A^{2}+B^{2}}\bigg[\cos(\omega t)[g_1^\prime(y) B-g_2^\prime(y) A+\sqrt{\dfrac{\omega}{2\nu_1}}(\sinh(e)\cos(e)+\sin(e)\cosh(e))(A^{2}+B^{2})]+\notag\\&\sin(\omega t)[-(g_1^\prime(y) A+g_2^\prime(y) B)+\sqrt{\dfrac{\omega}{2\nu_1}}(\sinh(e)\cos(e)-\sin(e)\cosh(e))(A^{2}+B^{2})]\bigg],\label{c1s}\\
\tau_{1t}(y,t)=&\sum_{m=1}^{\infty}(\dfrac{2\mu_1U_0\omega k_m^2}{\sqrt{\nu_1}(k_m^4+\omega^2)})\bigg[\dfrac{F_2(k_m)\cos(k_m\dfrac{y}{\sqrt{\nu_1}})+F_3(k_m)\sin(k_m\dfrac{y}{\sqrt{\nu_1}})}{F_4(k_m)}\bigg]\exp(-k_m^2 t),\label{c2s}
\end{align}
respectively. Here, the constants $A$ and $B$ are as defined in \eqref{6c} and \eqref{6d}. The functions $g_1(y)$ and $g_2(y)$ are as defined in \eqref{6e} and \eqref{6f}. And $e$ is as defined in \eqref{7e}. Also, $F_2(k_m)$, $F_3(k_m)$, and $F_4(k_m)$ are as defined in \eqref{e1}-\eqref{e3}. Here primes denote differentiation with respect to $y$.

Again, the steady periodic and transient shear stresses in the upper fluid can be obtained by plugging expressions \eqref{a2s} and \eqref{6t}, respectively, in formula \eqref{43e}, yielding
\begin{align}
	\tau_{2s}(y,t)=&\dfrac{\mu_2U_0}{A^{2}+B^{2}}\bigg[(g_4^\prime(y) A-g_3^\prime(y) B)\cos(\omega t)+(g_3^\prime(y) A+g_4^\prime(y) B)\sin(\omega t)\bigg],\label{a2ct}\\
	\tau_{2t}(y,t)=&\sum_{m=1}^{\infty}\dfrac{2\mu_2U_0 \omega k_m^2}{\sqrt{\nu_2}(k_m^4+\omega^2)}\bigg[\dfrac{\cos(k_m\dfrac{(H-y)}{\sqrt{\nu_2}})}{F_4(k_m)}\bigg]\exp(-k_m^2 t),\label{6c1t}
\end{align}
respectively. Here, the functions $g_3(y)$ and $g_4(y)$ are as defined in \eqref{6a} and \eqref{6b}.

We can now determine steady-state and transient shear stresses at the oscillating plate by substituting $y=0$ into equations \eqref{c1s} and \eqref{c2s}, respectively. And the results are
\begin{align}
\tau_{1ws}(0,t)=&\dfrac{\mu_1U_0}{A^{2}+B^{2}}\sqrt{\dfrac{\omega}{2\nu_1}}\bigg[\cos(\omega t)[(-K_1+K_2) B-(K_1+K_2) A]\notag\\&-\sin(\omega t)[(-K_1+K_2) A+(K_1+K_2) B]\bigg],\\
\tau_{1wt}(0,t)=&\sum_{m=1}^{\infty}(\dfrac{2\mu_1U_0\omega k_m^2}{\sqrt{\nu_1}(k_m^4+\omega^2)})\bigg[\dfrac{F_2(k_m)}{F_4(k_m)}\bigg]\exp(-k_m^2 t),
\end{align}
respectively. Where $K_1$ and $K_2$ are as defined \eqref{349} and \eqref{350}.

Again, We can evaluate steady-state and transient shear stresses at the stationary plate by putting $y=H$ in equations \eqref{a2ct} and \eqref{6c1t}, respectively, obtaining
\begin{align}
	\tau_{2ws}(H,t)=&-\dfrac{\mu_2U_0}{A^{2}+B^{2}}\sqrt{\dfrac{\omega}{2\nu_2}}\big[( A- B)\cos(\omega t)+( A+B)\sin(\omega t)\big],\\
	\tau_{2wt}(H,t)=&\sum_{m=1}^{\infty}\dfrac{2\mu_2U_0 \omega k_m^2}{\sqrt{\nu_2}(k_m^4+\omega^2)}\bigg[\dfrac{\exp(-k_m^2 t)}{F_4(k_m)}\bigg].
\end{align}
\subsubsection*{Special case: single-layer limit}
The steady periodic and transient velocity fields for oscillatory Couette flow for a single-layer fluid, \eqref{a2sc1} and \eqref{6tc1}, have been deduced earlier in this section as a special case. The velocity fields correspond to the case where the plate oscillates as $U_0  \cos(\omega t)$ (the cosine oscillations). We adopt the same procedure to deduce the steady-state and transient velocity fields for oscillatory Couette flow for a single-layer fluid related to the sine oscillations of the plate from the corresponding velocity fields for the lower fluid, \eqref{a1s} and \eqref{5t}. The steady periodic and transient velocity fields are
\begin{align}
	u_{ssc}(y,t)=&\dfrac{U_0}{A_1^{2}+B_1^{2}}\big[(f_2(y)A_1-f_1(y)B_1)\cos(\omega t)+(f_1(y)A_1+f_2(y)B_1)\sin(\omega t)\big],\label{a2sc}\\
	u_{tsc}(y,t)=&-\sum_{m=1}^{\infty}\dfrac{2U_0 \omega k_m \sqrt{\nu}}{k_m^4+\omega^2}\bigg[\dfrac{\sin(k_m\dfrac{(H-y)}{\sqrt{\nu}})}{H\cos(k_m\dfrac{(H)}{\sqrt{\nu}})}\bigg]\exp(-k_m^2 t),\label{6tc}
\end{align}
respectively. Here, the functions $f_1(y)$ and $f_2(y)$ are as defined in equations \eqref{6a1} and \eqref{6b1}. The constants $A_1$ and $B_1$ are given by equations \eqref{6a2} and \eqref{6b2}. Also, $k_m$ can be found from equation \eqref{km}. 

It is worth mentioning here that the results deduced above for the single-layer fluid are consistent with those that can be obtained from Khaled and Vafai\cite{KHALED2004795} for the same flow.

Note that the preceding steady-state and transient velocity fields for a single-layer fluid can also be deduced from the corresponding velocity fields for the upper fluid, \eqref{a2s} and \eqref{6t}, by adopting the procedure outlined earlier in this section.

We can now evaluate shear stresses at the oscillating and fixed plates related to the flow of the single-layer fluid. We can compute the shear stresses in the same manner as that we have adopted earlier in this section. The results for the steady periodic and transient shear stresses at the oscillating plate are
\begin{align}
	\tau_{cssc}(0,t)=&\dfrac{\mu U_0}{A_1^{2}+B_1^{2}}\big[(B_2A_1-A_2B_1)\cos(\omega t)+(A_2A_1+B_2B_1)\sin(\omega t)\big],\\
	\tau_{ctsc}(0,t)=&\sum_{m=1}^{\infty}\dfrac{2\mu U_0\omega k_m^2 }{H(k_m^4+\omega^2)}\exp(-k_m^2 t),
\end{align}
respectively.
Here, $A_2$ and $B_2$ are as defined in equations \eqref{scws1} and \eqref{scws2}, respectively.
The results for the steady-state and transient shear stresses at the stationary plate are
\begin{align}
	\tau_{cssc}(H,t)=&-\dfrac{\mu U_0}{A_1^{2}+B_1^{2}}\sqrt{\dfrac{\omega}{2\nu}}\big[(A_1-B_1)\cos(\omega t)+(A_1+B_1)\sin(\omega t)\big],\\
	\tau_{ctsc}(H,t)=&\sum_{m=1}^{\infty}\dfrac{2\mu U_0\omega k_m^2 }{k_m^4+\omega^2}\bigg[\dfrac{1}{H\cos(k_m\dfrac{H}{\sqrt{\nu}})}\bigg]\exp(-k_m^2 t),
\end{align}
respectively. 

We note here that, as far as we are aware, complete wall shear stresses related to Couette flow due to the cosine or the sine oscillations of the plate have not been previously reported in the literature. A complete shear stress is the sum of steady periodic and transient shear stresses.
\section{Results and illustrative examples}\label{s4}
In this work, we have studied Stokes’ second problem and oscillatory Couette flow for a two-layer fluid. In the Stokes’ problem case, the fluid is bounded only by a oscillating plate that causes the fluid motion. In the Couette flow case, the fluid is confined between two parallel plates, one of which oscillates and induce the fluid motion. In both the cases, we have considered both the cosine and the sine oscillations of the plate. For both the Stokes’ problem and the Couette flow, we have obtained analytical velocity fields consisting of transient and steady periodic parts for both the layers of fluids. The fluids have different viscosities, densities, and thicknesses. We have evaluated transient and steady-state shear stresses at the boundaries of the flows.

Consider the Stokes’ second problem and the oscillatory Couette flow for the two-layer fluid where a layer of corn oil (lighter) lies over a layer of water (heavier). In both the Stokes’ and Couette problems, the water rests on the oscillating plate. It should be noted that oil over water is encountered in many practical situations\cite{wang17}. We can utilize the analytical results obtained in the previous sections for Stokes’ second problem and oscillatory Couette flow for a two-layer fluid to get some physical insights into the particular flows we have considered here.  We show the effects of the forms and the oscillation frequency on the transient velocity fields(hence also on time $t$ to reach a steady-state flow of the lower or upper fluid),  and on the transient and steady-state wall shear stresses. We demonstrate oscillations in steady-state fluid velocities in both the water and oil layers. For the Stokes’ problem, we compare wall velocity with steady-state wall shear stress.

For the particular problems considered here, the values of the parameters (in the cm-gram-second (cgs) system) are as follows\cite{wang17}: the viscosity of the water $\mu_1=0.01$, the kinematic viscosity of the water $\nu_1=0.01$, the viscosity of the corn oil $\mu_2=0.2$, the kinematic viscosity of the corn oil $\nu_2=0.22$, the thickness of the water $h=0.2$ , the distance between the plates (Couette flow) $H=0.5$, and the plate velocity amplitude $U_0=2$. For the graphical representations of the results for the Couette flow, we have taken 20 terms of the infinite series representing the transient parts into account.

Note that henceforth, in this section, by the lower and upper fluids we mean the water and the corn oil, respectively.

\subsection{Stokes’ second problem for a two-layer fluid}
\begin{figure}[ht!]
	\begin{subfigure}{0.55\textwidth}
		\includegraphics[width=1\linewidth, height=8cm]{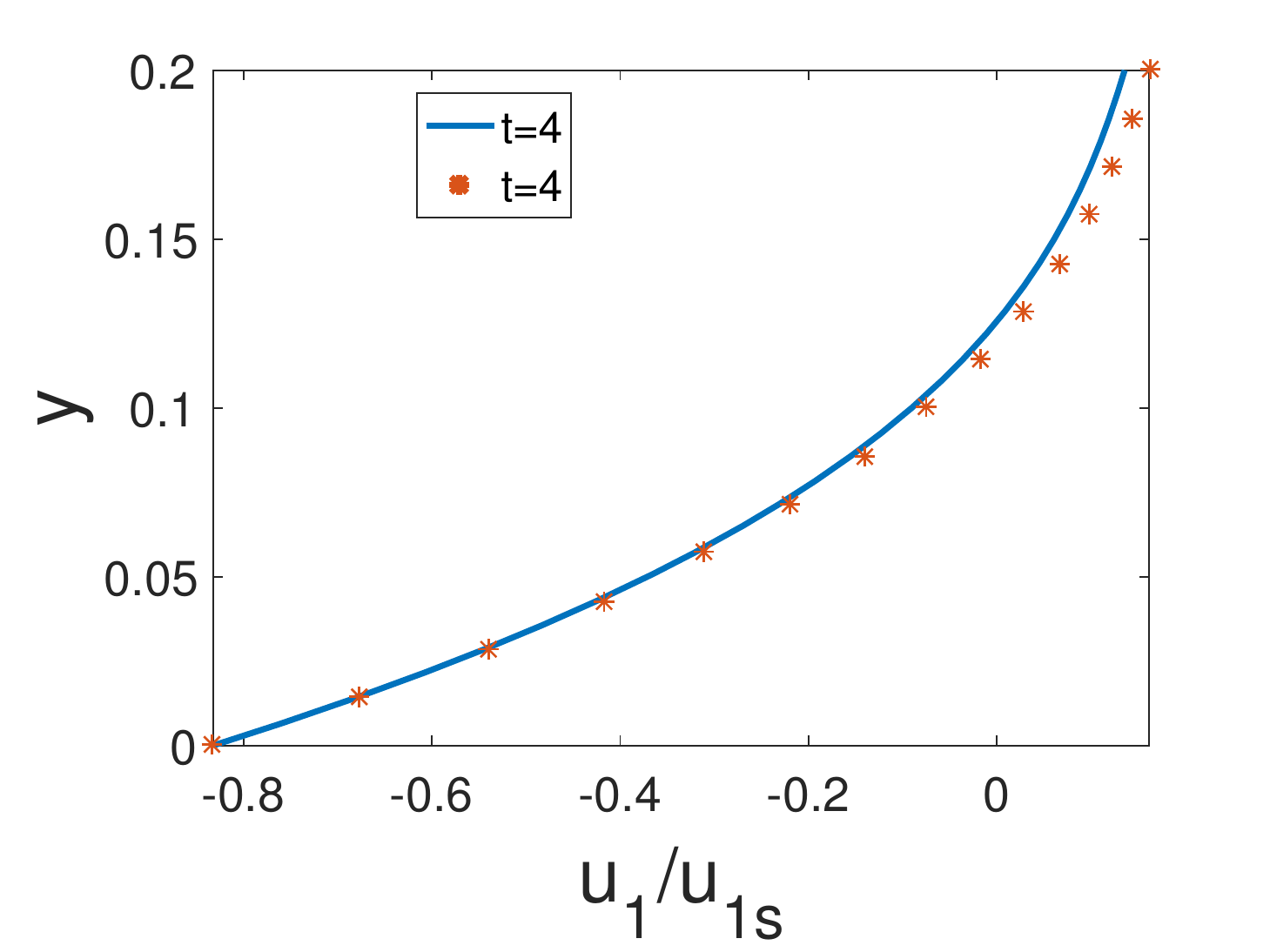} 
		\caption{}
		\label{}
	\end{subfigure}
	\begin{subfigure}{0.55\textwidth}
		\includegraphics[width=1\linewidth, height=8cm]{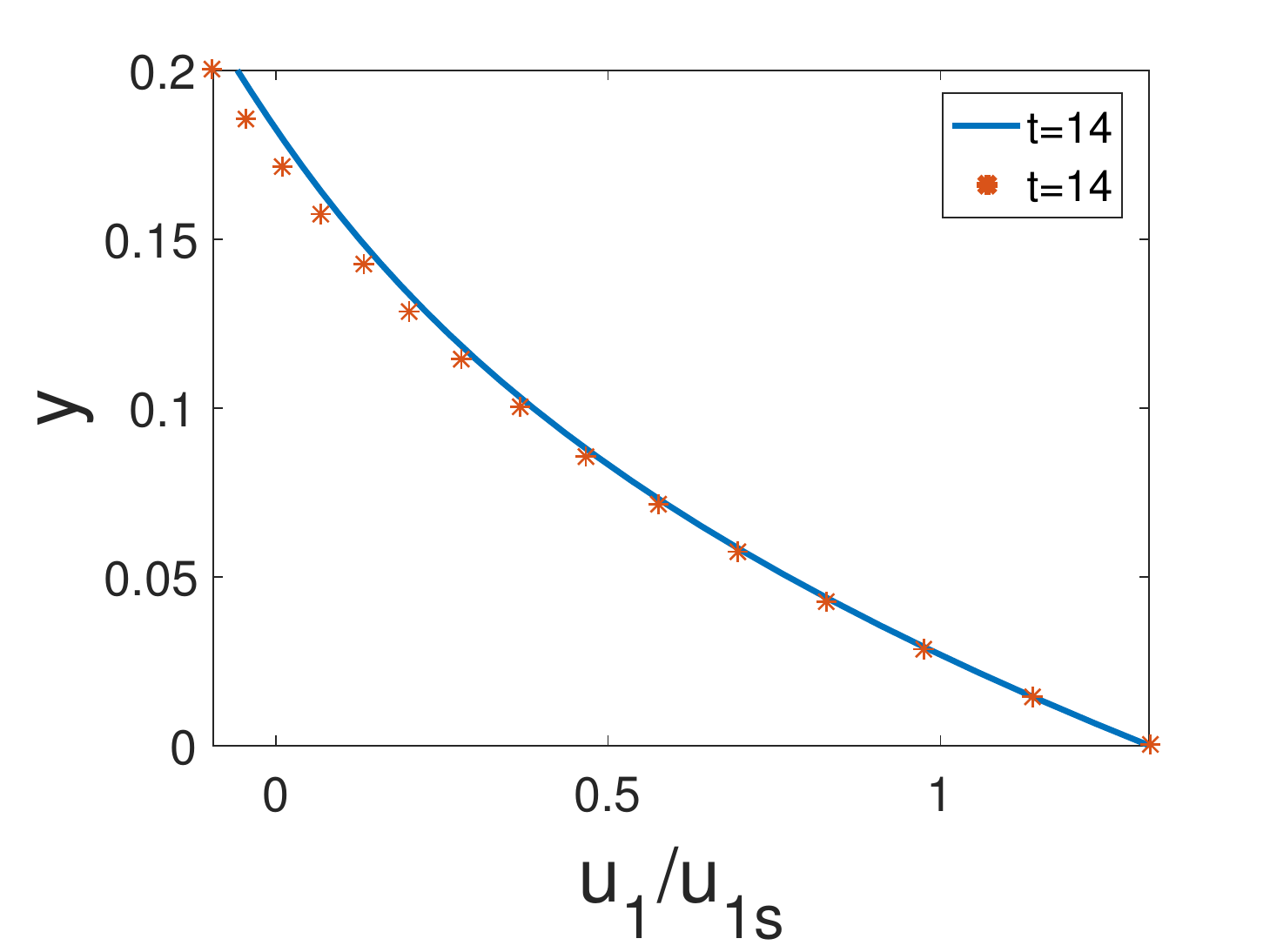}
		\caption{}
		\label{}
	\end{subfigure}
	\begin{subfigure}{0.55\textwidth}
	\includegraphics[width=1\linewidth, height=8cm]{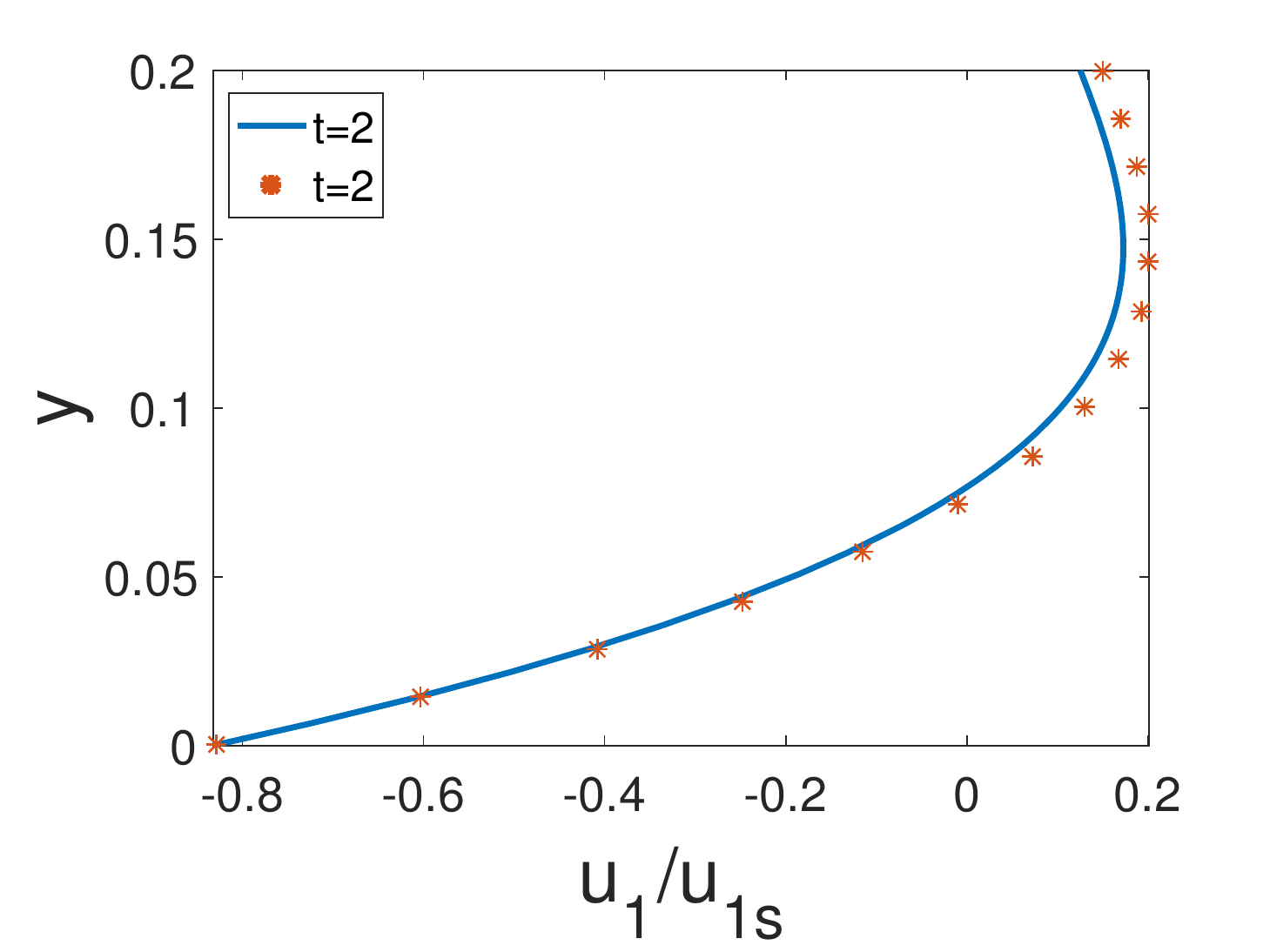} 
	\caption{}
	\label{}
\end{subfigure}
\begin{subfigure}{0.55\textwidth}
	\includegraphics[width=1\linewidth, height=8cm]{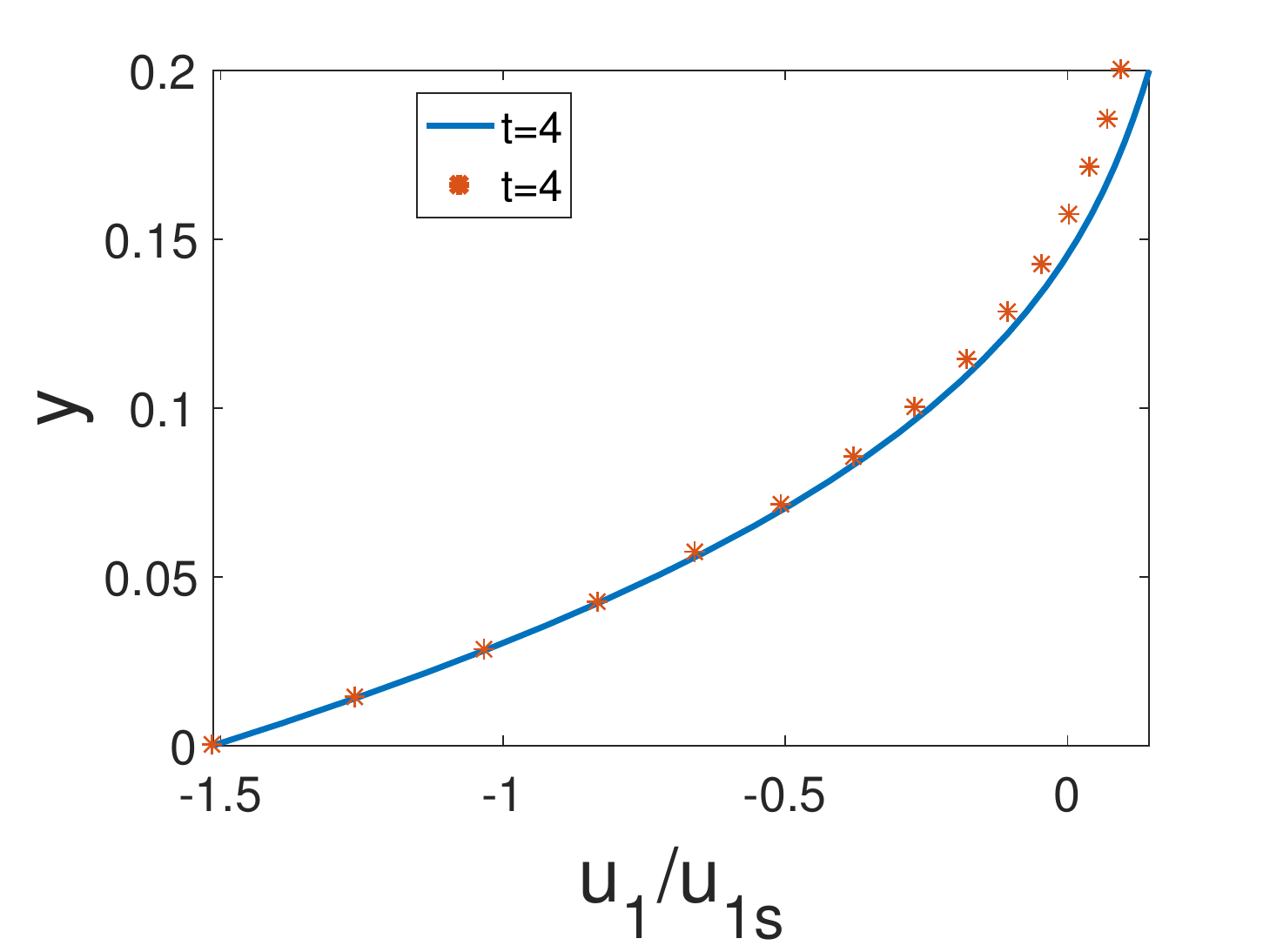}
	\caption{}
	\label{}
\end{subfigure}
	\caption{A profile for the starting velocity field (solid line) and a profile for the steady-state velocity field (line of asterisks) for the lower fluid when $U_0=2, h=0.2, \mu_1=0.01, \nu_1=0.01, \mu_2=0.2, \nu_2=0.22$: (a) the plate oscillates as $U_0\cos(\omega t)$, with $\omega=0.5$, (b) the plate oscillates as $U_0\sin(\omega t)$, with $\omega=0.5$, (c) the plate oscillates as $U_0\cos(\omega t)$, with $\omega=1$, and (d) the plate oscillates as $U_0\sin(\omega t)$, with $\omega=1$. (Stokes' problem)}
	\label{f2}
\end{figure}

\begin{figure}[h!]
	\begin{subfigure}{0.55\textwidth}
		\includegraphics[width=1\linewidth, height=8cm]{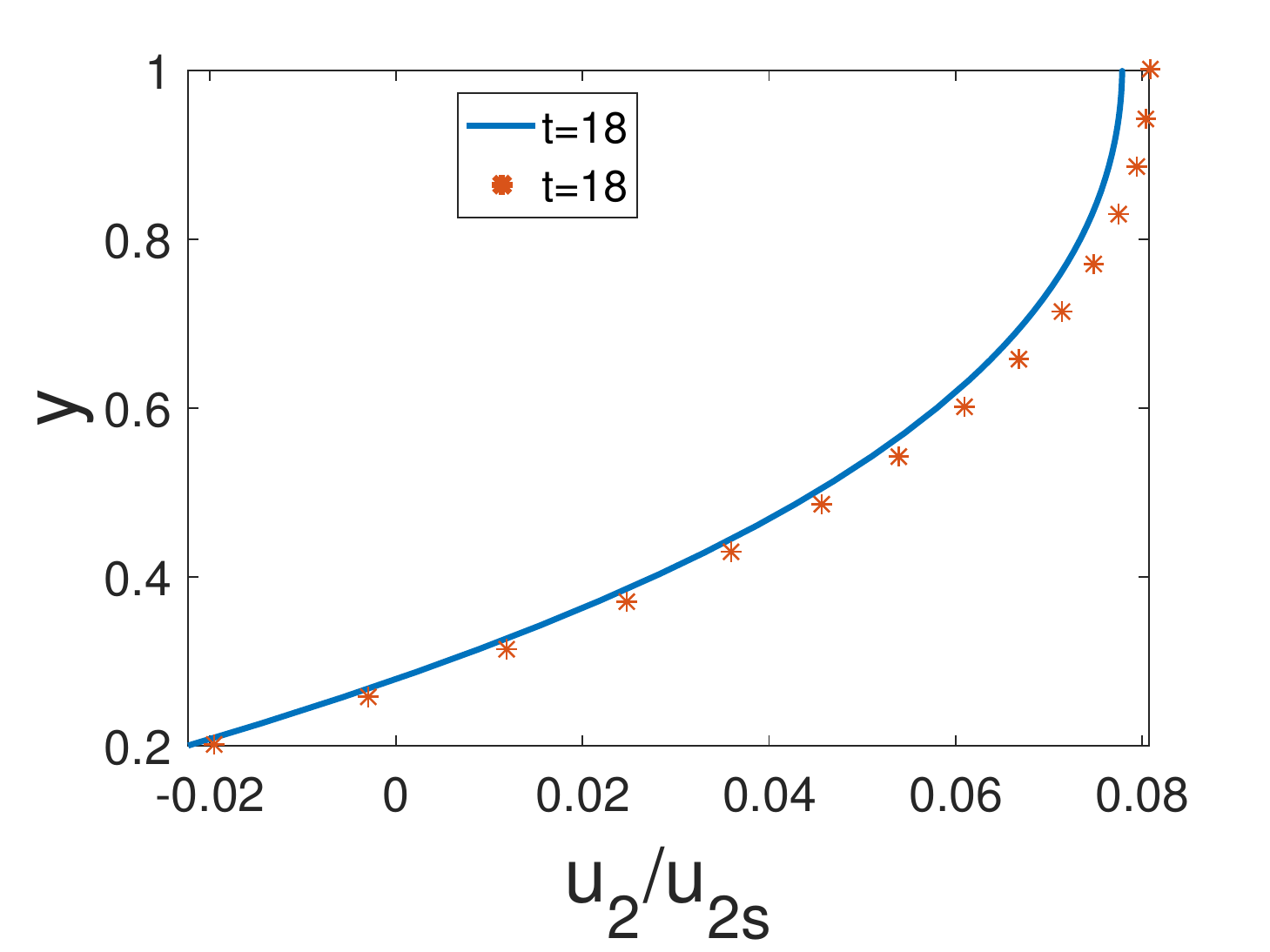} 
		\caption{}
		\label{}
	\end{subfigure}
	\begin{subfigure}{0.55\textwidth}
		\includegraphics[width=1\linewidth, height=8cm]{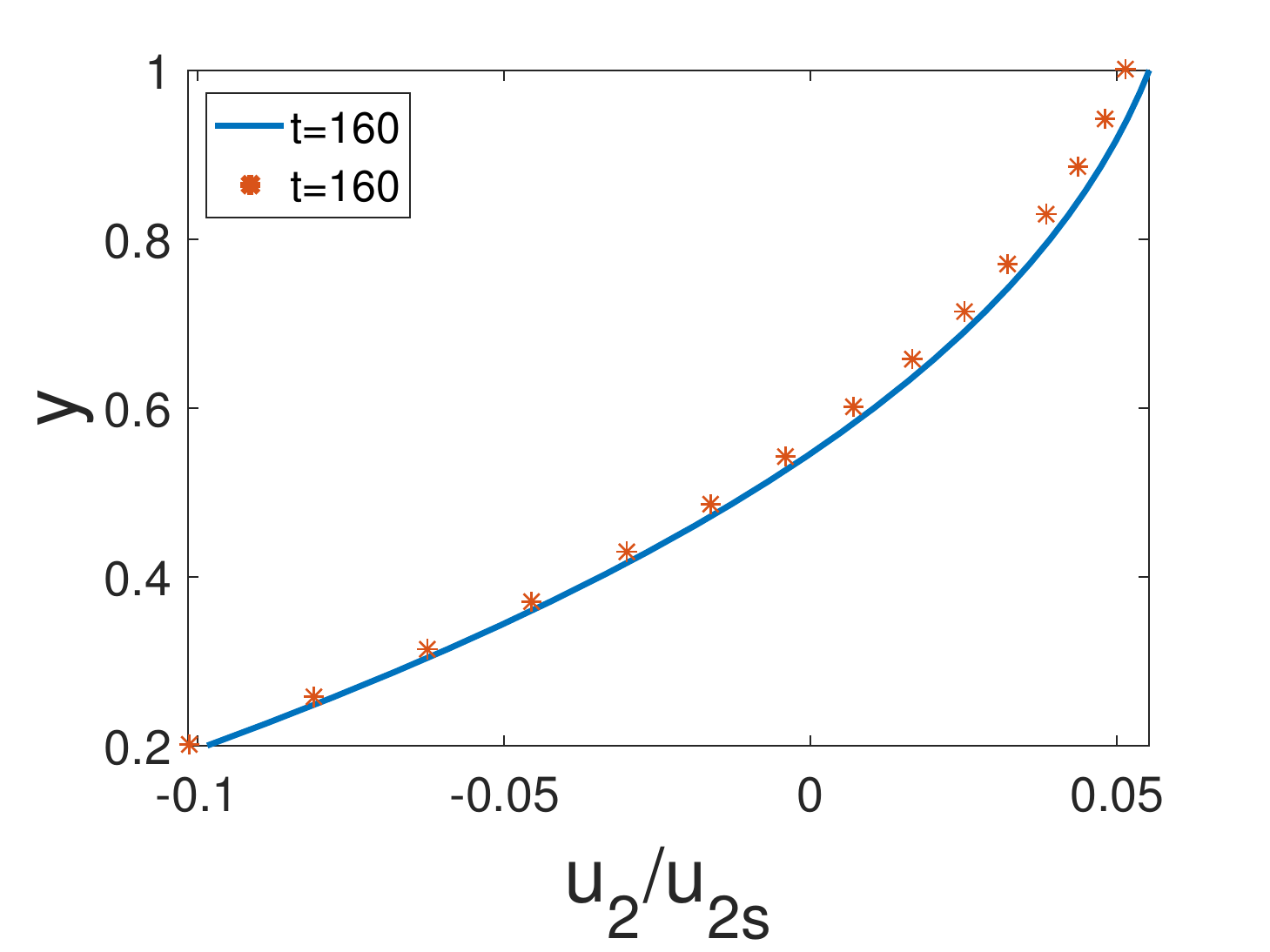}
		\caption{}
		\label{}
	\end{subfigure}
	\begin{subfigure}{0.55\textwidth}
		\includegraphics[width=1\linewidth, height=8cm]{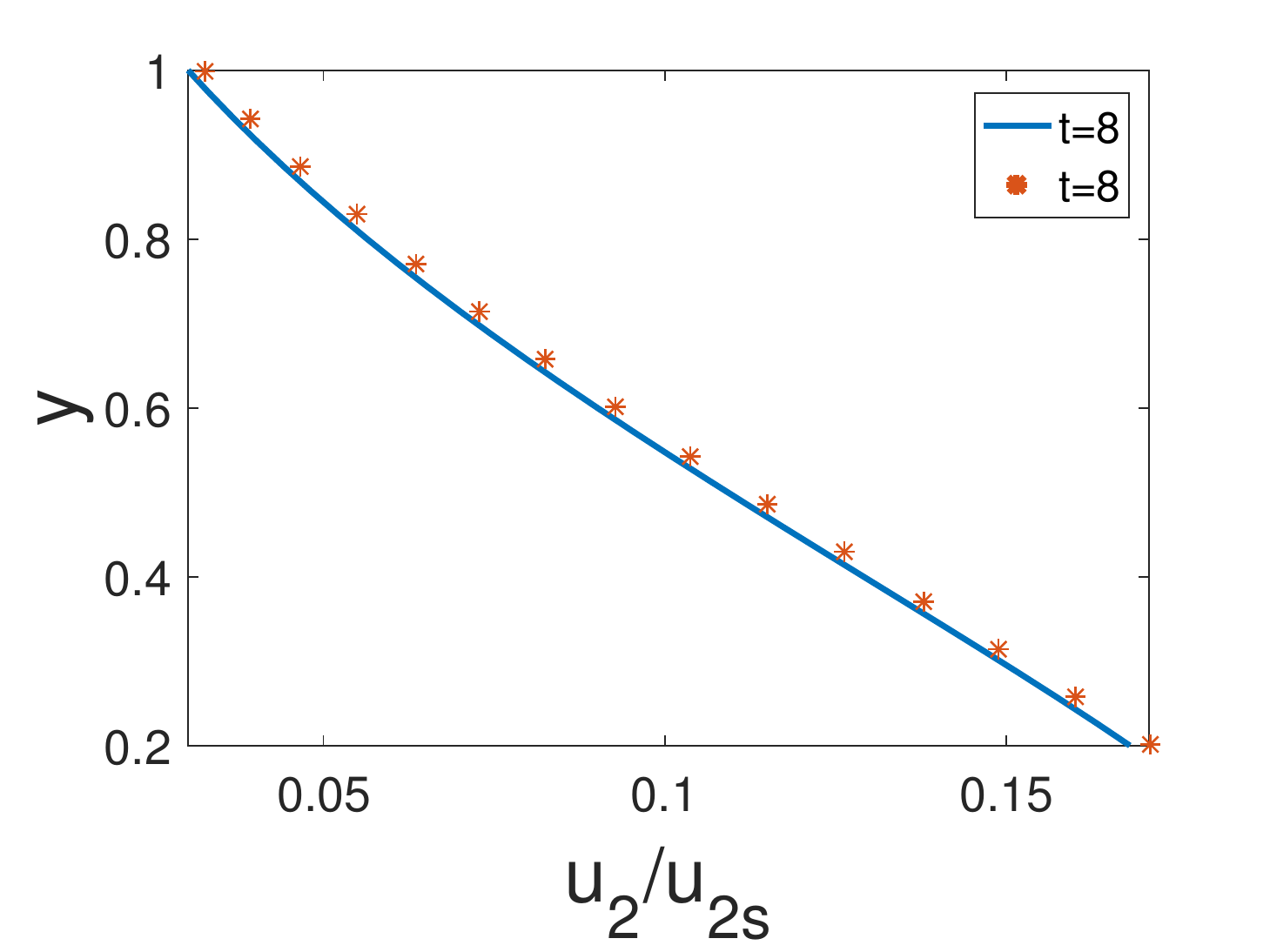} 
		\caption{}
		\label{}
	\end{subfigure}
	\begin{subfigure}{0.55\textwidth}
		\includegraphics[width=1\linewidth, height=8cm]{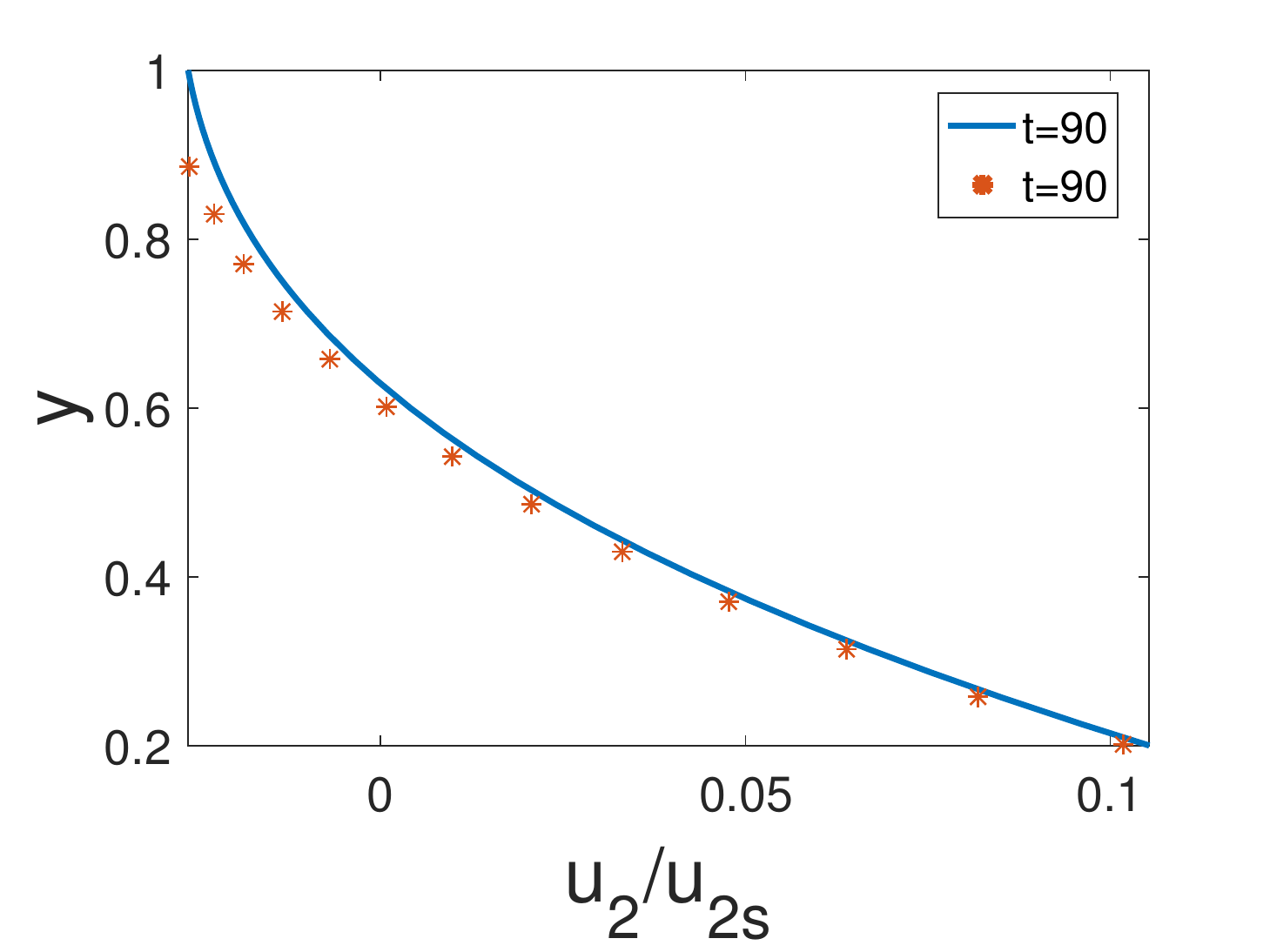}
		\caption{}
		\label{}
	\end{subfigure}
	\caption{A profile for the starting velocity field (solid line) and a profile for the steady-state velocity field (line of asterisks) for the upper fluid when $U_0=2, h=0.2, \mu_1=0.01, \nu_1=0.01, \mu_2=0.2, \nu_2=0.22$: (a) the plate oscillates as $U_0\cos(\omega t)$, with $\omega=0.5$, (b) the plate oscillates as $U_0\sin(\omega t)$, with $\omega=0.5$, (c) the plate oscillates as $U_0\cos(\omega t)$, with $\omega=1$, and (d) the plate oscillates as $U_0\sin(\omega t)$, with $\omega=1$. (Stokes' problem)}
	\label{f4a}
\end{figure}

  In Figs. \ref{f2} and \ref{f4a}, every panel depicts a starting velocity profile for a time and a steady-state velocity profile for the same time for a case of flow in the lower(water) or upper(corn oil) layer. We note here that a starting velocity field is the sum of steady-state and transient velocity fields. In each of the panels, the starting and steady-state velocity profiles are almost the same, implying that the transient has died out and the flow has attained steady-state. The figures show that the time required for a flow in the lower or upper layer to reach steady-state is much greater for the sine oscillations(the plate oscillates as $U_0\sin(\omega t)$) than that for the cosine oscillations. It is also noticed from the figures that for any given value of $\omega$, the oscillations frequency of the plate, the time needed to reach steady-state velocity in the lower fluid is much less than that in the upper fluid. The finding holds true for both forms of oscillations of the plate. As seen from Fig. \ref{f2}, when $\omega=0.5$, the fluid motion in the lower layer becomes steady periodic around $t=4$ when the plated is subjected to the cosine oscillations, and for the sine oscillations, the required time is $t=14$. Again, when $\omega=1$, the fluid flow in the lower layer attains steady-state around $t=2$ and $t=4$ for the cosine and the sine oscillations of the plate, respectively, as noticed from the same figure. Regarding the upper layer of fluid, Fig. \ref{f4a} shows  that when $\omega=0.5$, the flows corresponding to the cosine and the sine oscillations of the plate attain steady-state about $t=18$ and $t=160$, respectively. Again, when $\omega=1$, the upper fluid flow reaches steady-state around $t=8$ when the plate is subjected to the cosine oscillations, and for the sine oscillations of the plate, the required time is $t=90$, as noticed from the same figure.

\begin{figure}[h!]
	\begin{subfigure}{0.55\textwidth}
		\includegraphics[width=1\linewidth, height=8cm]{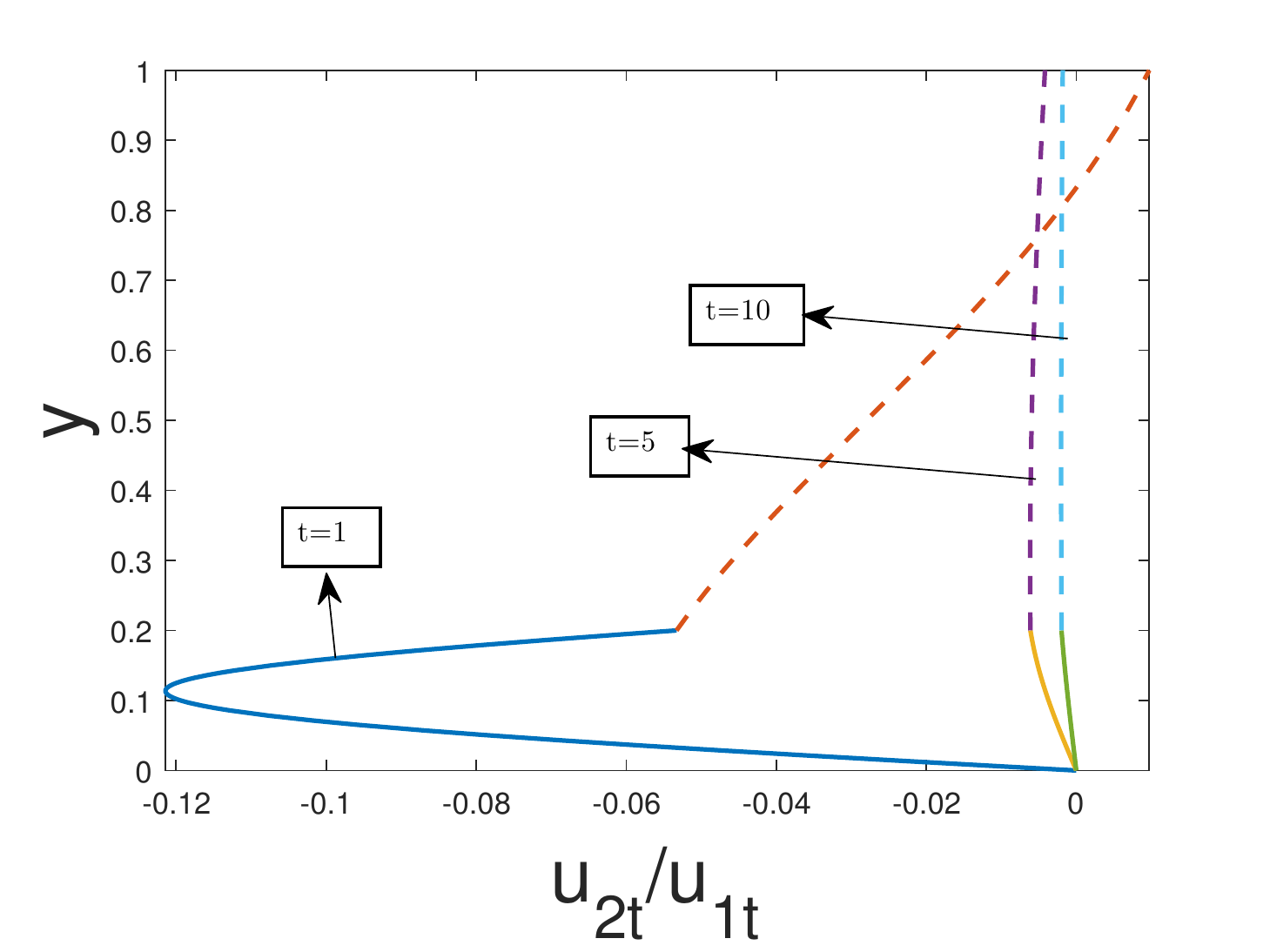} 
		\caption{}
		\label{f6a}
	\end{subfigure}
	\begin{subfigure}{0.55\textwidth}
		\includegraphics[width=1\linewidth, height=8cm]{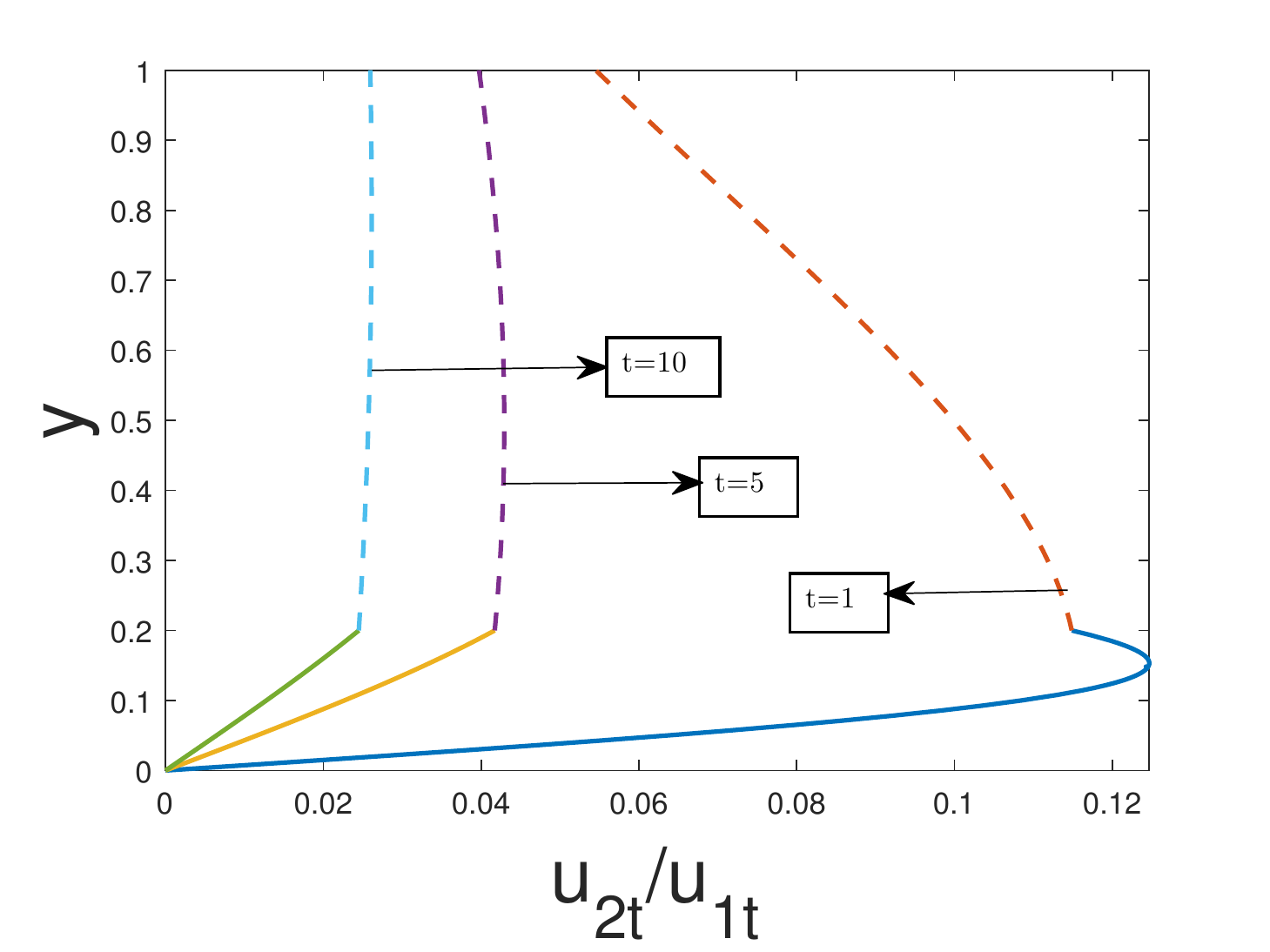}
		\caption{}
		\label{f6b}
	\end{subfigure}
	\caption{Profiles for the transient velocity fields for the lower (solid lines) and upper (broken lines) fluids when $U_0=2, h=0.2, \mu_1=0.01, \nu_1=0.01, \mu_2=0.2, \nu_2=0.22$, and $\omega=1$: (a) the plate oscillates as $U_0\cos(\omega t)$, and (b) the plate oscillates as $U_0\sin(\omega t)$. (Stokes' problem)}
	\label{f6}
\end{figure}

Fig. \ref{f6} gives transient velocity profiles for three given times for the lower and upper fluids. Fig. \ref{f6a} and Fig. \ref{f6b} illustrate profiles corresponding to the cosine and the sine oscillations of the plate, respectively. It is noticed from the figure that the transient velocity (absolute value) decreases rapidly at initial stages but after some time the rate of decreasing with respect to time slows down. It is observed for both forms of oscillations of the plate. Also, the following pieces of information are obtained from the figure. When the plate oscillates as $U_0 \cos(\omega t)$(Fig. \ref{f6a}), at $t=1$, the maximum transient velocity (absolute value) in the lower fluid is slightly greater than 0.12 and it occurs at just over $y=0.1$. At the same time, the maximum transient velocity (absolute value) in the upper fluid is slightly greater than 0.05 and it occurs at the interface of the fluids($y=0.2$). Again, when the plate oscillates as $U_0 \sin(\omega t)$(Fig. \ref{f6b}), at $t=1$, the maximum transient velocity (absolute value) in the lower fluid is 0.125, it occurs at $y=0.15$. At the same time, the maximum transient velocity (absolute value) in the upper fluid is slightly greater than 0.11 and it occurs at the interface of the fluids.

\begin{figure}[h!]
	\begin{subfigure}{0.55\textwidth}
		\includegraphics[width=1\linewidth, height=8cm]{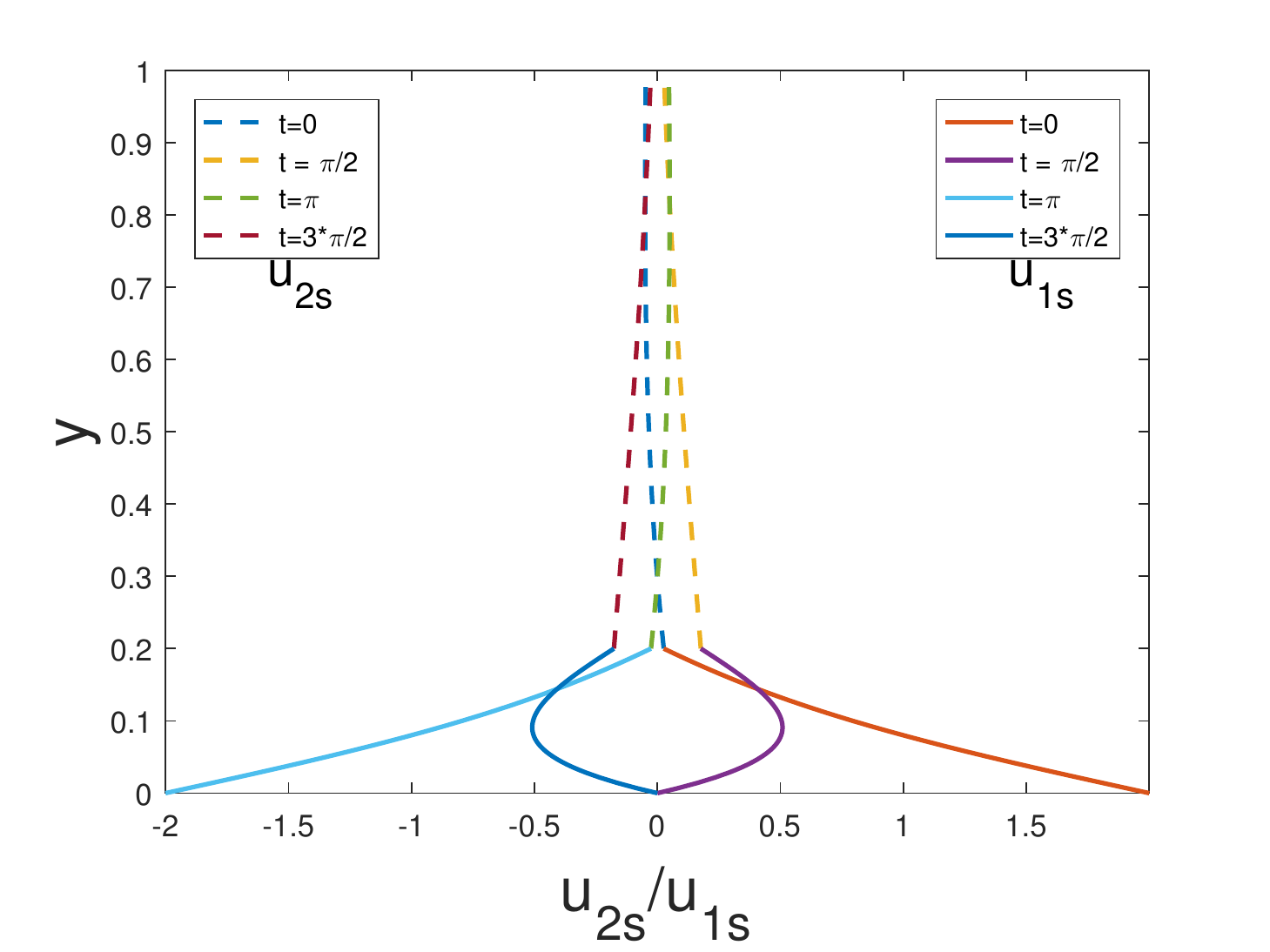} 
		\caption{}
		\label{f5a}
	\end{subfigure}
	\begin{subfigure}{0.55\textwidth}
		\includegraphics[width=1\linewidth, height=8cm]{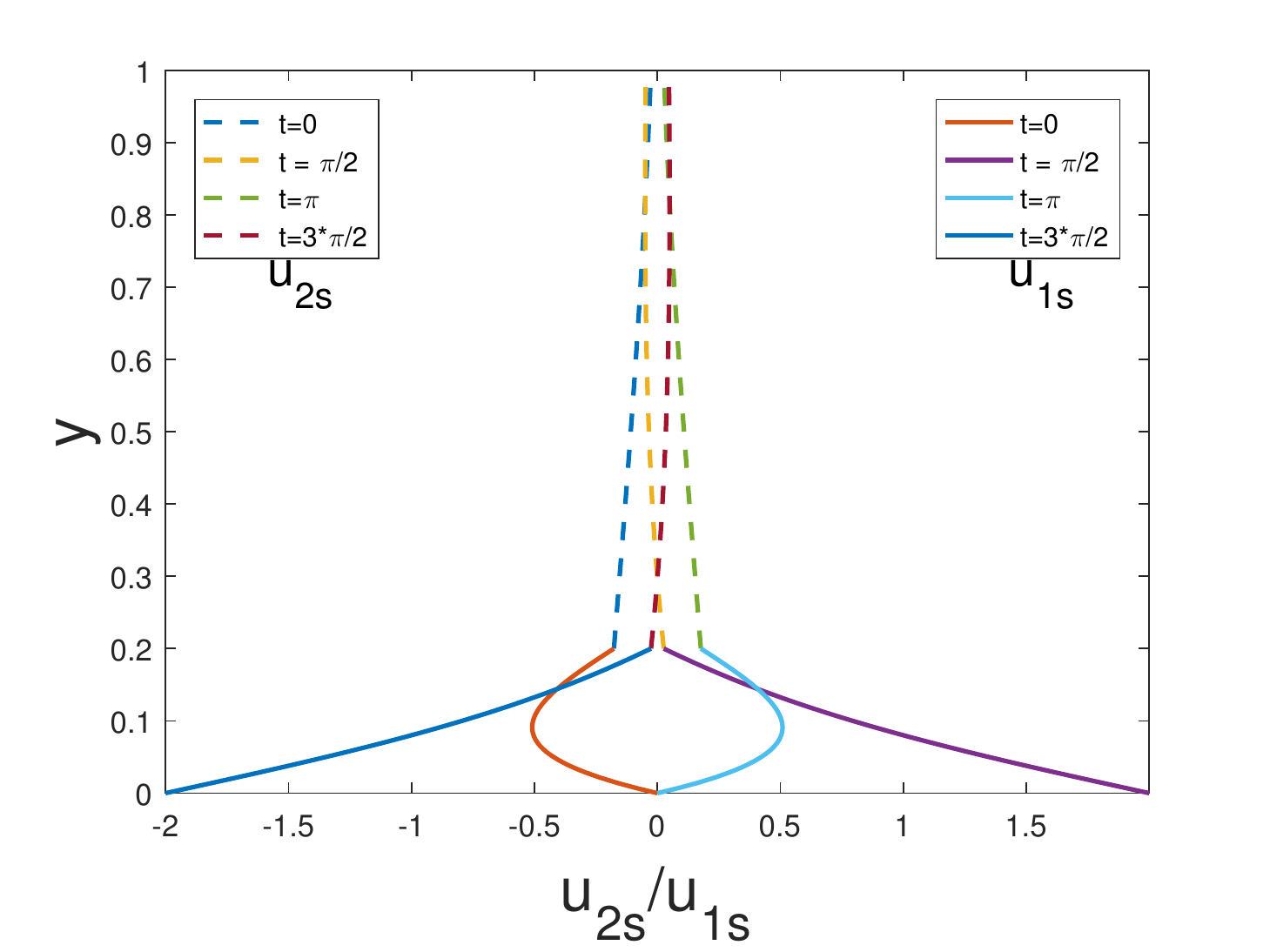}
		\caption{}
		\label{f5b}
	\end{subfigure}
	\caption{Steady-state velocity profiles for the lower (solid lines) and upper (broken lines) fluids when $U_0=2, h=0.2, \mu_1=0.01, \nu_1=0.01, \mu_2=0.2, \nu_2=0.22$, and $\omega=1$: (a) the plate oscillates as $U_0\cos(\omega t)$, and (b) the plate oscillates as $U_0\sin(\omega t)$. (Stokes' problem) (Use color in print)}
	\label{f5}
\end{figure}

Fig. \ref{f5}  shows steady periodic velocity profiles in the lower and upper fluids. Fig. \ref{f5a} and Fig. \ref{f5b} depict profiles corresponding to the cosine and the sine oscillations of the plate, respectively. Oscillations in the fluid velocities in both the layers are noticed from the figure, as expected.

\begin{figure}[ht!]
	\centering
	\includegraphics[width=.7\linewidth, height=8cm]{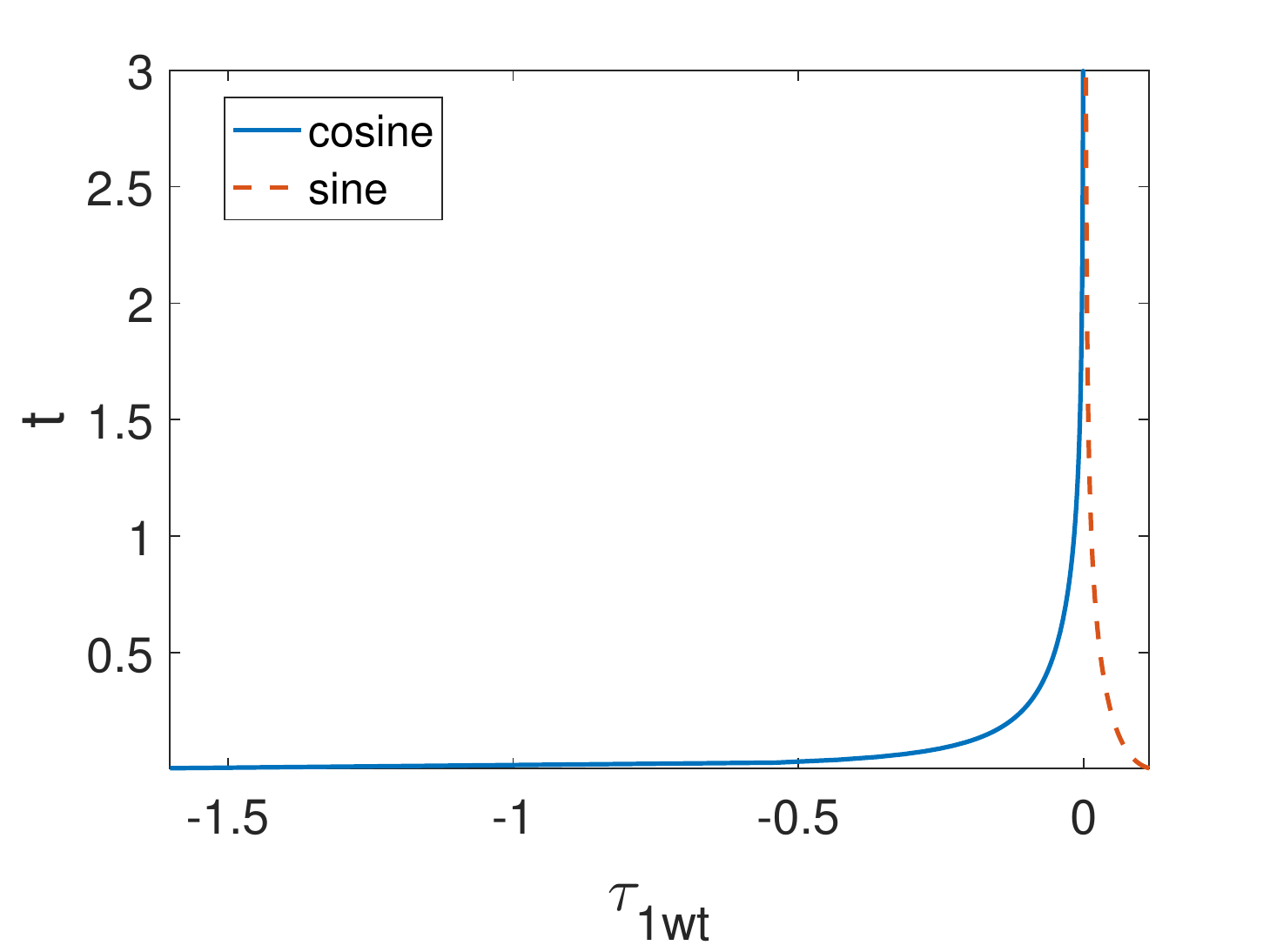} 
	\caption{The transient shear stress at the plate when $U_0=2, h=0.2, \mu_1=0.01, \nu_1=0.01, \mu_2=0.2, \nu_2=0.22$, and $\omega=1$. (Stokes' problem)}
	\label{f7}
\end{figure}

Fig. \ref{f7} illustrates transient wall shear stresses related to the cosine and the sine oscillations of the plate. The figure shows that at very small times the magnitude of transient wall shear stress for the cosine oscillations of plate is significantly bigger than that corresponding to the sine oscillations. However, in both the cases the transient wall shear stress dies out at around $t=1.5$.

\begin{figure}[ht!]
	\begin{subfigure}{0.55\textwidth}
		\includegraphics[width=0.95\linewidth, height=7cm]{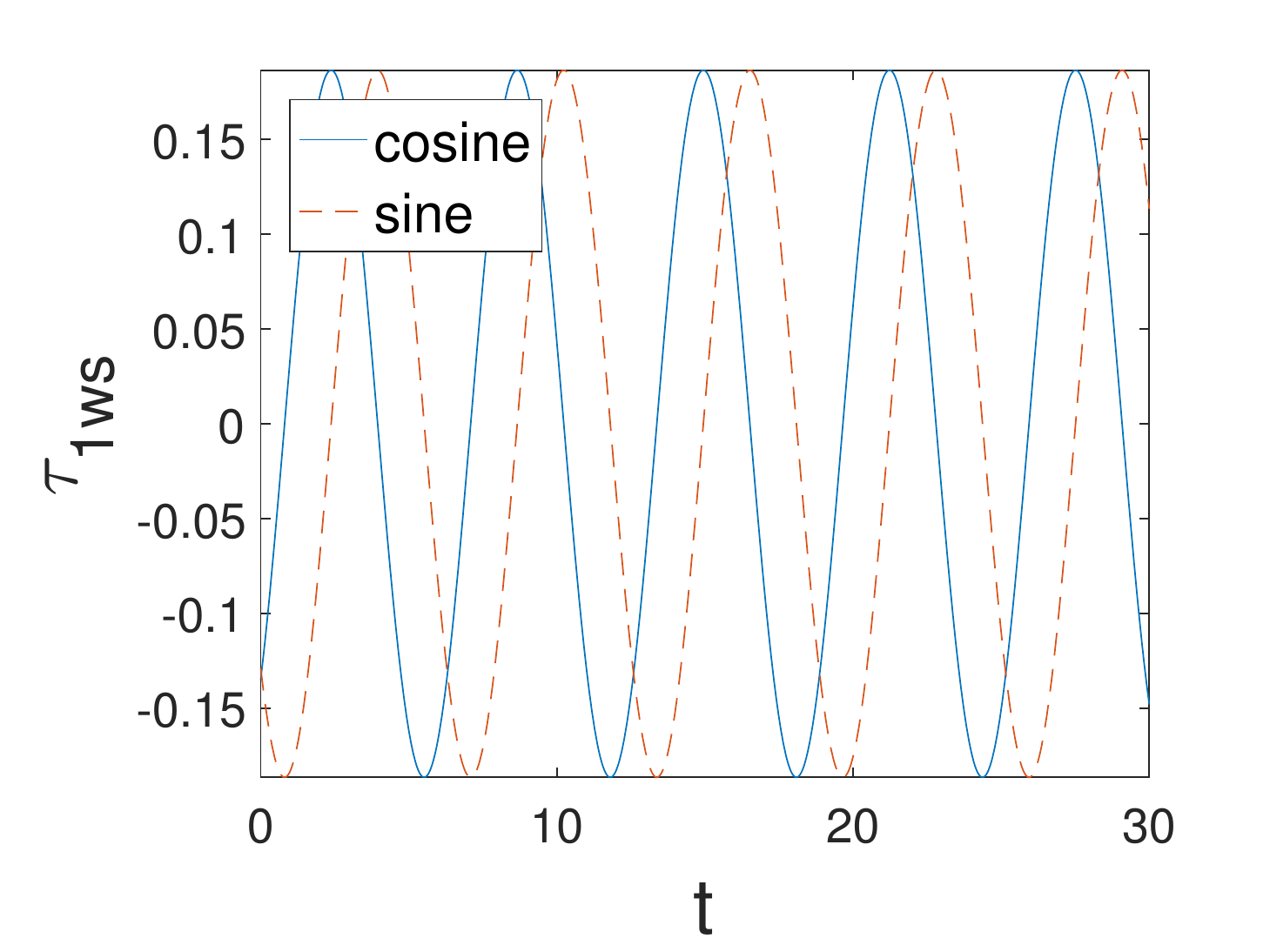} 
		\caption{}
		\label{}
	\end{subfigure}
	\begin{subfigure}{0.55\textwidth}
		\includegraphics[width=0.95\linewidth, height=7cm]{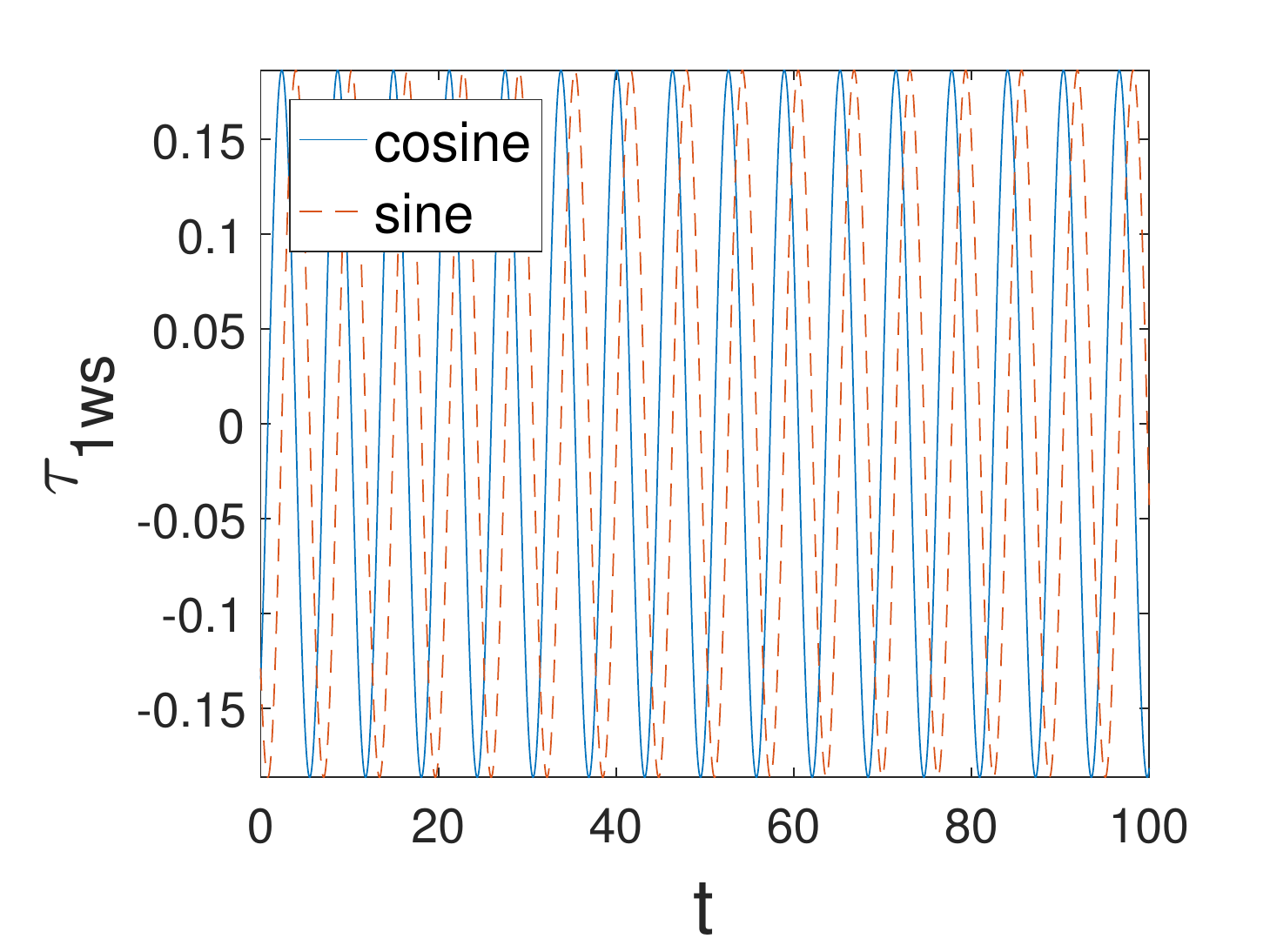}
		\caption{}
		\label{}
	\end{subfigure}
	\caption{The steady-state shear stress at the plate when $U_0=2, h=0.2, \mu_1=0.01, \nu_1=0.01, \mu_2=0.2, \nu_2=0.22$, and $\omega=1$: (a) the duration of motion $t\in[0,30]$, and (b) the duration of motion $t\in[0,100]$. (Stokes' problem)}
	\label{f45a}
\end{figure}

Fig. \ref{f45a} depicts steady-state wall shear stresses related to the cosine and the sine oscillations of the plate. Two intervals of time have been considered: a) the duration of motion $t\in[0,30]$, and b) the duration of motion $t\in[0,100]$. It is noticed from the figure that for all times, excepting for very small times, the steady-state wall shear stresses corresponding to the cosine and the sine oscillations of the plate have similar amplitudes with a phase difference.

\begin{figure}[h!]
	\begin{subfigure}{0.55\textwidth}
		\includegraphics[width=1\linewidth, height=8cm]{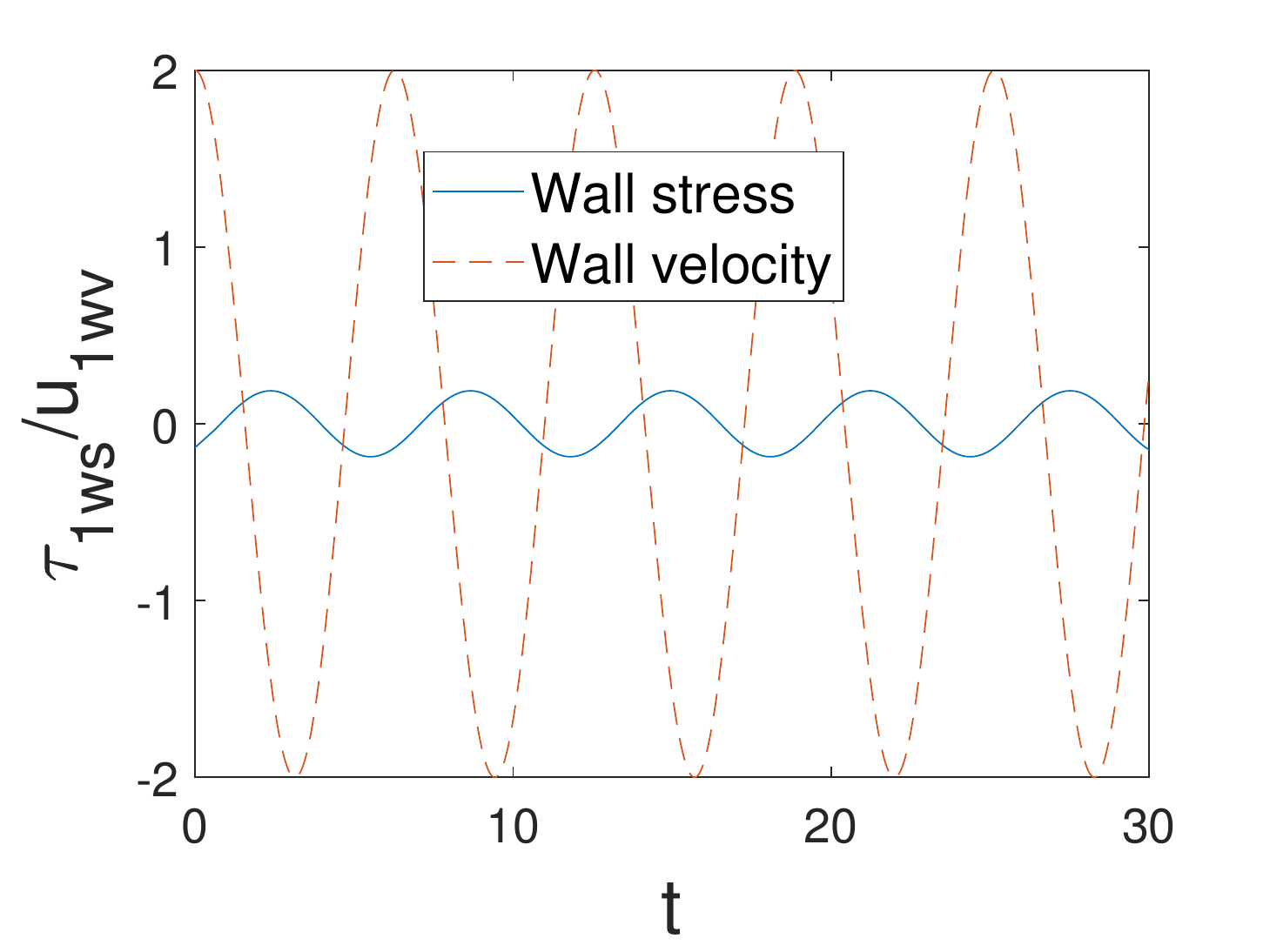} 
		\caption{}
		\label{f9a}
	\end{subfigure}
	\begin{subfigure}{0.55\textwidth}
		\includegraphics[width=1\linewidth, height=8cm]{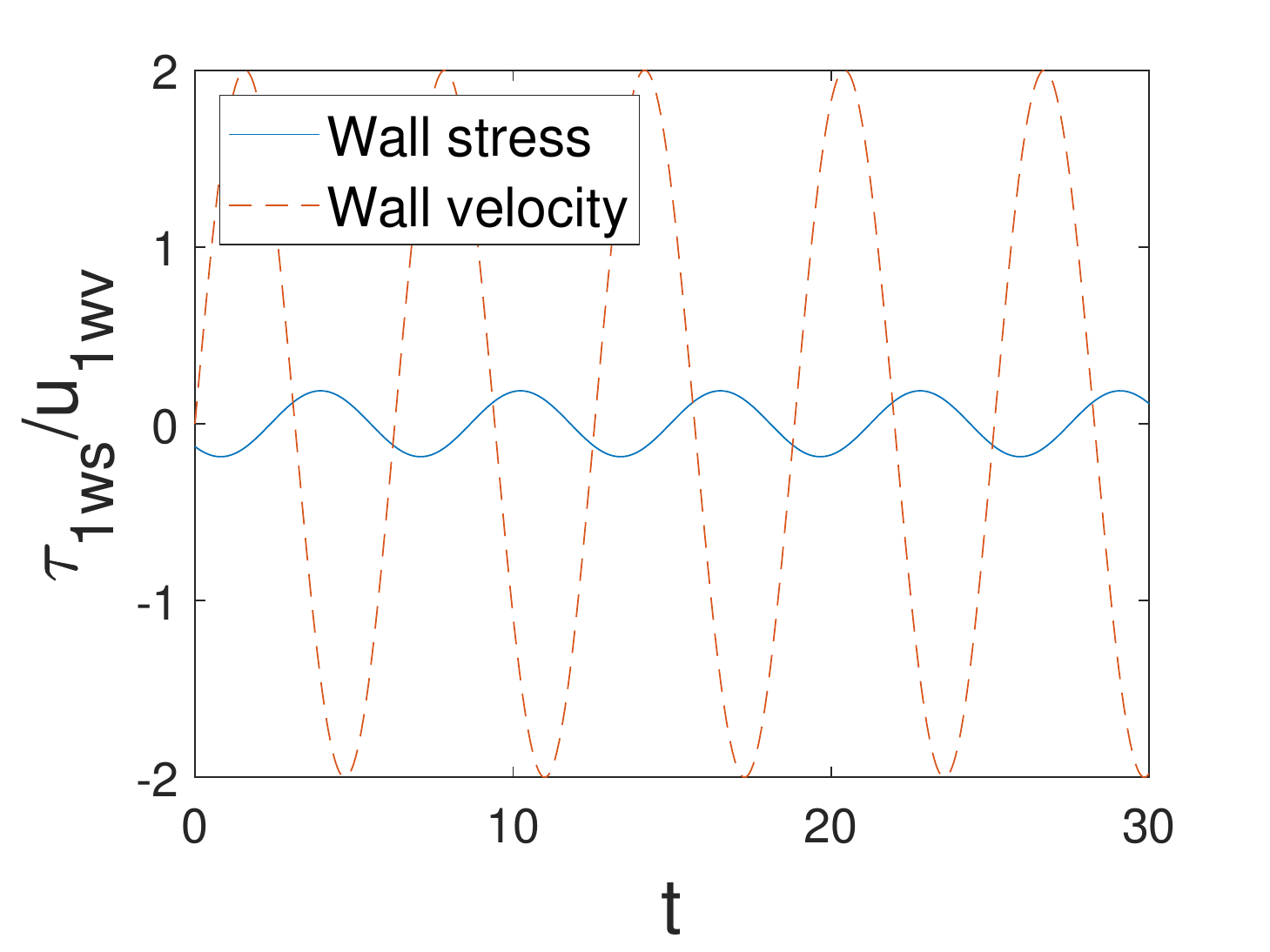}
		\caption{}
		\label{f9b}
	\end{subfigure}
	\caption{Wall velocity (solid line) and steady-state wall shear stress (broken line) when $U_0=2, h=0.2, \mu_1=0.01, \nu_1=0.01, \mu_2=0.2, \nu_2=0.22$, and $\omega=1$: (a) the plate oscillates as $U_0\cos(\omega t)$, and (b) the plate oscillates as $U_0\sin(\omega t)$. (Stokes' problem)}
	\label{f45b}
\end{figure}

Fig. \ref{f45b} compares steady-state wall shear stress with wall velocity. Fig. \ref{f9a} considers the cosine oscillations of the plate whereas Fig. \ref{f9b} do the sine oscillations of the plate. It is seen from the figure that for both the cosine and the sine oscillations of the plate, wall shear stress lags behind wall velocity. This can also be seen from expressions for steady-state wall shear stresses, which can be obtained from \eqref{43bc} and \eqref{43dc}, corresponding to wall velocity $U_0 \cos(\omega t)$ and $U_0 \sin(\omega t)$, respectively.



\clearpage
\subsection{Oscillatory Couette flow for a two-layer fluid}

\begin{figure}[h!]
	\begin{subfigure}{0.55\textwidth}
		\includegraphics[width=1\linewidth, height=8cm]{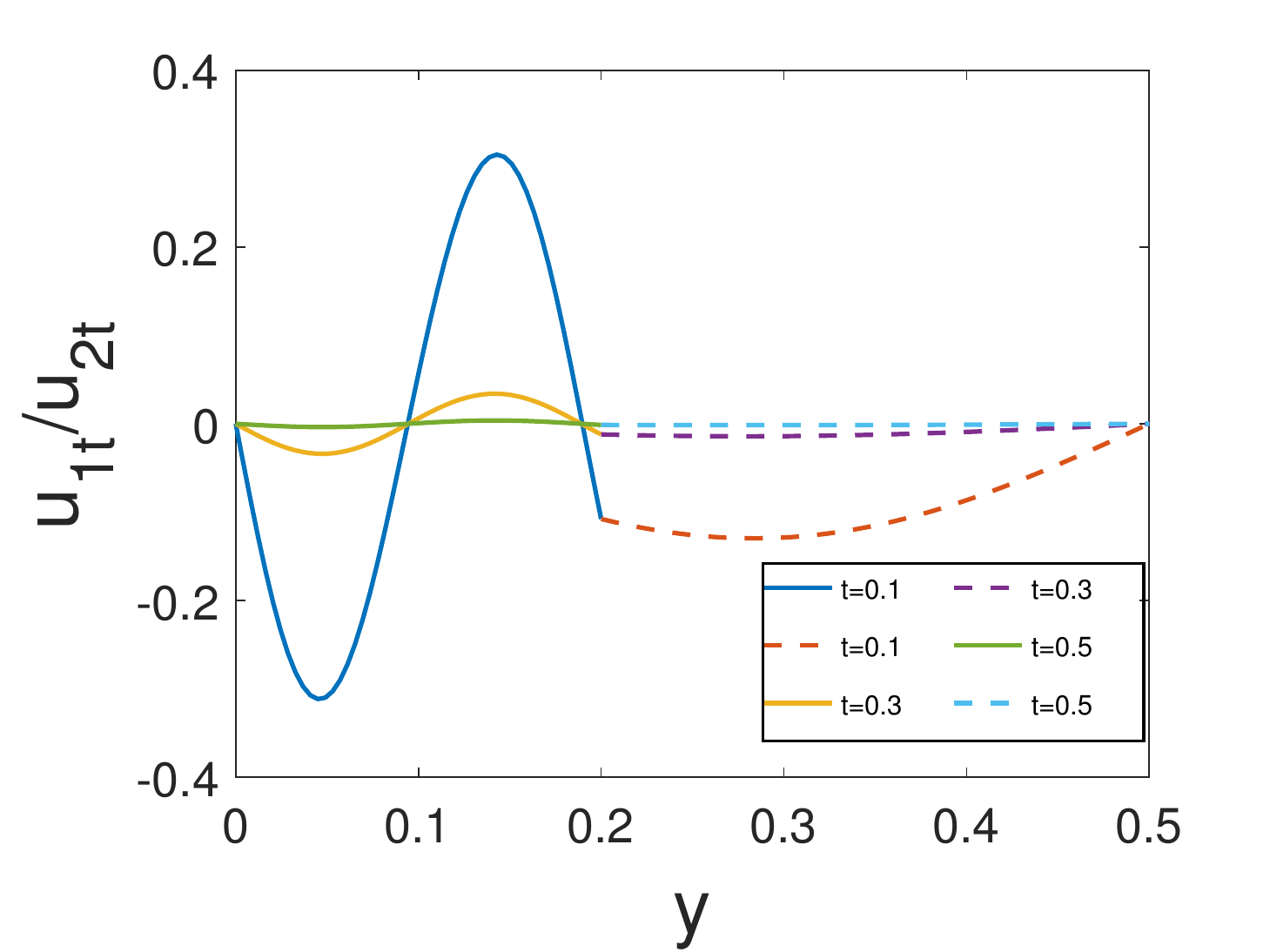} 
		\caption{}
		\label{f13a}
	\end{subfigure}
	\begin{subfigure}{0.55\textwidth}
		\includegraphics[width=1\linewidth, height=8cm]{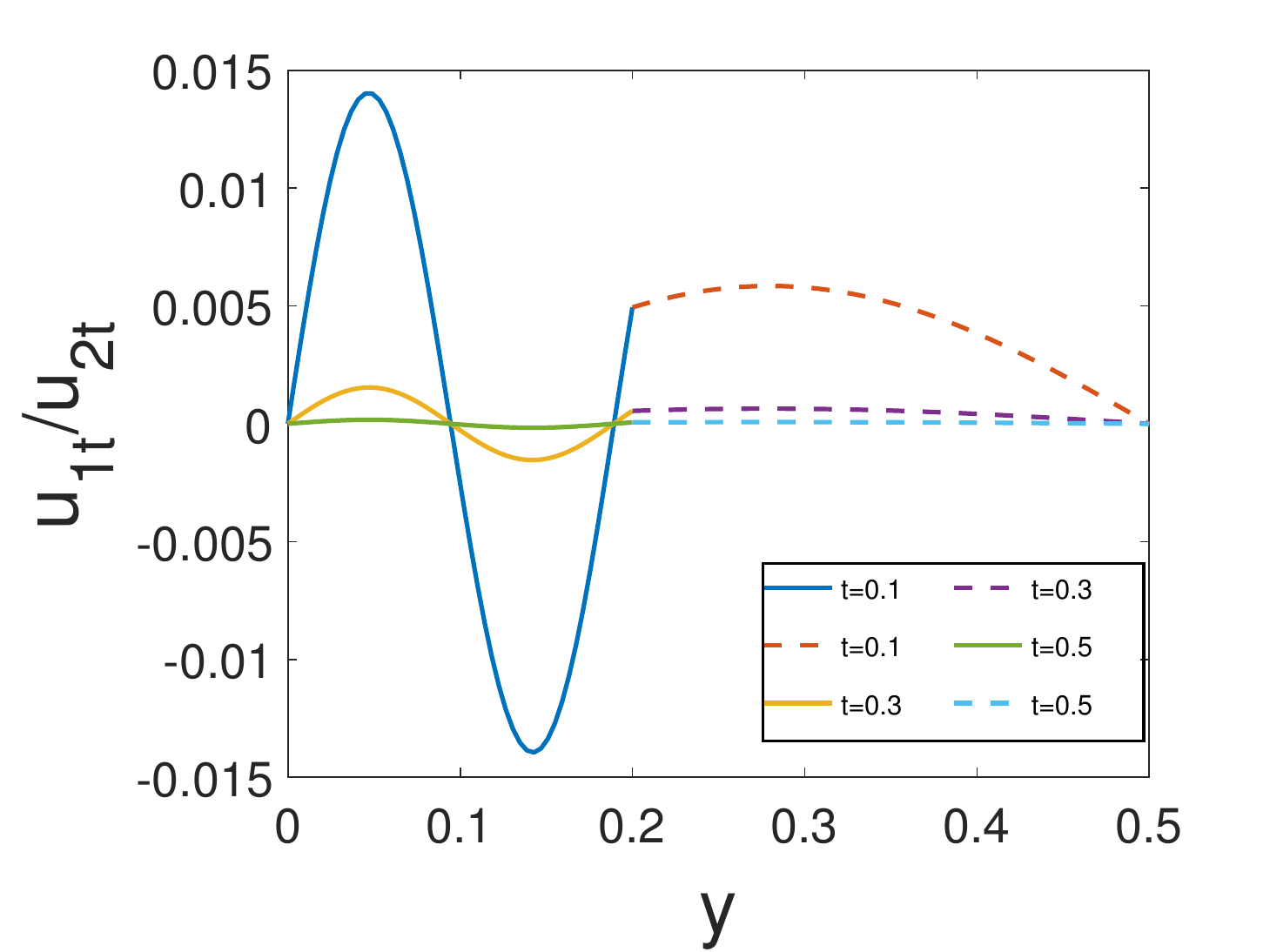}
		\caption{}
		\label{f13b}
	\end{subfigure}
	\caption{Profiles for the transient velocity fields for the lower (solid lines) and upper (broken lines) fluids when $U_0=2, h=0.2, H=0.5, \mu_1=0.01, \nu_1=0.01, \mu_2=0.2, \nu_2=0.22$, and $\omega=1$: (a) the plate oscillates as $U_0\cos(\omega t)$, and (b) the plate oscillates as $U_0\sin(\omega t)$. (Couette flow) (Use color in print)}
	\label{f13}
\end{figure}

Fig. \ref{f13} illustrates transient velocity profiles for the lower(water) and upper(corn oil) fluids for the cosine and the sine oscillations of the plate. Profiles corresponding to the cosine oscillations of the plate are presented in Fig. \ref{f13a}, and those related to the sine oscillations are depicted in Fig. \ref{f13b}. It is noticed from the figure that the transient velocities for both the lower and upper fluids die out very rapidly for both the forms of oscillations of the plate. The transient velocities disappear rapidly because of exponentials in their expressions, \eqref{5}, \eqref{6}, \eqref{5t}, and \eqref{6t}. On Fig. \ref{f13a}, at $t=0.1$, the maximum velocity (absolute value) in the lower fluid is about 0.3, and it occurs around $y=0.05$ and $y=0.14$. At the same time, the maximum velocity (absolute value) in the upper fluid is approximately 0.1, which occurs at $y=0.2$ (i.e. at the interface of the fluids). On Fig. \ref{f13b}, at $t=0.1$, the maximum velocity (absolute value) in the lower fluid is approximately 0.014, and it occurs about $y=0.05$ and $y=0.14$. At the same time, the maximum velocity (absolute value) in the upper fluid is 0.005, which occurs at the interface of the fluids.

\begin{figure}[h!]
	\begin{subfigure}{0.55\textwidth}
		\includegraphics[width=1\linewidth, height=8cm]{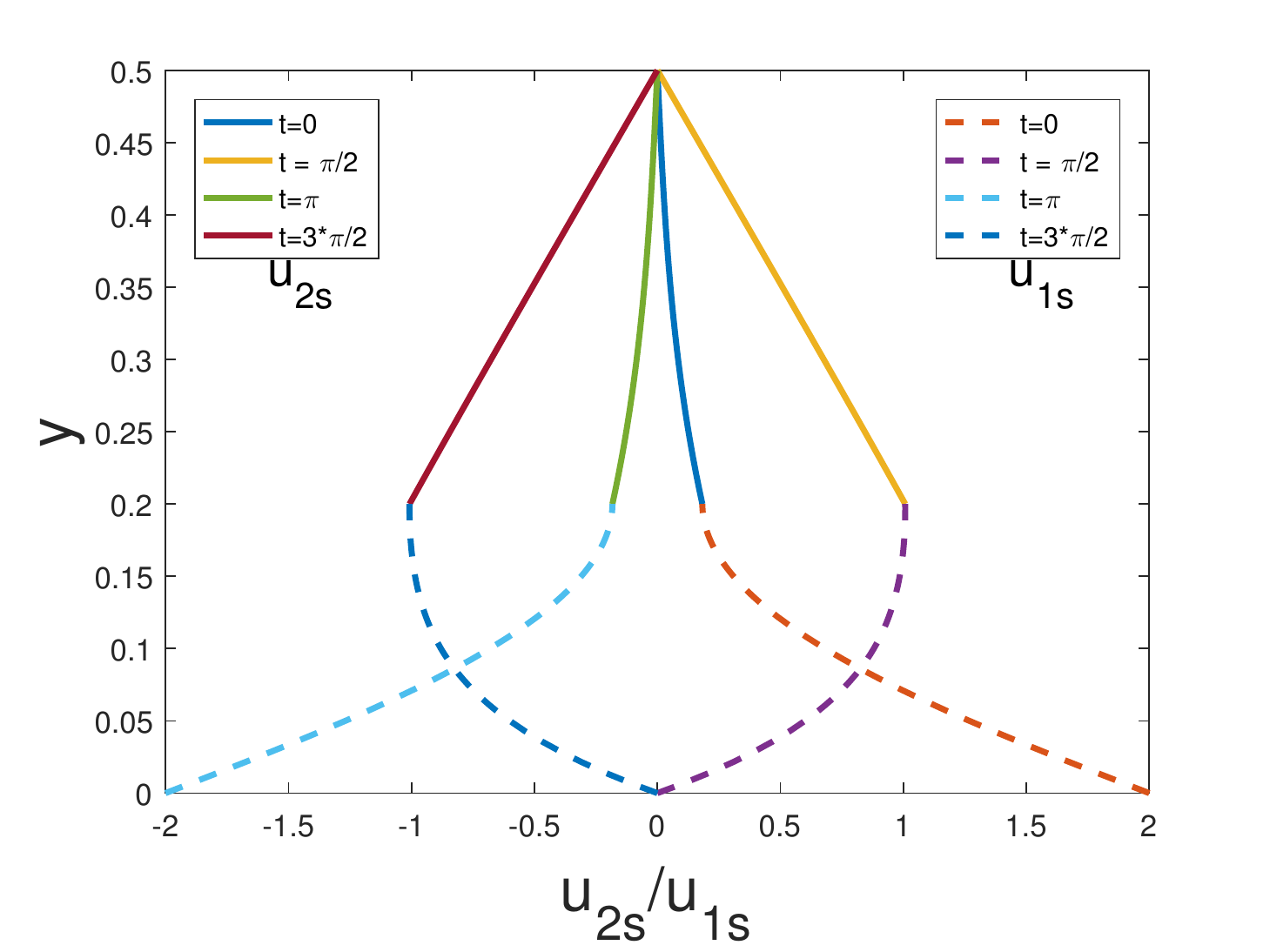} 
		\caption{}
		\label{f8a}
	\end{subfigure}
	\begin{subfigure}{0.55\textwidth}
		\includegraphics[width=1\linewidth, height=8cm]{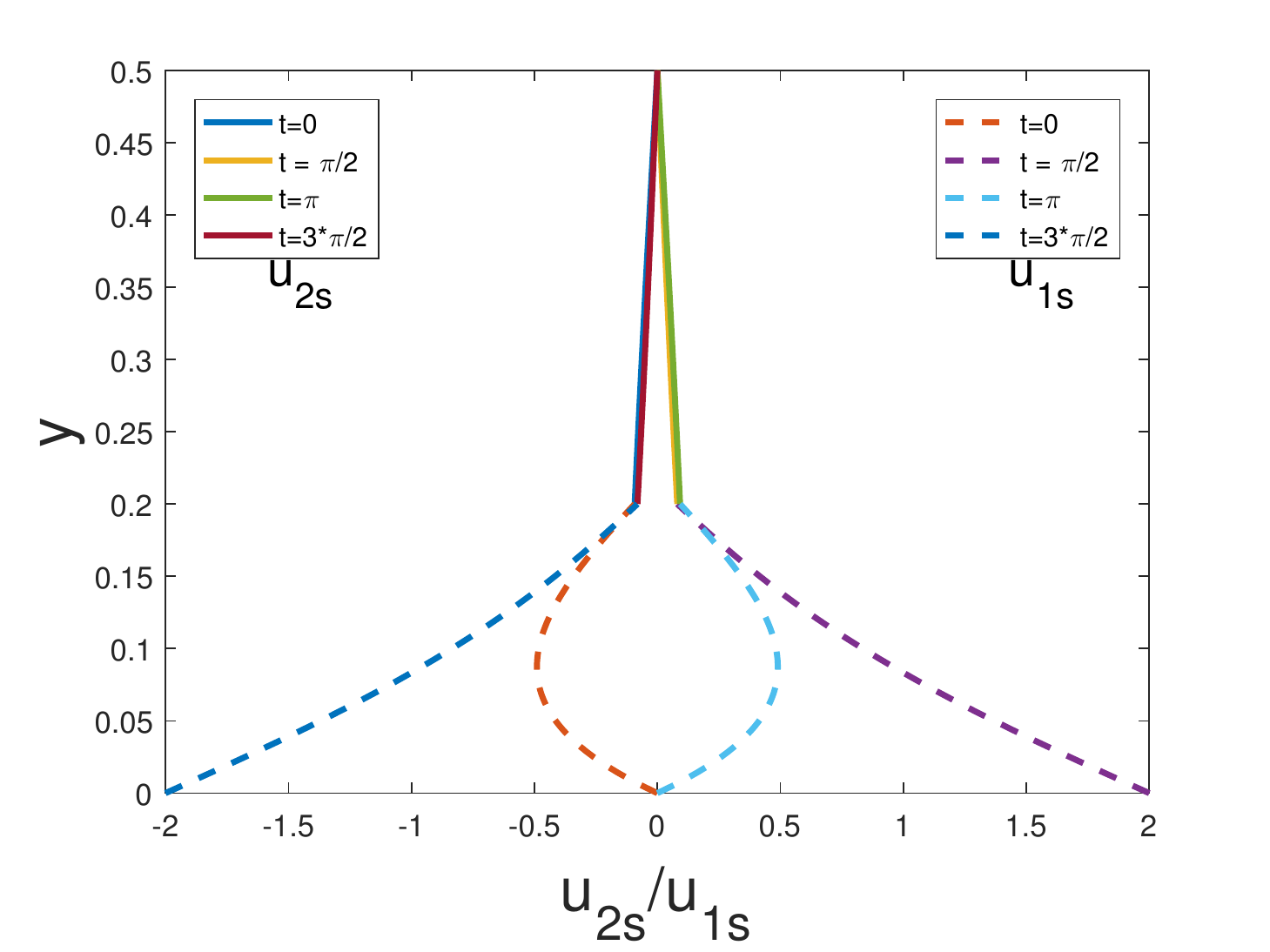}
		\caption{}
		\label{f8b}
	\end{subfigure}
	\caption{Steady-state velocity profiles for the lower (broken lines) and upper (solid lines) fluids when $U_0=2, h=0.2, H=0.5, \mu_1=0.01, \nu_1=0.01, \mu_2=0.2, \nu_2=0.22$, and $\omega=1$: (a) the plate oscillates as $U_0\cos(\omega t)$, and (b) the plate oscillates as $U_0\sin(\omega t)$. (Couette flow) (Use color in print)}
	\label{f8}
\end{figure}

Fig. \ref{f8} shows steady periodic velocity profiles for the lower and upper fluids. Fig. \ref{f8a} and Fig. \ref{f8b} illustrate profiles related to the cosine and the sine oscillations of the plate, respectively. Oscillations in fluid velocities for both the fluids are noticed from the figure, as expected.

\begin{figure}[h!]
	\begin{subfigure}{0.55\textwidth}
		\includegraphics[width=1\linewidth, height=8cm]{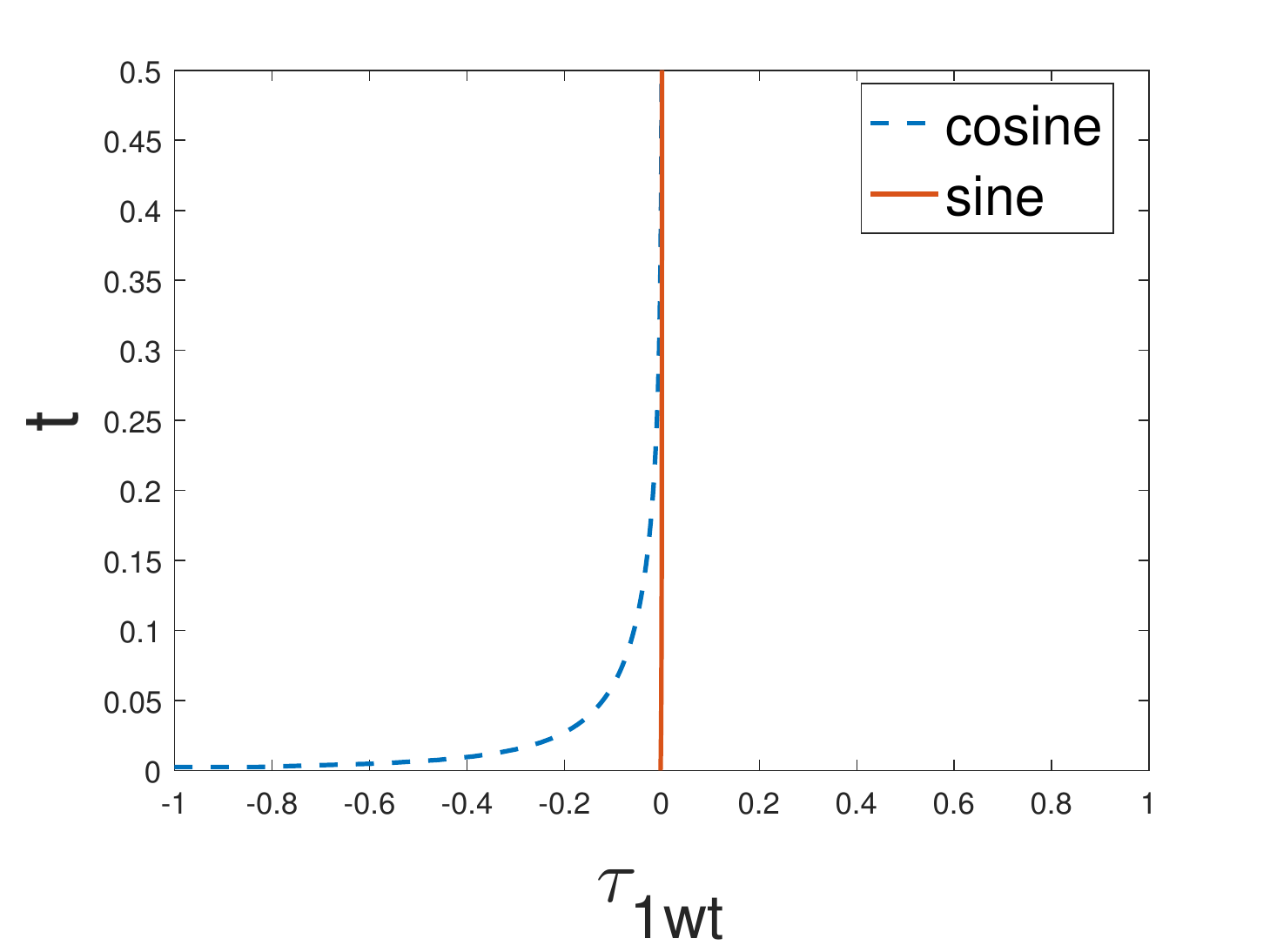} 
		\caption{}
		\label{f12a}
	\end{subfigure}
	\begin{subfigure}{0.55\textwidth}
		\includegraphics[width=1\linewidth, height=8cm]{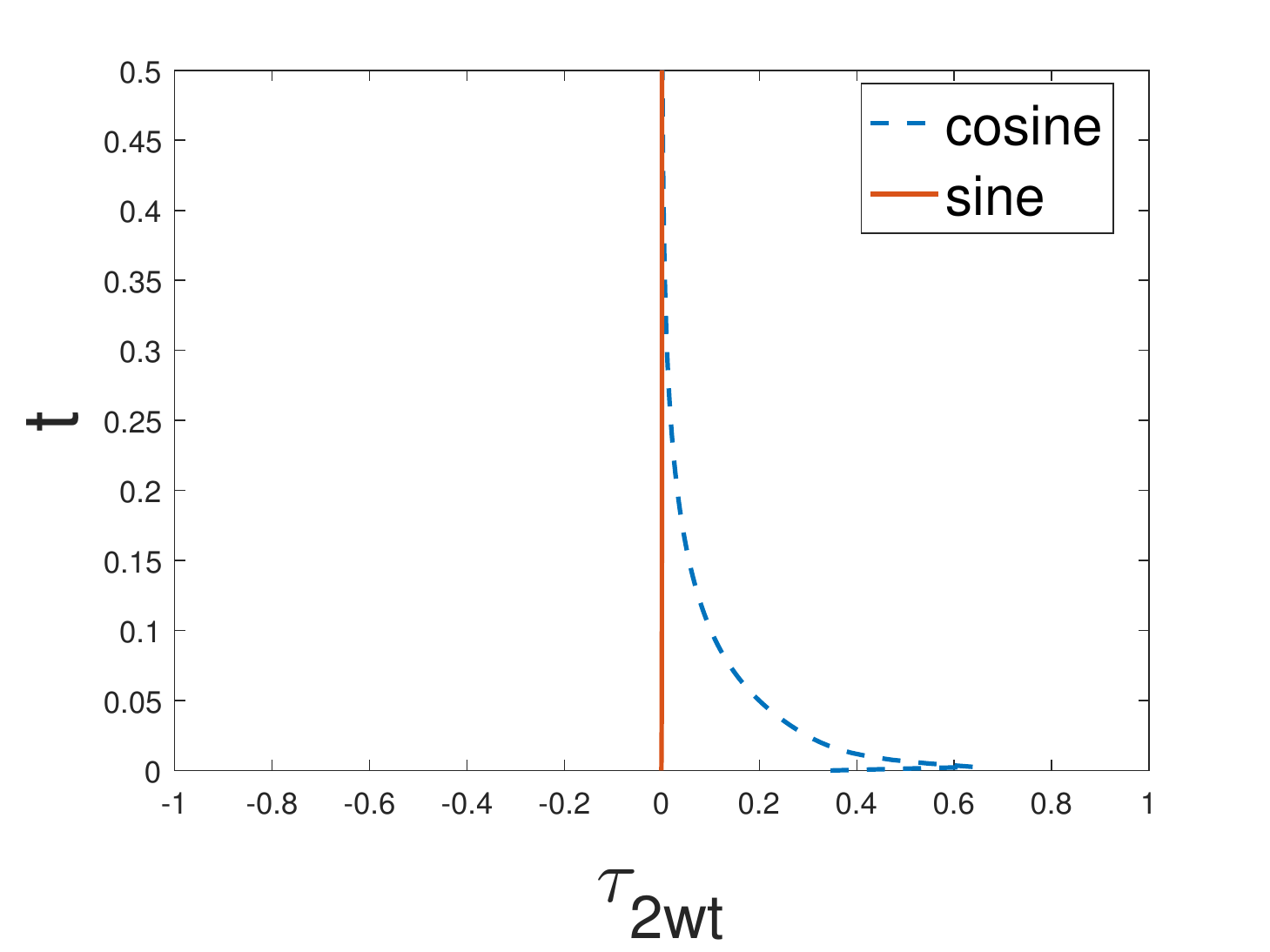}
		\caption{}
		\label{f12b}
	\end{subfigure}
	\caption{The transient shear stresses at the oscillating and fixed plates when $U_0=2, h=0.2, H=0.5, \mu_1=0.01, \nu_1=0.01, \mu_2=0.2, \nu_2=0.22$, and $\omega=1$: (a) oscillating plate, and (b) stationary plate. (Couette flow)}
	\label{f12}
\end{figure}

Fig. \ref{f12} illustrates transient shear stresses at the oscillating and fixed plates. Both the forms of oscillations of the plate are considered in the figure. While panel (a) illustrates transient shear stress at the oscillating plate, panel (b) depicts transient shear stress at the fixed plate. It is noticed from the figure that at both the plates, the transient shear stress corresponding to the sine oscillations of the plate is zero for all values of time $t$. Contrary to the sine oscillations, the magnitudes of the transient shear stresses at the plates related to the cosine oscillations of the plate are quite significant for very small values of time $t$. However, these transient shear stresses related to the cosine oscillations of the plate disappear very rapidly.

\begin{figure}[h!]
	\begin{subfigure}{0.55\textwidth}
		\includegraphics[width=1\linewidth, height=8cm]{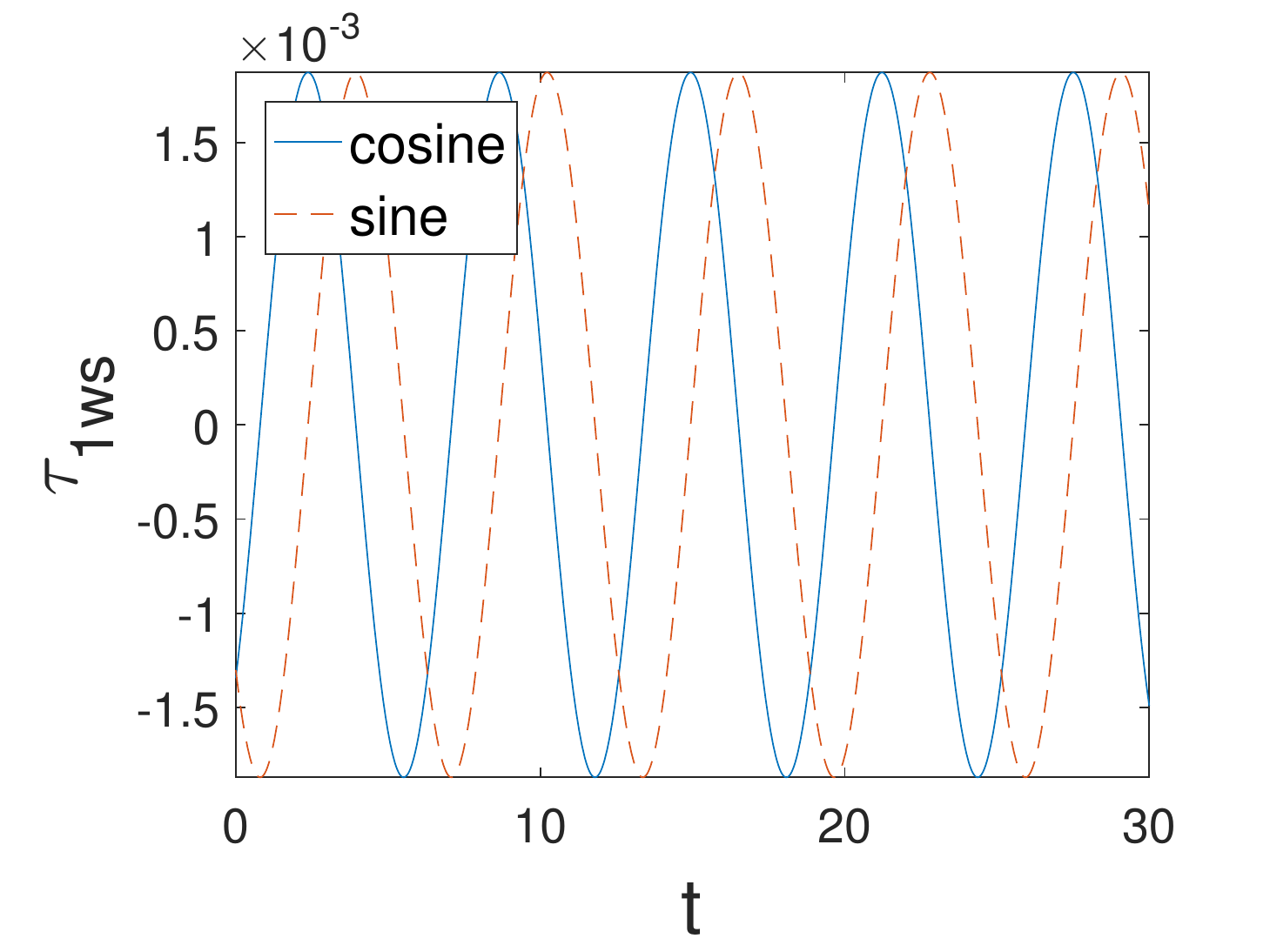} 
		\caption{}
		\label{}
	\end{subfigure}
	\begin{subfigure}{0.55\textwidth}
		\includegraphics[width=1\linewidth, height=8cm]{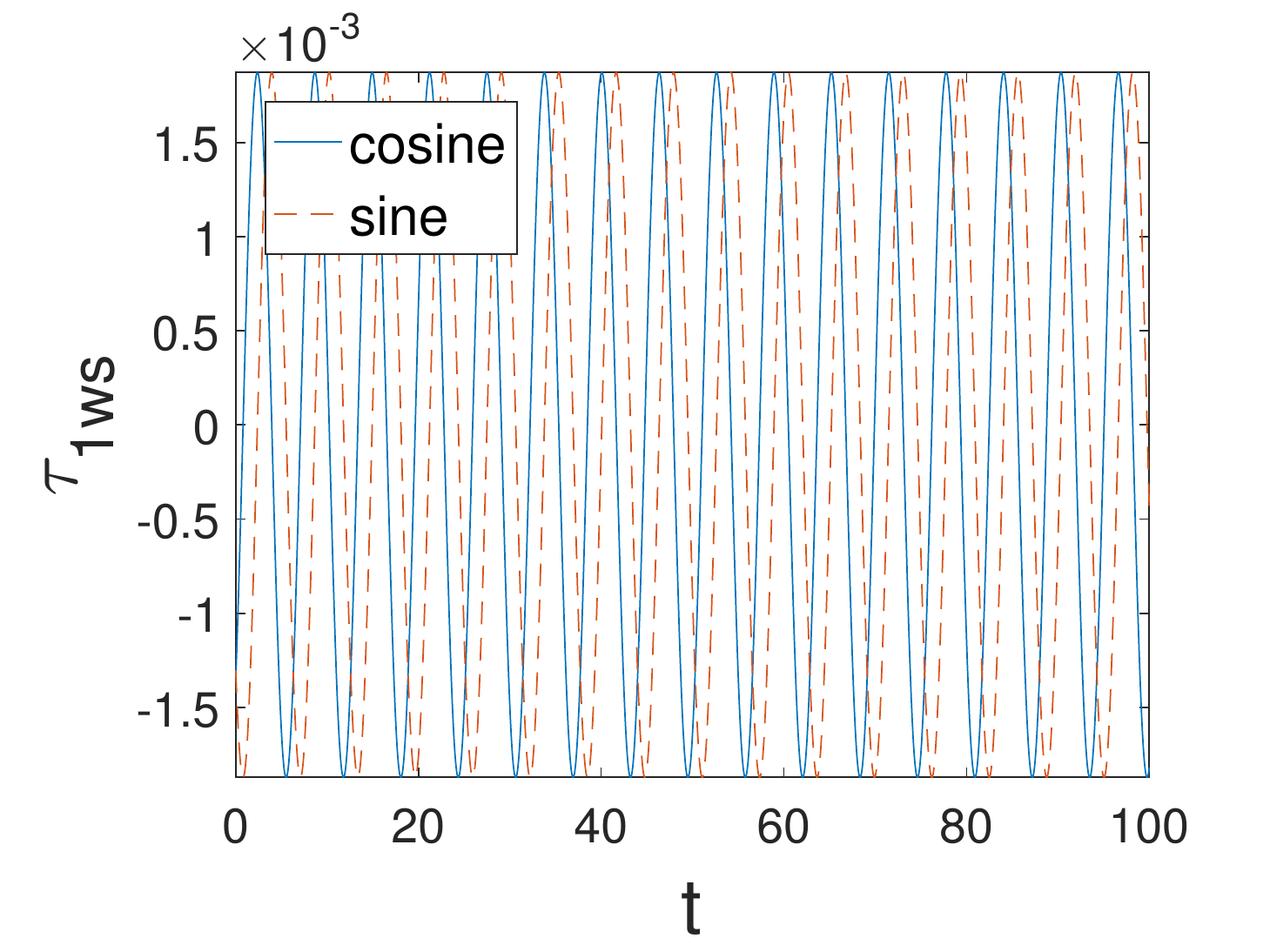}
		\caption{}
		\label{}
	\end{subfigure}
	\caption{The steady-state shear stress at the oscillating plate when $U_0=2, h=0.2, H=0.5, \mu_1=0.01, \nu_1=0.01, \mu_2=0.2, \nu_2=0.22$, and $\omega=1$: (a) the duration of motion $t\in[0,30]$, and (b) the duration of motion $t\in[0,100]$. (Couette flow)}
	\label{f10}
\end{figure}

\begin{figure}[h!]
	\begin{subfigure}{0.55\textwidth}
		\includegraphics[width=1\linewidth, height=8cm]{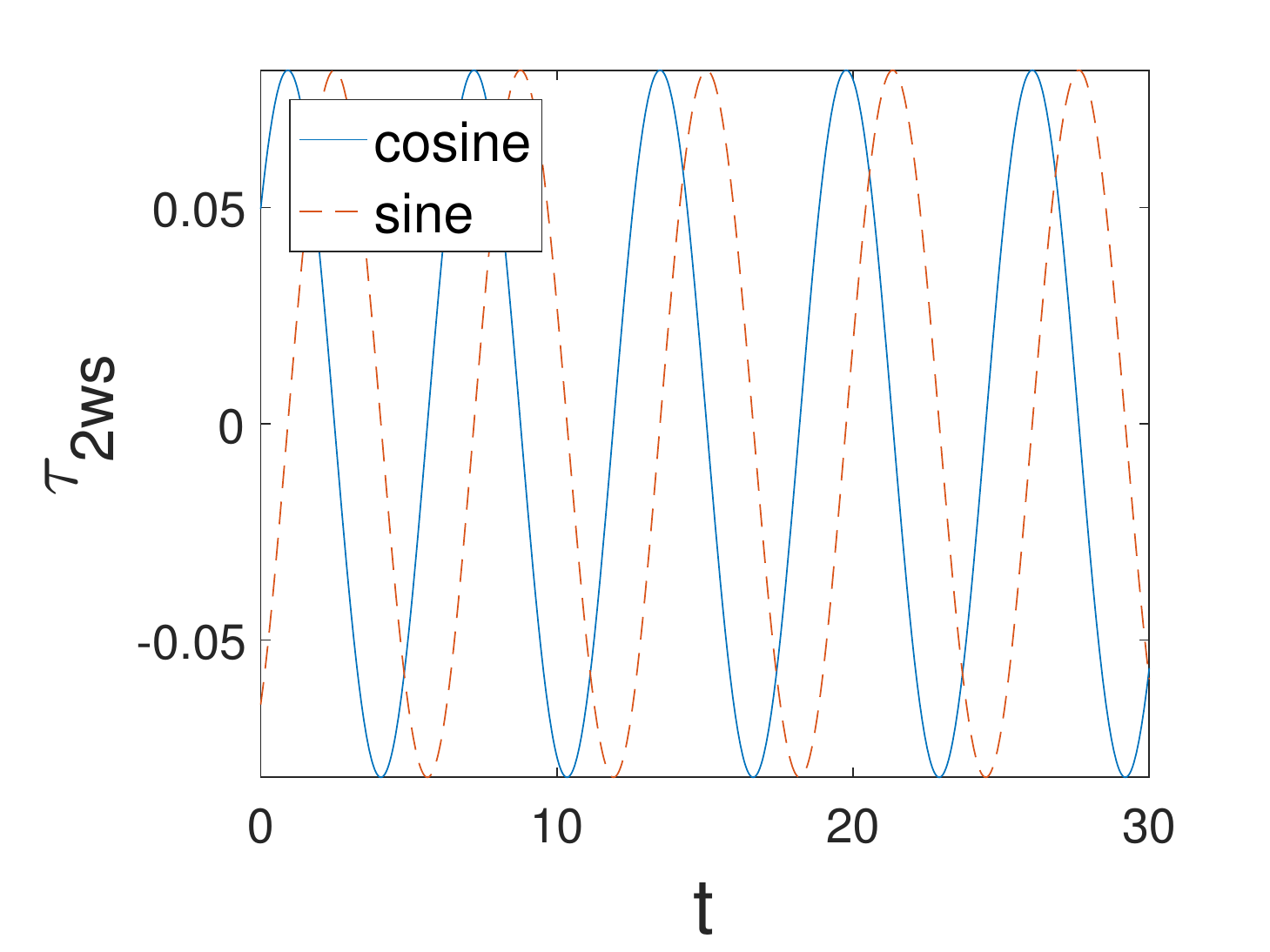} 
		\caption{}
		\label{}
	\end{subfigure}
	\begin{subfigure}{0.55\textwidth}
		\includegraphics[width=1\linewidth, height=8cm]{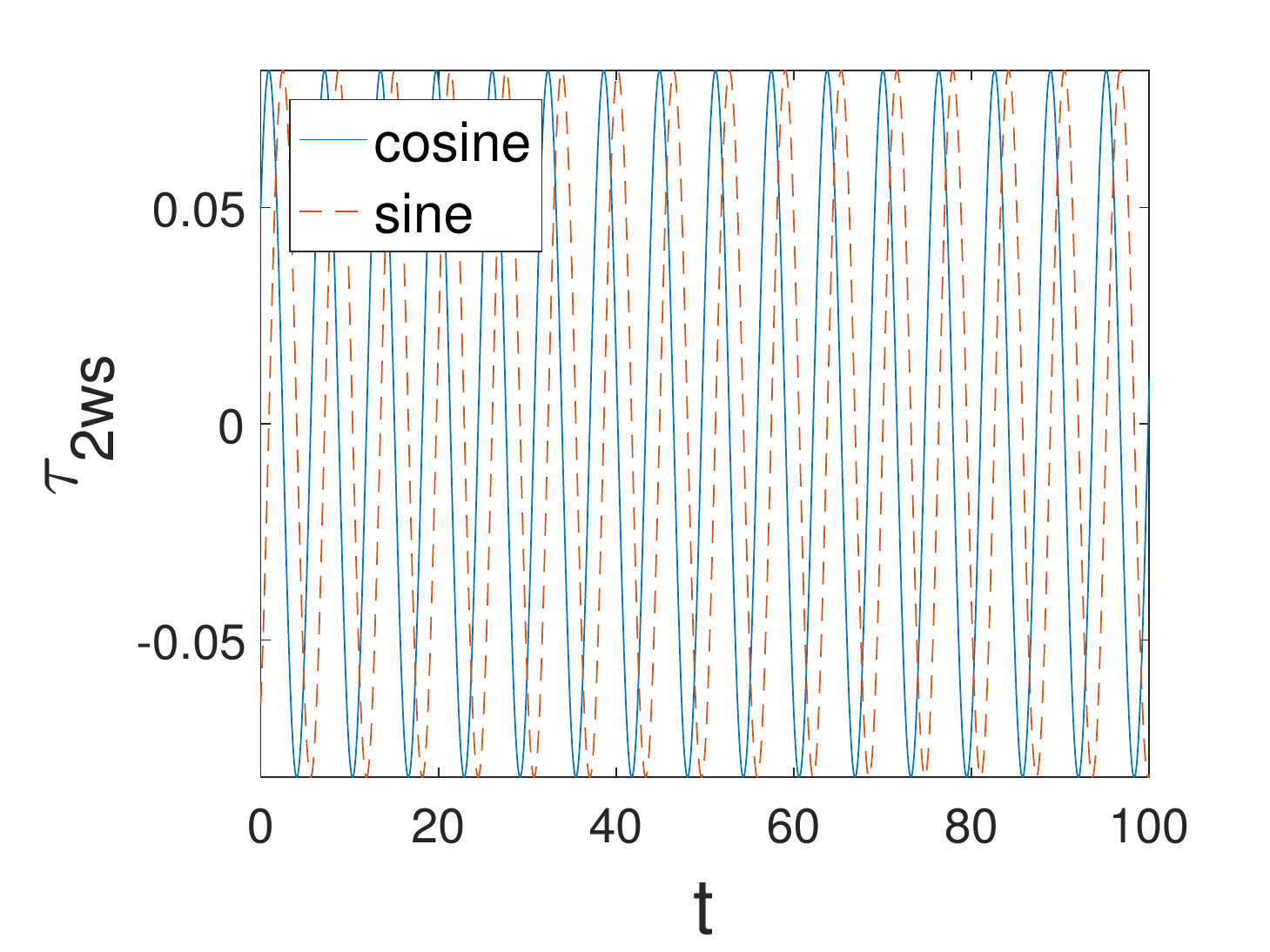}
		\caption{}
		\label{}
	\end{subfigure}
	\caption{The steady-state shear stress at the fixed plate when $U_0=2, h=0.2, H=0.5, \mu_1=0.01, \nu_1=0.01, \mu_2=0.2, \nu_2=0.22$, and $\omega=1$: (a) the duration of motion $t\in[0,30]$, and (b) the duration of motion $t\in[0,100]$. (Couette flow)}
	\label{f11}
\end{figure}

Figs. \ref{f10} and \ref{f11} illustrate steady-state shear stresses at the oscillating and stationary plates, respectively. Both the cosine and the sine oscillations of the plate are considered in the figures. In both the figures, two intervals of time $t$ have been considered: $t\in[0,30]$( panel (a)) and $t\in[0,100]$ (panel (b)). The steady periodic shear stresses at the oscillating plate for the cosine and the sine oscillations of the plate have similar amplitudes with a phase difference for all the times, with the exceptions for very small times, as seen from Fig. \ref{f10}. The same pattern is observed for steady-state shear stresses at the fixed plate for the cosine and the sine oscillations of the plate, as noticed from Fig. \ref{f11}.

\clearpage

\section{Conclusions}\label{s5}
In this paper, we have mathematically analyzed the unsteady motion of a two-layer fluid caused by oscillatory motion of a flat plate along its length. We have considered two cases: (i) the two-layer fluid is bounded only by the oscillating plate (Stokes’ second problem), and (ii) the two-layer fluid is confined between two parallel plates, one of which oscillates while the other is held stationary (oscillatory Couette flow). In each of the cases, both cosine and sine oscillations of the plate have been considered. It is assumed that the fluids are immiscible, and that the flat interface between the fluids remains flat for all times. Initially, the fluids and the plate have been at rest and then suddenly, the plate starts to oscillate along its length. The Laplace transform method has been employed to solve the associated initial-boundary value problems. And the Bromwich inversion integral and Cauchy’s residue theorem have been utilized to find inverse Laplace transforms of the velocity fields. For both the layers of fluid, we have obtained analytical expressions for starting and steady-state velocity fields. A starting velocity field is the sum of the transient and steady-state velocity fields and valid for small values of time $t$. The transient dies out gradually (or rapidly) as the time $t$ progresses. The steady-state velocity field represents the time periodic fluid motion and valid for large values of time $t$. Explicit expressions for transient and steady-state velocity fields have been presented. Also, we have computed transient and steady-state shear stresses at the boundaries of the flows. We have retrieved related previously known results for single-layer fluid flows from the results derived here. 

We have presented illustrative example of each of the Stokes’ problem and the Couette flow. In the examples, we have considered the two-layer fluid in which a layer of corn oil (lighter) lies over a layer of water (heavier), noting that the water rests on the oscillating plate. We have utilized the results obtained to get some physical insights into the particular problems considered. For the Stokes’ problem case, we have found that in both the layers of fluid, the time required to reach steady-state flow when the plate is subjected to cosine oscillations is much less than that when the plate is subjected to sine oscillations. The study has revealed that for a given oscillation frequency of the plate, the lower fluid which is adjacent to the plate attains steady-state flow much earlier than the upper fluid. It holds true for both the cosine and the sine oscillations of the plate. Again, irrespective of the form of oscillations of the plate and true for both the lower and upper fluids, when the frequency of oscillations increases, the time to reach steady-state flow decreases. For the Couette flow case, it is found that for both cosine and sine oscillations of the plate, the transient velocity disappears very rapidly in both the layers of fluids. Note that in the illustrative example, we have considered the case where the thickness of the lower fluid is less than that of the upper fluid.

It is our believe that the present study further our understanding of the motion of a two-layer viscous fluid caused by a vibrating wall in an engineering application. Again, the analytical results obtained in this paper could be used for validation of future numerical works dealing with problems similar to the current ones but consider wavy interface between the fluids. Note that in this paper, we have considered flat interface between the fluids. Moreover, this work may provide a basis for future researches on Stokes’ second problem and oscillatory Couette flow for a two-layer fluid where one or both the fluids are non-Newtonian. Furthermore, the present work may also be applicable to heat conduction in a two-layer composite solid subject to the following initial-boundary conditions. Initially, the composite solid has been kept at a uniform zero temperature and then suddenly, the surface of the solid comes into contact with a heat source with sinusoidal temperature variation. Again,  we hope that the present work will serve as a starting point for future works on Stokes’ second problem and oscillatory Couette flow for a two-layer fluid dealing with the effects of viscosity ratio, density ratio, and layer thickness ratio (for Couette flow) on the velocity fields and on other physical properties of interest.

\appendix
\section{Evaluation of inverse Laplace transform using Bromwich inversion integral}\label{a}
 Here we evaluate 
$\mathcal{L}^{-1}\bigg(\dfrac{s\exp(-a\sqrt{s})}{s^{2}+\omega^{2}}\bigg)$, $a>0$,
 where $\mathcal{L}^{-1}$ is inverse Laplace transform operator. The inverse transform has been utilized to obtain the velocity fields for the lower and upper fluids, \eqref{43} and \eqref{43a}, in section \ref{221}.
  
The Laplace transform of a function $f(t)$ is defined as follows:  
\begin{equation}
F(s)=\mathcal{L}(f(t))=\int_{0}^{\infty} f(t)\exp(-st) dt,\label{ae1}
\end{equation}
where $\mathcal{L}$ is Laplace transform operator, and $s$ is transform variable.

For time $t>0$, the inverse Laplace transform of $F(s)$ is given by the following formula: 

\begin{equation}
	f(t)=\mathcal{L}^{-1}(F(s))=\dfrac{1}{2\pi i}\int_{\gamma-i\infty}^{\gamma+i\infty} F(s)\exp(st) ds.\label{ae1c}
\end{equation}
The inversion formula is called the Bromwich inversion integral\cite{king_billingham_otto_2003,sos}. In the formula, $\gamma$ is a real number, and it must be chosen such that all the singularities of $F(s)$ ( poles, branch points or essential singularities) lie to the left of the line $s=\gamma$ in the complex s-plane. The integration in the formula is to be evaluated along the line $s=\gamma$. However, in practice, the integration is performed along a closed contour composed of  the line $s=\gamma$ and a circular arc on the left of the line. This is done in order to facilitate the use of Cauchy's residue theorem\cite{Brown}. 

\begin{figure}[h!]
	\centering
	\includegraphics[width=0.45\linewidth, height=0.3\textheight]{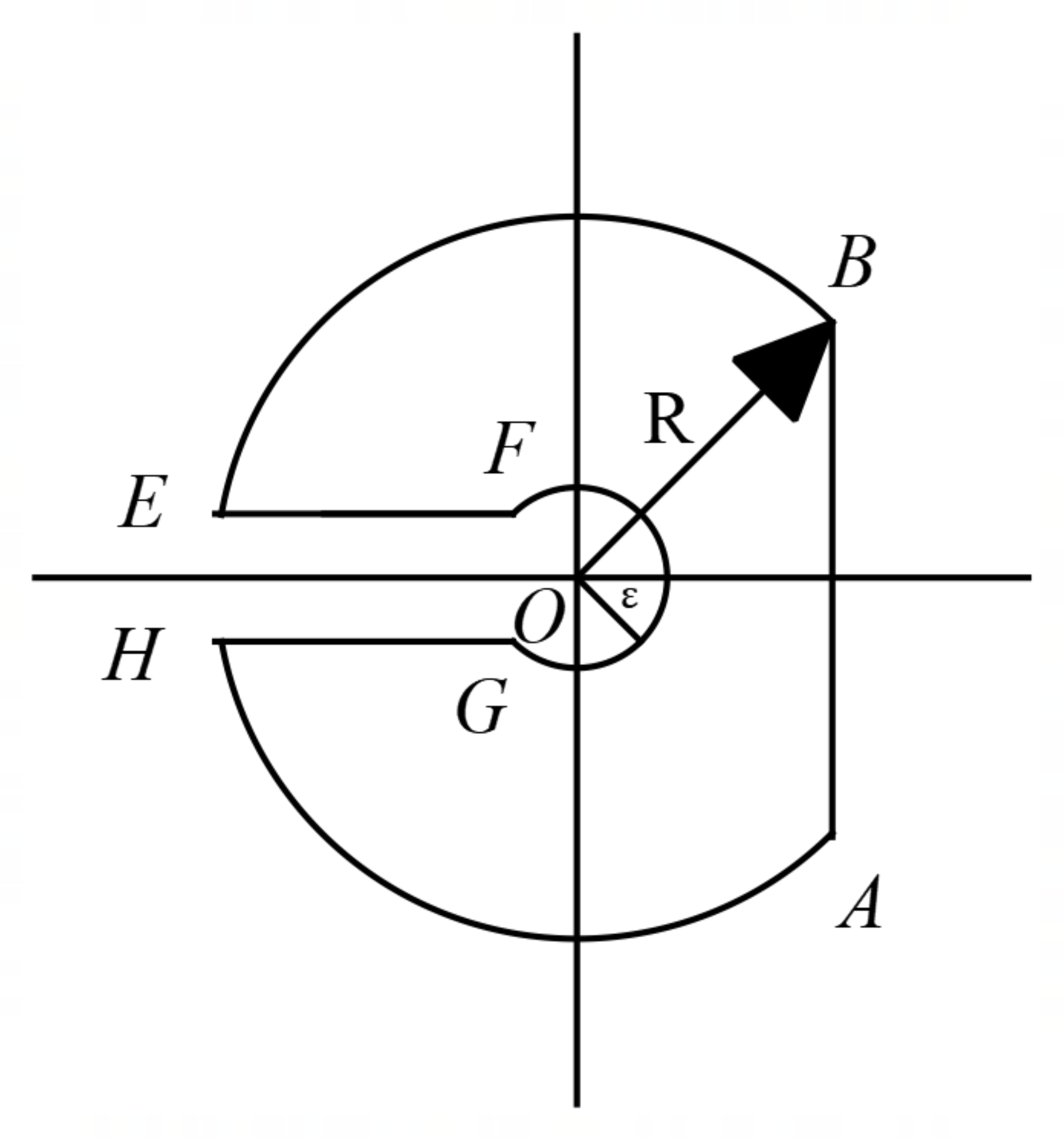}
	\caption{Bromwich contour integral}
	\label{fig:bc}
\end{figure}

For the case in hand,
\begin{equation}
F(s)=\bigg(\dfrac{s\exp(-a\sqrt{s})}{s^{2}+\omega^{2}}\bigg)\label{a3}.
\end{equation}
$F(s)$ has simple poles at $s=-i\omega$ and $s=i\omega$. Also, since $F(s)$ contains a fractional power, $s^\frac{1}{2}$, the point $s=0$ is a branch point.
We now consider the contour integral 
\begin{equation}
	\dfrac{1}{2\pi i}\oint_C F(s)\exp(st) ds,
\end{equation}
where $C$ is the contour of Fig. \eqref{fig:bc}. The contour $C$ is a keyhole contour. We have drawn the keyhole contour in order to exclude the branch cut along the negative real axis. The contour $C$ is composed of line $AB$, circular arc $BE$, line $EF$, a small circle around the origin $O$ of radius $\epsilon$, line $GH$, and circular arc $HA$. Arcs $BE$ and $HA$ are arcs of a circle of radius $R$ with center at the origin $O$.
Now, it follows from formula \eqref{ae1c} that
\begin{equation}
	f(t)=\lim_{R\to\infty}\dfrac{1}{2\pi i}\int_{\gamma-iL}^{\gamma+iL} F(s)\exp(st) ds,\label{ae1a}
\end{equation}
since $L=\sqrt{R^2-\gamma^2}$.
Again, it follows from \eqref{ae1a} that
\begin{align}
	f(t)=\lim_{\begin{subxarray}R&\to\infty\\\epsilon&\to 0\end{subxarray}}\bigg[&\dfrac{1}{2\pi i}\oint_C F(s)\exp(st) ds\notag\\&-\dfrac{1}{2\pi i}\{\int_{BE}+\int_{EF}+\int_{FG}+\int_{GH}+\int_{HA}\}F(s)\exp(st) ds\bigg]\label{a6}.
\end{align}
Now, on the arcs $BE$ and $HA$, $F(s)\to 0$ exponentially as $R\to\infty$, since real part of $s^\frac{1}{2}$ is positive. Therefore,
\begin{equation}
\int_{BE} F(s)\exp(st)ds=0\quad\text{and}\quad\int_{HA} F(s)\exp(st)ds=0\label{a7}.
\end{equation}
We now compute 
\begin{equation*}
	\int_{FG} F(s)\exp(st)ds.
\end{equation*}
We use $s=\epsilon e^{i\theta}$, where $\theta$ runs from $\pi$ to $-\pi$, to parameterize the small circle $FG$. On this circle,
$s^\frac{1}{2}=\epsilon^\frac{1}{2} e^{\frac{i\theta}{2}}$, so that
\begin{align*}
		\int_{FG} \dfrac{s\exp(st-as^\frac{1}{2})}{s^2+\omega^2} ds
		=&\int_{\pi}^{-\pi} \dfrac{i\epsilon^2 e^{2i\theta}\exp(\epsilon e^{i\theta}t-a\epsilon^\frac{1}{2} e^{\frac{i\theta}{2}})}{\epsilon^2 e^{2i\theta}+\omega^2}d\theta.
\end{align*}
Therefore, as $\epsilon\to 0$, 
\begin{align}
	\int_{FG} \dfrac{s\exp(st-as^\frac{1}{2})}{s^2+\omega^2} ds=0\label{a8}.
\end{align}
To evaluate the integrals along the lines $EF$ and $GH$, we parameterize the lines using $s=r\exp(i\pi)=-r$ and $s=r\exp(-i\pi)=-r$, respectively.  Along the line $EF$,$s^{\frac{1}{2}}=r^{\frac{1}{2}}\exp(\frac{i\pi}{2})=ir^{\frac{1}{2}}$, and along the line $GH$, $s^{\frac{1}{2}}=r^{\frac{1}{2}}\exp(\frac{-i\pi}{2})=-ir^{\frac{1}{2}}$. Therefore, as $\epsilon\to 0$ and $R\to \infty$,
\begin{align*}
	\int_{EF} \dfrac{s\exp(st-as^\frac{1}{2})}{s^2+\omega^2} ds=&-\int_{0}^{\infty} \dfrac{r\exp(-rt-iar^\frac{1}{2})}{r^2+\omega^2} dr
\end{align*}
and
\begin{align*}
	\int_{GH} \dfrac{s\exp(st-as^\frac{1}{2})}{s^2+\omega^2} ds=\int_{0}^{\infty} \dfrac{r\exp(-rt+iar^\frac{1}{2})}{r^2+\omega^2} dr.
\end{align*}
Therefore, we have
\begin{align}
	 \{\int_{EF}+\int_{GH}\}\dfrac{s\exp(st-as^\frac{1}{2})}{s^2+\omega^2} ds=2i\int_{0}^{\infty} \dfrac{r\exp(-rt)\sin(a\sqrt{r})}{r^2+\omega^2} dr\label{a9}.
\end{align}
 Using \eqref{a7}, \eqref{a8}, and \eqref{a9} in \eqref{a6}, we have
\begin{align}
	f(t)=\lim_{\begin{subxarray}R&\to\infty\\\epsilon&\to 0\end{subxarray}}\bigg[&\dfrac{1}{2\pi i}\oint_C F(s)\exp(st) ds\bigg]-\frac{1}{\pi}\int_{0}^{\infty} \dfrac{r\exp(-rt)\sin(a\sqrt{r})}{r^2+\omega^2} dr\label{a10}.
\end{align} 
Now, as $R\to\infty$, all the poles of the integrand of the integration around the closed contour $C$ lie within $C$. The integrand has simple poles at $s=-i\omega$ and $s=i\omega$, noting that $F(s)$ is given by equation \eqref{a3}. Therefore, according to Cauchy's residue theorem\cite{Brown}, 
 \begin{align}
 	\oint_C F(s)\exp(st) ds=&\oint_C\dfrac{s\exp(st-as^\frac{1}{2})}{s^2+\omega^2} ds\notag\\=&2\pi i\bigg[\lim_{s\to-i\omega}\bigl\{(s+i\omega)\dfrac{s\exp(st-as^\frac{1}{2})}{s^2+\omega^2}\bigr\}\notag\\&+\lim_{s\to i\omega}\bigl\{(s-i\omega)\dfrac{s\exp(st-as^\frac{1}{2})}{s^2+\omega^2}\bigr\}\bigg]\notag\\=&2\pi i\exp\bigg(-a\sqrt{\dfrac{\omega}{2}}\bigg)\cos\bigg(\omega t -a\sqrt{\dfrac{\omega}{2}}\bigg)\label{a11}.
 \end{align}
We now use \eqref{a11} in \eqref{a10} to obtain the following result:
\begin{align}
	\mathcal{L}^{-1}\bigg(\dfrac{s\exp(-a\sqrt{s})}{s^{2}+\omega^{2}}\bigg)=\exp\bigg(-a\sqrt{\dfrac{\omega}{2}}\bigg)\cos\bigg(\omega t -a\sqrt{\dfrac{\omega}{2}}\bigg)-\frac{1}{\pi}\int_{0}^{\infty} \dfrac{r\exp(-rt)\sin(a\sqrt{r})}{r^2+\omega^2} dr,
\end{align}
for $a>0$.
\section{Deduction of velocity field for oscillatory Couette flow for a single-layer fluid as a special case, when the plate oscillates as $U_0  \cos(\omega t)$}\label{ab} 
\subsection{Deduction from the velocity field for the lower fluid}
Here we provide hints about the deductions of the steady periodic and transient velocity fields for oscillatory Couette flow for a single-layer fluid, \eqref{a2sc1} and \eqref{6tc1}, respectively, in section \ref{321}. The deductions are made from the steady periodic and transient velocity fields for the lower fluid, \eqref{a1} and \eqref{5}, in the same section. The steady periodic velocity field for the lower fluid, \eqref{a1}, contains $A,B,g_1(y)$, and $g_2(y)$, which are defined in equations \eqref{6c}, \eqref{6d}, \eqref{6e}, and \eqref{6f}, respectively. If we let $h=H$, $\mu_1=\mu_2=\mu$(say the viscosity of the single-layer fluid), and $\nu_1=\nu_2=\nu$(say the kinematic viscosity of the single-layer fluid) in equations \eqref{6c}, \eqref{6d}, \eqref{6e}, and \eqref{6f}, we have
\begin{align}  
	&A=\cos b_1\sinh b_1,\label{b3}\\                               
	&B=\sin b_1\cosh b_1,\label{b4}\\
	&g_1(y)=-\sin b_2\cosh b_2 \sin b_1\sinh b_1+\cos b_2\sinh b_2 \cos b_1\cosh b_1, \label{b5}\\                               
	&g_2(y)=\sin b_2\cosh b_2 \cos b_1\cosh b_1+\cos b_2\sinh b_2 \sin b_1\sinh b_1,\label{b6} \\                                  
	 &\text{where}\quad b_1=\sqrt{\dfrac{\omega}{2\nu}}H,\quad \text{and}\quad b_2=\sqrt{\dfrac{\omega}{2\nu}}y\label{b7}.
\end{align}
If we substitute \eqref{b3}-\eqref{b6} into equation \eqref{a1}, we obtain the steady periodic velocity field for oscillatory Couette flow for the single-layer fluid, \eqref{a2sc1}.

Again, the transient velocity field for the lower fluid, \eqref{5}, contains $F_2 (k_m ),F_3 (k_m )$, and $F_4 (k_m )$, which are defined in equations \eqref{e1}-\eqref{e3}, respectively. If we put $h=H$, $\mu_1=\mu_2=\mu$(say the viscosity of the single-layer fluid), and $\nu_1=\nu_2=\nu$(say the kinematic viscosity of the single-layer fluid)  in equations \eqref{e1}-\eqref{e3}, we obtain
\begin{align}
	&F_2(k_m)=\cos(k_m \frac{H}{\sqrt{\nu}}),\label{b9}\\
		&F_3(k_m)=\sin(k_m \frac{H}{\sqrt{\nu}}),\label{b10}\\
		& F_4(k_m)=\frac{H}{\sqrt{\nu}}\cos(k_m \frac{H}{\sqrt{\nu}})\label{b11}.
\end{align}
Note that here, $k_m$ is as defined in \eqref{km}. Substituting equations \eqref{b9}-\eqref{b11} into equation \eqref{5}, we obtain the transient velocity field for oscillatory Couette flow for the single-layer fluid, \eqref{6tc1}.
\subsection{Deduction from the velocity field for the upper fluid}\label{ab2}
Here we give hints on the deductions of the steady periodic and transient velocity fields for oscillatory Couette flow for a single-layer fluid, equations \eqref{a2sc1} and \eqref{6tc1} in section \eqref{321}. The deductions are made from the corresponding results for the upper fluid, equations \eqref{a2} and  \eqref{6} in the same section. The steady-state velocity field for the upper fluid, \eqref{a2}, contains $A,B,g_3(y)$, and $g_4(y)$, which are defined in equations \eqref{6c}, \eqref{6d}, \eqref{6a}, and \eqref{6b}, respectively. If we let $h=0$(meaning that the lower fluid ceases to exist), $\mu_1=\mu_2=\mu$(say the  viscosity of the single-layer fluid), and $\nu_1=\nu_2=\nu$(say the kinematic viscosity of the single-layer fluid)  in equations \eqref{6c}, \eqref{6d}, \eqref{6a}, and \eqref{6b}, $A$ and $B$ reduce to those defined in equations \eqref{b3} and \eqref{b4}, and $g_3(y)$ and $g_4(y)$ reduce to as follows:
\begin{align}
	&g_3(y)=\cos(b_1-b_2)\sinh(b_1-b_2),\label{b1}\\                                       
	&g_4(y)=\cosh(b_1-b_2)\sin(b_1-b_2),\label{b2}
\end{align}
where $b_1$ and $b_2$ are as defined in \eqref{b7}.

 Substitutions of equations \eqref{b3}, \eqref{b4}, \eqref{b1}, and \eqref{b2} into equation \eqref{a2} results in the steady periodic velocity field for oscillatory Couette flow for the single-layer fluid, \eqref{a2sc1}.

Again, the transient velocity field for the upper fluid, \eqref{6}, contains $F_4 (k_m )$, which is defined in \eqref{e3}. If we put $h=0$(meaning that the lower fluid ceases to exist), $\mu_1=\mu_2=\mu$(say the viscosity of the single-layer fluid), and $\nu_1=\nu_2=\nu$(say the kinematic viscosity of the single-layer fluid)  in equation \eqref{e3}, $F_4 (k_m )$ reduces to that defined in equation \eqref{b11}. Substitution of equation \eqref{b11} into \eqref{6} yields the transient velocity field for oscillatory Couette flow for the single-layer fluid, \eqref{6tc1}.
\bibliographystyle{unsrtnat}
\bibliography{references} 

\end{document}